\documentclass{aa}
\usepackage[stable]{footmisc}
\usepackage{amsmath}
\usepackage[varg]{txfonts}
\usepackage{placeins}
\usepackage{natbib}
\bibpunct{(}{)}{;}{a}{}{,}
\usepackage{graphicx}
\usepackage{epstopdf}
\usepackage{ifthen}
\usepackage{mathtools}
\usepackage[breaklinks, colorlinks, citecolor=blue]{hyperref}
\usepackage{fixltx2e}
\usepackage{url}
\usepackage{hyperref}
\usepackage{enumitem}
\usepackage[table,usenames,dvipsnames]{xcolor}
\usepackage{comment} 
\usepackage{datetime} 
\usepackage{soul}
\usepackage{hhline}
\usepackage{arydshln}

\def\setsymbol#1#2{\expandafter\def\csname #1\endcsname{#2}}
\def\getsymbol#1{\csname #1\endcsname}

\def\Planck{\textit{Planck}}

\newbox\tablebox    \newdimen\tablewidth
\def\leaderfil{\leaders\hbox to 5pt{\hss.\hss}\hfil}
\def\endPlancktable{\tablewidth=\columnwidth 
    $$\hss\copy\tablebox\hss$$
    \vskip-\lastskip\vskip -2pt}

\def\tablenote#1 #2\par{\begingroup \parindent=0.8em
    \abovedisplayshortskip=0pt\belowdisplayshortskip=0pt
    \noindent
    $$\hss\vbox{\hsize\tablewidth \hangindent=\parindent \hangafter=1 \noindent
    \hbox to \parindent{$^#1$\hss}\strut#2\strut\par}\hss$$
    \endgroup}
\def\doubleline{\vskip 3pt\hrule \vskip 1.5pt \hrule \vskip 5pt}

\def\L2{\ifmmode L_2\else $L_2$\fi}

\def\DeltaT{\ifmmode \Delta T\else $\Delta T$\fi}
\def\deltat{\ifmmode \Delta t\else $\Delta t$\fi}
\def\fknee{\ifmmode f_{\rm knee}\else $f_{\rm knee}$\fi}
\def\Fmax{\ifmmode F_{\rm max}\else $F_{\rm max}$\fi}
\def\solar{\ifmmode{\rm M}_{\mathord\odot}\else${\rm M}_{\mathord\odot}$\fi}
\def\Msolar{\ifmmode{\rm M}_{\mathord\odot}\else${\rm M}_{\mathord\odot}$\fi}
\def\Lsolar{\ifmmode{\rm L}_{\mathord\odot}\else${\rm L}_{\mathord\odot}$\fi}
\def\inv{\ifmmode^{-1}\else$^{-1}$\fi}
\def\mo{\ifmmode^{-1}\else$^{-1}$\fi}
\def\sup#1{\ifmmode ^{\rm #1}\else $^{\rm #1}$\fi}
\def\expo#1{\ifmmode \times 10^{#1}\else $\times 10^{#1}$\fi}
\def\,{\thinspace}
\def\lsim{\mathrel{\raise .4ex\hbox{\rlap{$<$}\lower 1.2ex\hbox{$\sim$}}}}
\def\gsim{\mathrel{\raise .4ex\hbox{\rlap{$>$}\lower 1.2ex\hbox{$\sim$}}}}

\def\simprop{\mathrel{\raise .4ex\hbox{\rlap{$\propto$}\lower 1.2ex\hbox{$\sim$}}}}
\def\deg{\ifmmode^\circ\else$^\circ$\fi}
\def\pdeg{\ifmmode $\setbox0=\hbox{$^{\circ}$}\rlap{\hskip.11\wd0 .}$^{\circ}
          \else \setbox0=\hbox{$^{\circ}$}\rlap{\hskip.11\wd0 .}$^{\circ}$\fi}
\def\arcs{\ifmmode {^{\scriptstyle\prime\prime}}
          \else $^{\scriptstyle\prime\prime}$\fi}
\def\arcm{\ifmmode {^{\scriptstyle\prime}}
          \else $^{\scriptstyle\prime}$\fi}
\newdimen\sa  \newdimen\sb
\def\parcs{\sa=.07em \sb=.03em
     \ifmmode \hbox{\rlap{.}}^{\scriptstyle\prime\kern -\sb\prime}\hbox{\kern -\sa}
     \else \rlap{.}$^{\scriptstyle\prime\kern -\sb\prime}$\kern -\sa\fi}
\def\parcm{\sa=.08em \sb=.03em
     \ifmmode \hbox{\rlap{.}\kern\sa}^{\scriptstyle\prime}\hbox{\kern-\sb}
     \else \rlap{.}\kern\sa$^{\scriptstyle\prime}$\kern-\sb\fi}
\def\ra[#1 #2 #3.#4]{#1\sup{h}#2\sup{m}#3\sup{s}\llap.#4}
\def\dec[#1 #2 #3.#4]{#1\deg#2\arcm#3\arcs\llap.#4}
\def\deco[#1 #2 #3]{#1\deg#2\arcm#3\arcs}
\def\rra[#1 #2]{#1\sup{h}#2\sup{m}}

\def\dots{\relax\ifmmode \ldots\else $\ldots$\fi}
\def\WHzsr{\ifmmode $W\,Hz\mo\,sr\mo$\else W\,Hz\mo\,sr\mo\fi}
\def\mHz{\ifmmode $\,mHz$\else \,mHz\fi}
\def\GHz{\ifmmode $\,GHz$\else \,GHz\fi}
\def\mKs{\ifmmode $\,mK\,s$^{1/2}\else \,mK\,s$^{1/2}$\fi}
\def\muKs{\ifmmode \,\mu$K\,s$^{1/2}\else \,$\mu$K\,s$^{1/2}$\fi}
\def\muKRJs{\ifmmode \,\mu$K$_{\rm RJ}$\,s$^{1/2}\else \,$\mu$K$_{\rm RJ}$\,s$^{1/2}$\fi}
\def\muKHz{\ifmmode \,\mu$K\,Hz$^{-1/2}\else \,$\mu$K\,Hz$^{-1/2}$\fi}
\def\MJysr{\ifmmode \,$MJy\,sr\mo$\else \,MJy\,sr\mo\fi}
\def\MJysrmK{\ifmmode \,$MJy\,sr\mo$\,mK$_{\rm CMB}\mo\else \,MJy\,sr\mo\,mK$_{\rm CMB}\mo$\fi}
\def\microns{\ifmmode \,\mu$m$\else \,$\mu$m\fi}

\def\muK{\ifmmode \,\mu$K$\else \,$\mu$\hbox{K}\fi}
\def\microK{\ifmmode \,\mu$K$\else \,$\mu$\hbox{K}\fi}
\def\muW{\ifmmode \,\mu$W$\else \,$\mu$\hbox{W}\fi}
\def\kms{\ifmmode $\,km\,s$^{-1}\else \,km\,s$^{-1}$\fi}
\def\kmsMpc{\ifmmode $\,\kms\,Mpc\mo$\else \,\kms\,Mpc\mo\fi}
\providecommand{\sorthelp}[1]{}

\def\NHUNIT{\ifmmode {\rm \,cm^{-2}} \else $\rm \,cm^{-2}$ \fi} 

\def\aap{{A\&A}}

\def\apjs{{ApJ Supp.}}
\def\muKcmb{\ifmmode \,\mu$K$_{\rm CMB}$\else \,$\mu$K$_{\rm CMB}$\fi}

\newcommand{\planck}{\Planck}

\newcommand{\OmegaM}{\ifmmode\Omega_{\rm M}\else $\Omega_{\rm M}$\fi}

\providecommand{\Planck}{\textit{Planck}}
\providecommand{\planck}{\Planck}

\providecommand{\text}[1]{\rm{#1}}

\newcommand{\Msun}{M_\odot}

\renewcommand{\d}{\text{d}}

\providecommand{\muK}{\mu\rm{K}}

\newcommand{\begm}{\begin{pmatrix}}
\newcommand{\enm}{\end{pmatrix}}

\def\pmb#1{\setbox0=\hbox{#1}%
    \kern-.025em\copy0\kern-\wd0
    \kern.05em\copy0\kern-\wd0
    \kern-.025em\raise.0433em\box0}

\def\p2Y{\;_2Y}
\def\m2Y{\;_{-2}Y}
\def\beglet{
  \addtocounter{equation}{1}%
  \setcounter{parentequation}{\value{equation}}%
  \setcounter{equation}{0}%
  \def\theequation{\arabic{parentequation}\alph{equation}}%
  \ignorespaces
}
\def\endlet{
  \setcounter{equation}{\value{parentequation}}%
  \def\theequation{\arabic{equation}}%
}
\providecommand{\beglet}{\begin{subequations}}
\providecommand{\endlet}{\end{subequations}}

\newcommand{\mksym}[1]{\ifmmode {\rm #1}\else #1\fi}

\providecommand{\text}[1]{\rm{#1}}

\renewcommand{\d}{\text{d}}

\providecommand{\muK}{\mu\rm{K}}

\newcommand\ba{\begin{eqnarray}}
\newcommand\ea{\end{eqnarray}}
\newcommand\bea{\begin{eqnarray}}
\newcommand\eea{\end{eqnarray}}

\newcommand\be{\begin{equation}}
\newcommand\ee{\end{equation}}

\newcommand{\camb}{{\tt camb}}

\newcommand{\ghz}{\rm GHz}

\newcommand{\shot}{\rm SN}
\newcommand{\cib}{\rm CIB}

\begin{document}

\title{A halo model approach to describe clustering and emission of the two main star forming galaxy populations for Cosmic Infrared Background studies}

\author{
G. Zagatti\orcid{0009-0003-9595-1158}\inst{1,2}~\thanks{\email{giorgia.zagatti@unife.it}}
\and
E. Calabrese\orcid{0000-0003-0837-0068}\inst{3}
\and
C. Chiocchetta\inst{1,2}
\and
M. Gerbino\orcid{0000-0002-3538-1283}\inst{2,1}
\and
M. Negrello\inst{3}
\and
L. Pagano\orcid{0000-0003-1820-5998}\inst{1,2,4}
}

\institute{\small Dipartimento di Fisica e Scienze della Terra, Universit\`a degli Studi di Ferrara, via Saragat 1, I-44122 Ferrara, Italy\goodbreak
\and
Istituto Nazionale di Fisica Nucleare, Sezione di Ferrara, via Saragat 1, I-44122 Ferrara, Italy\goodbreak
\and
School of Physics and Astronomy, Cardiff University, The Parade, CF24 3AA Cardiff, Wales, UK\goodbreak
\and
Institut d'Astrophysique Spatiale, CNRS, Univ. Paris-Sud, Universit\'{e} Paris-Saclay, B\^{a}t. 121, 91405 Orsay cedex, France}

\date{\vglue -1.5mm \today \vglue -5mm}

\abstract{The \textit{Cosmic Infrared Background} (CIB), traced by the emission from dusty star-forming galaxies, provides a crucial window into the phases of star formation throughout cosmic history. These galaxies, although challenging to detect individually at high redshifts due to their faintness, cumulatively contribute to the CIB which then becomes a powerful probe of galaxy formation, evolution and clustering. Here, we introduce a physically-motivated model for the CIB emission spanning a wide range of frequency and angular resolution, employing a halo model approach and distinguishing, within dark matter halos, between two main populations of star forming galaxies, i.e. normal late-type spiral and irregular galaxies and the progenitors of early-type galaxies. The emission from two galaxy populations maps into different regimes in frequency/resolution space, allowing us to constrain the clustering parameters of the model -- $M_{\text{min}}$, the mass of a halo with 50\% probability of having a central galaxy and $\alpha$, the power law index regulating the number of satellite galaxies --  through a fit to \planck\,and \textit{Herschel}-SPIRE CIB anisotropy measurements. We find that, while being able to place constraints on some of the clustering parameters, the \planck\,frequency and multipole coverage cannot effectively disentangle the contributions from the two galaxy populations. On the other side, the \textit{Herschel}-SPIRE measurements separate out and constrain the clustering of both populations. Our work, though, highlights an inconsistency of the results between the two datasets, partially already reported in other literature and still not understood.
}


\authorrunning{Zagatti et al.}

\titlerunning{A halo model approach to describe clustering and emission of the two main SFG populations for CIB studies}

\maketitle
\section{Introduction}\label{sec:intro}
The \textit{Cosmic Infrared Background} (CIB) represents the emission from dusty star-forming galaxies across the cosmic history, appearing as a diffuse, cosmological background. The CIB was firstly detected by \cite{Puget:1996fx} from \textit{FIRAS} on the \textit{COBE} satellite~\citep{Fixsen:1996nj}, and partially resolved into light from individual galaxies using SCUBA on the James Clerk Maxwell Telescope \citep{Smail:1997wn, Hughes:1998sx, Eales:1998fn}.

Detailed studies of CIB anisotropies have been conducted using e.g., power spectra measured from the South Pole Telescope (SPT, \citealp{Hall:2009rv}), \planck\, \citep{planck2013-pip56} and \textit{Herschel}-SPIRE \citep{Viero:2012uq} data. These observations provide for valuable cosmological and astrophysical information that both complements and extends beyond observations of individual sources, since at high redshifts most of the light originates from sources at or below the confusion limit. Moreover, there is a relation, which can be modeled through a halo model formalism, between the dusty star-forming galaxies and the dark matter halos in which they reside. Particularly, anisotropies in this cosmological background trace the underlying dark matter halo distribution, thus probing the clustering properties of galaxies. Starting from the perturbations in the initial density field and through the spherical collapse model, the halo model predicts how dark matter halos are distributed in the Universe. Integrating this framework with information on how galaxies populate dark matter halos links observations of the CIB with the large-scale structure (LSS) of the Universe.

Knowing how dark matter halos are distributed in the Universe, by incorporating a model of galaxy distribution within these halos, along with precise CIB observations over broad frequency and multipole ranges, it is possible to constrain the clustering parameters of galaxies. The \planck\, and \textit{Herschel}-SPIRE experiments have been instrumental in this regard, offering insights into the frequency-dependent distribution of cosmic infrared light and its anisotropies. \planck, characterized by a broad frequency coverage (in particular with the high end of the microwave range, 143-857 GHz), and \textit{Herschel}-SPIRE, focused on higher-resolution submillimeter wavelengths, have both contributed to a more comprehensive understanding of the CIB and its underlying physical drivers. 

A crucial aspect at this point is the choice of the specific model for the galaxies distribution. Previous works have used SPT, \planck\, and SPIRE CIB data to model clustering properties in various ways, including empirical models \citep{Lagache:2002xq}, power-law representations of the CIB power spectrum \citep{Mak:2016ykk}, as well as the common approach of populating dark matter halos with a single galaxy population \citep{Maniyar:2018xfk, Maniyar:2020tzw}. In particular, for the latter scenario many attempts have been made in the context of the halo model, also studying series of halo models that differ by the treatment of the Spectral Energy Distribution (SED) \citep{Viero:2012uq}. Other works, based on the observed dichotomy between elliptical and spiral galaxies, have expanded the model from single to two galaxy populations~\citep{Cai:2013wna, Xia:2011dt}.

In this study, we adopt the approach proposed by \cite{Cai:2013wna} (hereafter C13), which describes star forming galaxies with two populations, and we import it in a full halo model framework. We populate dark matter halos with late-type (LT) galaxies and with the progenitors of early-type (ET) galaxies. 
LT galaxies are characterized by young stellar populations ($\lsim7-8\,$Gyr), corresponding to formation redshifts $z\lsim1-1.5$, low to moderate star formation rates ($\lsim10\,M_{\sun}\,$yr$^{-1}$), and they reside in less massive dark matter halos. 
On the other hand, ET galaxies exhibit old stellar population ($\gsim8-9\,$Gyr), indicative of formation redshifts $z\gtrsim 1.5$, low to null star formation rates, and reside in more massive halos. However, during their formation phase, ET galaxies experience intense star formation activity (SFR up to 100-1000\,M$_{\sun}\,$yr$^{-1}$) and are characterised by high dust content, thus showing up as dust-obscured star forming objects, which dominate the peak of the cosmic star formation activity in the Universe at $z\sim2$. It is this early, highly star forming, dust-obscured phase, that we will consider in here for ET galaxies, although, for simplicity, we will keep using the acronym ET to refer to the {\it proto-spheroidal} galaxies. 
The choice of populating dark matter halos with ET and LT galaxies, together with the fact that the halo model formalism naturally predicts the correlation between the two galaxy populations at the level of the galaxy power spectrum - and, consequently, of the CIB - is crucial to achieve a more realistic and coherent description of the CIB emission over a wide range of frequencies and angular scales.

Our analysis extends the halo model to account for the specific luminosity functions and clustering properties of these galaxy populations, leading to a refined prediction of the CIB power spectrum.

This paper is organized as follows: Section \ref{sec:model} delves into the halo model formalism, detailing the derivation of both dark matter and galaxy power spectra in the nonlinear regime. Section \ref{sec:CIB} describes the models for galaxy emission, tailored to ET and LT galaxies, and their implications for the CIB. Section \ref{sec:dataset} describes the three datasets employed in this study, while Section \ref{sec:results} discusses the results from fitting our model predictions to the observational data from \planck\,and SPIRE, focusing on the inferred clustering parameters of the two galaxy populations. The software developed for this analysis is publicly available\footnote{ \url{https://github.com/giorgiazagatti/CIB_halomodel}}.

\section{The Halo Model}\label{sec:model}
As mentioned above, in this work we use the halo model approach (see e.g., \citealp{Asgari:2023mej,Cooray:2002dia} for a complete review) to describe the non-linear behaviour of the matter density field. 
One of the main assumptions of the halo model is that all matter in the Universe resides within virialized structures, the dark matter halos. This assumption allows to construct the dark matter power spectrum by adding together the contributions coming from different halos. In this section, we summarize the fundamentals of the halo model, beginning with its most general building blocks, which will then be extended to include dark matter and galaxies.

A real-space field \textit{u} can be represented as follows:
\begin{equation}
    \theta_u(\Vec{x}) = \displaystyle\sum_i N_i W_{u,i}(|\Vec{x} - \Vec{x}_i|) ,
\end{equation}
where the sum runs over the volume elements, with $N_i = {0,1}$ denoting whether the volume element is occupied ($i=1$) or not ($i=0$) by a halo, and $W_{u,i}$ representing the density profile of the field under study. In our case, the general field \textit{u} could be represented by dark matter, halo or galaxy.
It is convenient to express the above equation in Fourier space:
\begin{equation}
    \hat{\theta}_u(\Vec{k}) = \displaystyle\sum_i e^{-i \Vec{k} \cdot \Vec{x_i}} N_i \hat{W}_{u,i}(M,k) , 
\end{equation}
where $\hat{W}_{u,i}(M,k)$ is the Fourier transform of the halo profile, assuming spherical symmetry and a mass-independent halo. Additionally, it is convenient to split the expression of the Fourier transform of the halo profile in two contributions, $\hat{W}_{u}(M,k) = W(M)u(M,k)$, where $W(M)$ contains the information about the amplitude of the profile and $u(k,M)$ is the normalized Fourier transform of the halo profile.

The correlation between two fields \textit{u} and \textit{v} is given by:
\begin{equation}
    \Bigl<\hat{\theta}_u(\Vec{k})\hat{\theta}_v^*(\Vec{k}^\prime)\Bigr> = \Bigl<\displaystyle\sum_{i,j} e^{-i \Vec{k} \cdot \Vec{x_i}} e^{i \Vec{k}^\prime \cdot \Vec{x_j}} N_i N_j \hat{W}_{u,i}(M,k) \hat{W}_{v,i}(M,k^\prime)\Bigr> . 
\end{equation}
Splitting the above sum in two parts, i.e, one when $i=j$ and one when $i\neq j$, we obtain the two main contributions to the power spectra in the framework of the halo model -- the one-halo and the two-halo terms. Moving from the ensemble average and the sum to a continuous integral, the one-halo term describes the contribution from $i=j$, i.e., the power coming from regions belonging to the same dark matter halo:
\begin{equation}\label{1halo}
    P_{uv}^{1h}(k) = \int_0^\infty \hat{W}_u(m,k)\hat{W}_v(m,k)\dfrac{\d n}{\d m}\d m ;
\end{equation}
the term with $i\neq j$ is the two-halo term and describes the contributions to the power spectra coming from different dark matter halos:
\begin{equation}\label{2halo}
    P_{uv}^{2h} = P_{mm}^{lin}(k)\prod_{n=u,v}\Biggl[\int_o^\infty \hat{W}_n(m,k)b(m)\dfrac{\d n}{\d m}\d m\Biggr] .
\end{equation}
In both expressions ${\d n}/{\d m}$ is the halo mass function and $b(m)$ in Eq. \eqref{2halo} is the halo bias. $P_{mm}^{lin}(k)$ is the linear matter power spectrum\footnote{We take the linear matter power spectrum from the output of the Boltzmann solver \camb~\citep{Lewis:1999bs,Howlett:2012mh}.}. More details on the halo mass function and the halo bias will be given later in this section.

The total power spectrum of the two fields is finally given by:
\begin{equation}\label{totPS}
    P_{uv}(k) = P_{uv}^{1h}(k) + P_{uv}^{2h}(k) .
\end{equation}

\subsection{The ingredients of the halo model}

\subsubsection{The halo mass function}
The halo mass function ${\d n}/{\d m}(m,z)$ is the comoving number density of bound objects of mass \textit{m} at redshift \textit{z}. It is obtained from the initial matter density field by relating its properties to halos that form later. The relation is usually set by the peak height~\citep{Jenkins:2000bv}:
\begin{equation}\label{nu}
    \nu = \dfrac{\delta_c}{\sigma(m)}.
\end{equation}
Here $\delta_c$ is the critical linear overdensity needed for a region to start the collapse (for an Einstein-De Sitter background, $\delta_c = 1.686$), and $\sigma(m)$ is the variance in the linear matter density field as a function of the mass. 
The halo mass function is usually parameterized in terms of $\nu$ or $\sigma$, rather than explicitly in terms of the mass. The reason is that the $\nu$ or $\sigma$ parametrizations show a close-to-universal behaviour as a function of cosmology and redshift~\citep{Press:1973iz,Bond:1990iw,Sheth:1999mn,Tinker:2008ff}. The relation between the halo mass function and the peak height reads as:
\begin{equation}\label{hmf}
    \dfrac{m^2\dfrac{\d n}{\d m}(m,z)}{\overline{\rho}}\dfrac{\d m}{m} = \nu f(\nu) \dfrac{\d \nu}{\nu}, 
\end{equation}
where $\overline{\rho}$ is the comoving background density and $\nu f(\nu)$ is the number density of peaks. 

The literature presents various formulations of the halo mass function, expressed in terms of either $\nu$ or $\sigma$ (e..g, \citealp{Sheth:1999mn, Jenkins:2000bv, Warren:2005ey, Reed:2006rw, Peacock:2007cw, Tinker:2008ff, Tinker:2010my, Crocce:2009mg, Bhattacharya:2010wy,Courtin:2010gx,Watson:2012mt,Despali:2015yla,McClintock:2018uyf,Bocquet:2020tes}). In this study we adopt the halo mass function presented by~\cite{Tinker:2008ff} since it is better aligned to the halo mass function evaluated from N-body simulations and is parameterized in terms of $\sigma$. In particular, expressing Eq. \eqref{hmf} in terms of $\sigma$, rather than $\nu$, we obtain:
\begin{equation}\label{hmf tinker}
    \dfrac{\d n}{\d\ln m} = -\dfrac{1}{2}f(\sigma,z)\dfrac{\overline{\rho}}{m}\dfrac{\d\ln\sigma^2}{\d\ln m}, 
\end{equation}
where, following \cite{Tinker:2008ff}, $f(\sigma,z)$ takes the following form:
\begin{equation}\label{func hmf}
    f(\sigma,z)=A(z)\left[\left(\dfrac{\sigma}{b(z)}\right)^{-a(z)} + 1\right]e^{-c(z)/\sigma^2} .
\end{equation}
Here the parameters \textit{A, a, b} and \textit{c}, are the amplitude of the mass function, the amplitude and the slope of the low mass part of the halo mass function and the cut-off scale, respectively, and they are functions of the density contrast at the collapse and the redshift. The density contrast, $\Delta$, is defined as the ratio between the critical overdensity and the comoving background density, i.e., $\Delta = \delta_c/\overline{\rho}$. The redshift evolution of these parameters is due to two contributions; the first one comes from a weak dependence of the density contrast on cosmology, while the second is the explicit redshift dependence parameterized by \cite{Tinker:2008ff}.

\subsubsection{The halo bias}
Dark matter halos are biased tracers of the underling matter distribution. For this reason, when using the halo model to describe the overall matter power spectrum, we have to include a bias term capable of relating the overdensity of halos to the mass overdensity. At the lowest order, the relation between the two overdensities is linear~\citep{Mo:1995cs}:
\begin{equation}
    \delta_h(m,z_1|M,V,z_0) = b(m,z_1)\delta ,
\end{equation}
where $\delta_h(m,z_1|M,V,z_0)$ is the number density of halos of mass \textit{m} formed at redshift $z_1$ in a cell of comoving volume \textit{V}, mass \textit{M}, today, i.e. $z_0$, $\delta$ is the mass overdensity, and $b(m,z_1)$ is the halo bias. In this work, we adopt the model proposed in \cite{Tinker:2009jp} for the halo bias which shows a good agreement with simulations for a wide range of overdensities ($200<\Delta<3200$). The equation of the bias term used in \cite{Tinker:2009jp} is:
\begin{equation}\label{halo bias}
    b(m,z_1) = 1 - A\dfrac{\nu^a}{\nu^a+\delta_c^a} + B\nu^b + C\nu^c ,
\end{equation}
where the constants in the equation above, for a $\Delta=200$ halo, are:
\begin{align*}
    A=1.04,\,a=0.132,\,B=0.183,\,b=1.5,
    C=0.262,\,c=2.4.  
\end{align*}

\subsubsection{The halo density profile}
The last ingredient of the halo model is the halo density profile describing the radial distribution of the density of dark matter halos. The inner structure of dark matter halos depends on their formation time and on the initial density distribution of the collapsed region. Since we define the halos as peaks in the initial density field, the higher the peak the more massive the corresponding halo is. Furthermore, the density around a higher peak is shallower with respect to the density around a lower peak. Consequently, smaller halos are more concentrated. The most common model for the density profile of dark matter halos is the Navarro-Frenk-White (NFW) profile~\citep{Navarro:1996gj}:
\begin{equation}\label{NFW profile}
    \rho(r)=\dfrac{\rho_s}{r/r_s(1+r/r_s)^2} ,
\end{equation}
where $r_s$ and $\rho_s$ are the scale radius and the scale density, respectively. To prevent divergence at large radii, the profile is typically truncated at the halo radius, $r_h$, evaluated as:
\begin{equation}
    M=\dfrac{4}{3}\pi r_h^3 \Delta_h\overline{\rho} .
\end{equation}
The ratio between the scale radius and the halo radius is the concentration parameter:
\begin{equation}
    c = \dfrac{r_h}{r_s} .
\end{equation}
Simulations show that the concentration of halos of the same mass is described by a log-normal distribution:
\begin{equation}
    p(c|m,z) \d c = \dfrac{\d\ln c}{\sqrt{2\pi \sigma_{\ln c}^2}}\exp\Biggl\{\frac{\ln^2[c/\overline{c}(m,z)]}{2\sigma_{\ln c}^2}\Biggr\} ,
\end{equation}
where $\overline{c}$ is the mean concentration parameter, which depends on the halo mass and redshift, and $\sigma_c^2$ is the width of the distribution. Several formulations exist for the mean concentration parameter (e.g.,~\citealp{Navarro:1996gj,Bullock:1999he,Eke:2000av,Neto:2007vq,Maccio:2008pcd,Duffy:2008pz,Prada:2011jf,Kwan:2012nd,Ludlow:2013vxa,Diemer:2014gba,Correa:2015dva,Okoli:2015dta,Ludlow:2016ifl,Child:2018skq,Diemer:2018vmz}). In this work, we adopt the parametrization by \cite{Bullock:1999he}:
\begin{equation}\label{conc param}
    \overline{c}(m,z) = \dfrac{9}{1+z}\Biggl[\dfrac{m}{m_*(z)}\Biggr]^{-0.13}, 
\end{equation}
where $m_*(z)$ is a characteristic mass scale where $\nu(m,z)=1$ ($\nu$ has been defined in Eq.~\eqref{nu}).

\subsubsection{Validation tests}
All the ingredients listed in the previous subsections represent the building blocks for computing the non-linear matter and galaxy power spectra within the framework of the halo model.

To ensure the robustness of our code, we validated it by comparing our predictions with the output from both {\tt TheHaloMod} \citep{Murray:2020dcd} and {\tt pyhalomodel} \citep{Asgari:2023mej}. All the individual predictions of the different terms in our code are in agreement with those of {\tt TheHaloMod} and {\tt pyhalomodel}. These comparisons confirm that our code produces reliable results and can be used for further analysis.
Specifically, we compared the halo mass functions at different redshifts from our code for a given formalism (Eqns.~\eqref{hmf tinker} - \eqref{func hmf}) with the predictions from the two publicly available codes. To evaluate the robustness of the computation of the NFW profile (Eq.~\eqref{NFW profile}), we also compared the implementation of the concentration parameter (Eq.~\eqref{conc param}). After testing the accuracy of the concentration parameter computation, we focused on the implementation of the Fourier transform of the NFW profile by conducting various tests. In particular, we compared the predictions of our code with those of {\tt TheHaloMod} and {\tt pyhalomodel} for fixed halo masses at different redshifts and for fixed redshifts at different halo masses. Additionally, we tested the predictions for the Fourier transform of the NFW profile with and without normalization. The final component in the computation of the non-linear power spectra that we tested by comparing the outputs from the different codes is the halo bias (Eq.~\eqref{halo bias}).  

\subsection{Matter power spectrum}
The ingredients described previously are needed to compute the matter power spectrum in its non-linear regime which we assemble here. 
We first need the Fourier transform of the halo density profile as a function of the mass of the halo:
\begin{equation}
    \hat{W}_m(m,k) = \dfrac{m}{\overline{\rho}} u(k,m) ,
\end{equation}
where $m/\overline{\rho}$ is the amplitude of the matter profile, $\overline{\rho}$ represents the mean comoving cosmological matter density and $u(k,m)$ is the normalized Fourier transform of the halo density profile (Eq.~\eqref{NFW profile}).
Incorporating this relation into the expressions for the 1h (Eq.~\eqref{1halo}) and 2h (Eq.~\eqref{2halo}) terms yields:
\begin{align}\label{1h halo mm}
    P_{mm}^{1h}(k) & = \int_0^{\infty} \mathrm{\d }m \, \dfrac{\d n}{\d m} \left(\dfrac{m}{\overline{\rho}}\right)^2|u(k,m)|^2, \\ \label{2h halo mm}
    P_{mm}^{2h}(k) & = P_{mm}^{lin}(k)\Biggl[\int_0^{\infty} \d m\, b(m) \dfrac{\d n}{\d m} \dfrac{m}{\overline{\rho}}|u(k,m)|\Biggr]^2. 
\end{align}

\subsection{Galaxy power spectrum}
Once we know how dark matter halos are distributed in the Universe, we can extend the halo model formalism also to galaxies, ending up with the non linear galaxy power spectrum. One of the predictions of the halo model is the fact that baryonic gas cools and, consequently, forms stars only in virialized dark matter halos. Thus, the number of galaxies within a dark matter halo is related to the mass of the hosting halo itself. Furthermore, we work under the assumption that the first galaxy forms at the centre of the hosting halo and we refer to it as central galaxy, and all the other are satellite galaxies.

In order to understand how central and satellite galaxies populate a given dark matter halo of mass $M$, we have to specify the Halo Occupation Distribution (HOD). There are many models for the HOD in literature (see e.g.,~\citealp{SDSS:2004oes,Scoccimarro:2000gm,Berlind:2002rn}), in this study we use the one proposed in \cite{Tinker:2009mx}, which models the number of central galaxies as:
\begin{equation}\label{N_cent}
    <N_{\text{cent}}|M> = \dfrac{1}{2}\Biggl[1+\text{erf}\Biggl(\dfrac{\log M-\log M_{\text{min}}}{\sigma_{\log M}}\Biggr)\Biggr] ,
\end{equation}
and the number of satellite galaxies as:
\begin{equation}\label{N_sat}
    <N_{\text{sat}}|M> = \dfrac{1}{2}\Biggl[1+\text{erf}\Biggl(\dfrac{\log M-\log2M_{\text{min}}}{\sigma_{\log M}}\Biggr)\Biggr]\Biggl(\dfrac{M}{M_{\text{sat}}}\Biggr)^\alpha .
\end{equation}
In the equations above, $M_{\text{min}}$ represents the mass of a halo with $50\%$ probability of having a central galaxy. The factor of 2 before $M_{\text{min}}$ in Eq.~\eqref{N_sat} is to avoid dark matter halos having a larger probability of hosting satellites than central galaxies; $\sigma_{\log M}$ regulates the transition in the dark matter halo from having no galaxies to having one central galaxy; $M_{\text{sat}}$ is the minimum mass for a dark matter halo in order to host a satellite galaxy and $\alpha$ is the power law index regulating the number of satellite galaxies.

We populate dark matter halos with two galaxy populations: early-type (ET) and late-type (LT) galaxies. As mentioned before, ET galaxies have star formation redshifts $z_{\text{SF}}^{\text{ET}} \gtrapprox 1.5$, while LT galaxies have star formation redshifts $z_{\text{SF}}^{\text{LT}} \lsim1-1.5$. Values of the clustering parameters of the two galaxy populations estimated in different analyses from \cite{planck2011-6.6} (hereafter P11), \cite{Xia:2011dt} (hereafter X12) and C13, are reported in Table \ref{tab: literature values clust}. The main difference between these previous works is in the modeling of the galaxy populations. P11 fits for one single galaxy population, allowing the clustering parameters to vary depending on the frequency channel (this frequency dependence explains the reason why in Table \ref{tab: literature values clust}, for P11, we report a range of values for the clustering parameters rather than a single estimate). X12 and C13 consider two galaxy populations but use different frequency ranges to constrain the clustering parameters. Specifically, X12 works in the 150-1200 GHz frequency range, while C13 in the 600-3000 GHz frequency range. Pushing to higher frequencies, C13 was capable of constraining the minimum mass for the LT galaxies. Both works keep the $\alpha_{\text{LP}}$ clustering parameter fixed to unity to reduce the number of free parameters in the fit. As stated in P11, the other clustering parameters which we do not report in Table \ref{tab: literature values clust}, i.e., $\sigma_{\log M}$ and $M_{\text{sat}}$, are not critical parameters for the fit. We tested this assumption by taking a wide range of possible values for $\sigma_{\log M}$ and $M_{\text{sat}}$~\citep{planck2011-6.6, Tinker:2009jp} and calculating the CIB power spectrum for different frequencies and found no visible impact from these two clustering parameters. For this reason, we also fix them to $\sigma_{\log M} = 0.1$ and $M_{\text{sat}} = 20M_{\text{min}}^{\text{EP/LP}}$ respectively, for both galaxy populations.
\begin{table}[t!]
	\begingroup
	\nointerlineskip
	\vskip -3mm
	\footnotesize
	\setbox\tablebox=\vbox{
		\newdimen\digitwidth
		\setbox0=\hbox{\rm 0}
		\digitwidth=\wd0
		\catcode`*=\active
		\def*{\kern\digitwidth}
		\newdimen\signwidth
		\setbox0=\hbox{+}
		\signwidth=\wd0
		\catcode`!=\active
		\def!{\kern\signwidth}
		\halign{\hbox to 0.8in{#\leaderfil}\tabskip=1em&
			\hfil#\hfil\tabskip=10pt&
			\hfil#\hfil\tabskip=10pt&
			\hfil#\hfil\tabskip=10pt&
			\hfil#\hfil\tabskip=10pt&
			\hfil#\hfil\tabskip=10pt&
			\hfil#\hfil\tabskip=0pt\cr
			\noalign{\doubleline}
			\noalign{\vskip 3pt}
            \omit   & X12 & P11 & C13\\[1ex]\cr
			\noalign{\doubleline}
 $\log(M_{\text{min}}^{\text{EP}}/\Msun   h^{-1})$ & $12.09\pm0.06$ & $11.95\pm2.10 -$ & $12.00\pm0.04$\cr
                     \omit           &                & $12.21\pm0.51$   &  \cr
 $\alpha_{\text{EP}}$ & $1.81\pm0.04$ & $1.02\pm0.87 -$ & $1.55\pm0.05$\cr
       \omit   &               & $1.30\pm1.16$   & \cr
 $\log(M_{\text{min}}^{\text{LP}}/\Msun   h^{-1})$ & $\equiv 10.85$ & - & $10.85\pm0.06$\cr
 $\alpha_{\text{LP}}$ & $\equiv 1$ & - & $\equiv 1$\cr
	\noalign{\vskip 5pt\hrule\vskip 3pt}}}
 \endPlancktable
\vskip 3pt
\caption{Constraints on key clustering parameters from previous analyses in X12, P11 and C13.}
\label{tab: literature values clust}
 \endgroup
\end{table}

We now extend the halo model formalism to obtain the galaxy power spectrum. First of all, since galaxies are a discrete field, the most general expression for the 1h term (see Eq.~\eqref{1halo}) must be slightly modified in order to allow for the fact that a discrete tracer auto-correlates with itself at $r=0$. When moving to Fourier space, this auto-correlation translates into a constant term across all the wave numbers. The 1h term for a single discrete field within the same dark matter halo reads:
\begin{equation}
    \begin{aligned}
        P_{\text{gal}}^{1h}(k) &= \dfrac{1}{\overline{n}_{\text{gal}}^2}\int_0^\infty \d m [N_{\text{gal}}(m)(N_{\text{gal}}(m)-1)u_{\text{gal}}(m,k)u_{\text{gal}}(m,k) \\
        &\quad +N_{\text{gal}}(m)]\dfrac{\d n}{\d m} ,
    \end{aligned}
\end{equation}
where $\overline{n}_{\text{gal}}$ is the average number density of galaxies and $N_{\text{gal}} = N_{\text{cent}} + N_{\text{sat}}$ is the total number of galaxies for a specific galaxy population. 
In this study we work under the common approximation of $u_{\text{gal}}(k,m) = u(k,m)$, which can be justified by the fact that we are populating dark matter halos with a central galaxy and all remaining are satellite galaxies tracing the halo mass~\citep{Peacock:2000qk, Seljak:2000gq}. The second term in the square brackets is the Poisson noise, i.e., the auto-correlation between a discrete field with itself. When modeling the galaxy power spectrum, the Poisson noise represents a spurious contribution that does not reflect the true clustering properties of the galaxy distribution but rather the discreteness of the sample. By subtracting this noise, we ensure that the resulting power spectrum more accurately represents the physical clustering of galaxies. When expressing $N_{\text{gal}}$ in terms of the number of central and satellite galaxies, and neglecting the Poisson noise, the 1h term reduces to:
\begin{equation}
    \begin{aligned}
        P_{\text{gal}}^{1h}(k) = \dfrac{1}{\overline{n}_{\text{gal}}^2} \int_0^\infty \d m \dfrac{\d n}{\d m}[2N_{\text{cent}}N_{\text{sat}}u(m,k)+N_{\text{sat}}^2u(k,m)^2] .
    \end{aligned}
\end{equation}
The expression for the 2h term in case of discrete fields does not change with respect to the most general Eq.~\eqref{2halo}, which in case of galaxies reads as:
\begin{equation}
    P_{\text{gal}}^{2h}(k) = P_{mm}^{lin}(k)\Biggl[\int_0^\infty \d m \dfrac{\d n}{\d m} b(m) \dfrac{N_{\text{gal}}(m)}{\overline{n}_{\text{gal}}} u(k,m)\Biggr]^2 .
\end{equation}
The expressions for the 1h and the 2h terms apply individually to the ET and the LT galaxies. In order to compute the full galaxy power spectrum, we have to include also the contributions coming from the correlation between the two galaxy populations. When dealing with two different discrete fields, there is no Poisson noise contribution, and the expressions for both the 1h and the 2h terms directly follow from the most general Eqns.~\eqref{1halo} - \eqref{2halo}:
\begin{equation}
    \begin{aligned}
        P_{\text{mix}}^{1h}(k) &= \dfrac{1}{\overline{n}_{\text{gal}}^{\text{ET}}\overline{n}_{\text{gal}}^{\text{LT}}}\int_0^\infty \d m\dfrac{\d n}{\d m}[(N_{\text{cent}}^{\text{ET}}N_{\text{sat}}^{\text{LT}}+N_{\text{sat}}^{\text{ET}}N_{\text{cent}}^{\text{LT}})u +\\
        & N_{\text{sat}}^{\text{ET}}N_{\text{sat}}^{\text{LT}}u^2] ,
    \end{aligned}
\end{equation}
\begin{equation}
    \begin{aligned}
        P_{\text{mix}}^{2h}(k) &= P_{mm}^{lin}\Biggl[\int_0^\infty \d m\dfrac{\d n}{\d m}b(m)\dfrac{N_{\text{gal}}^{\text{ET}}}{\overline{n}_{\text{gal}}^{\text{ET}}}u\Biggr]\times \\
        & \Biggl[\int_0^\infty \d m\dfrac{\d n}{\d m}b(m)\dfrac{N_{\text{gal}}^{\text{LT}}}{\overline{n}_{\text{gal}}^{\text{LT}}}u\Biggr] .
    \end{aligned}
\end{equation}
The galaxy power spectrum is finally given as the sum of the 1h and the 2h terms;
\begin{equation}\label{gg PS}
    P_{\text{gal}}(k) = P_{\text{gal}}^{1h}(k) + P_{\text{gal}}^{2h}(k) ,
\end{equation}
where both the 1h and the 2h terms include the contributions from ET galaxies, LT galaxies and the mixing between the two populations:
\begin{equation}
    P_{\text{gal}}^{1h}(k) = P_{\text{ET}}^{1h}(k) + P_{\text{LT}}^{1h}(k) + P_{\text{mix}}^{1h}(k) , 
\end{equation}
\begin{equation}
    P_{\text{gal}}^{2h}(k) = P_{\text{ET}}^{2h}(k) + P_{\text{LT}}^{2h}(k) + P_{\text{mix}}^{2h}(k) .
\end{equation}
The different terms required for the evaluation of the galaxy power spectrum are as follows:
\begin{equation}
    \begin{aligned}
        P_{\text{ET}}^{1h}(k) &= \dfrac{1}{\left(\overline{n}_{\text{gal}}^{\text{ET}}\right)^2} \int_0^\infty \d m \dfrac{\d n}{\d m}[2N_{\text{cent}}^{\text{ET}}N_{\text{sat}}^{\text{ET}}u(m,k)+\\
        &\quad \left(N_{\text{sat}}^{\text{ET}}\right)^2u(k,m)^2] ,
    \end{aligned}
\end{equation}
\begin{equation}
    \begin{aligned}
        P_{\text{LT}}^{1h}(k) &= \dfrac{1}{\left(\overline{n}_{\text{gal}}^{\text{LT}}\right)^2} \int_0^\infty \d m \dfrac{\d n}{\d m}[2N_{\text{cent}}^{\text{LT}}N_{\text{sat}}^{\text{LT}}u(m,k)+\\
        &\quad \left(N_{\text{sat}}^{\text{LT}}\right)^2u(k,m)^2] ,
    \end{aligned}
\end{equation}
\begin{equation}
    P_{\text{ET}}^{2h}(k) = P_{mm}^{lin}(k)\Biggl[\int_0^\infty \d m \dfrac{\d n}{\d m} b(m) \dfrac{N_{\text{gal}}^{\text{ET}}}{\overline{n}_{\text{gal}}^{\text{ET}}} u(k,m)\Biggr]^2 ,
\end{equation}
\begin{equation}
    P_{\text{LT}}^{2h}(k) = P_{mm}^{lin}(k)\Biggl[\int_0^\infty \d m \dfrac{\d n}{\d m} b(m) \dfrac{N_{\text{gal}}^{\text{LT}}}{\overline{n}_{\text{gal}}^{\text{LT}}} u(k,m)\Biggr]^2 .
\end{equation}

\begin{figure}[ht]
\centering
\includegraphics[width=.5\textwidth]{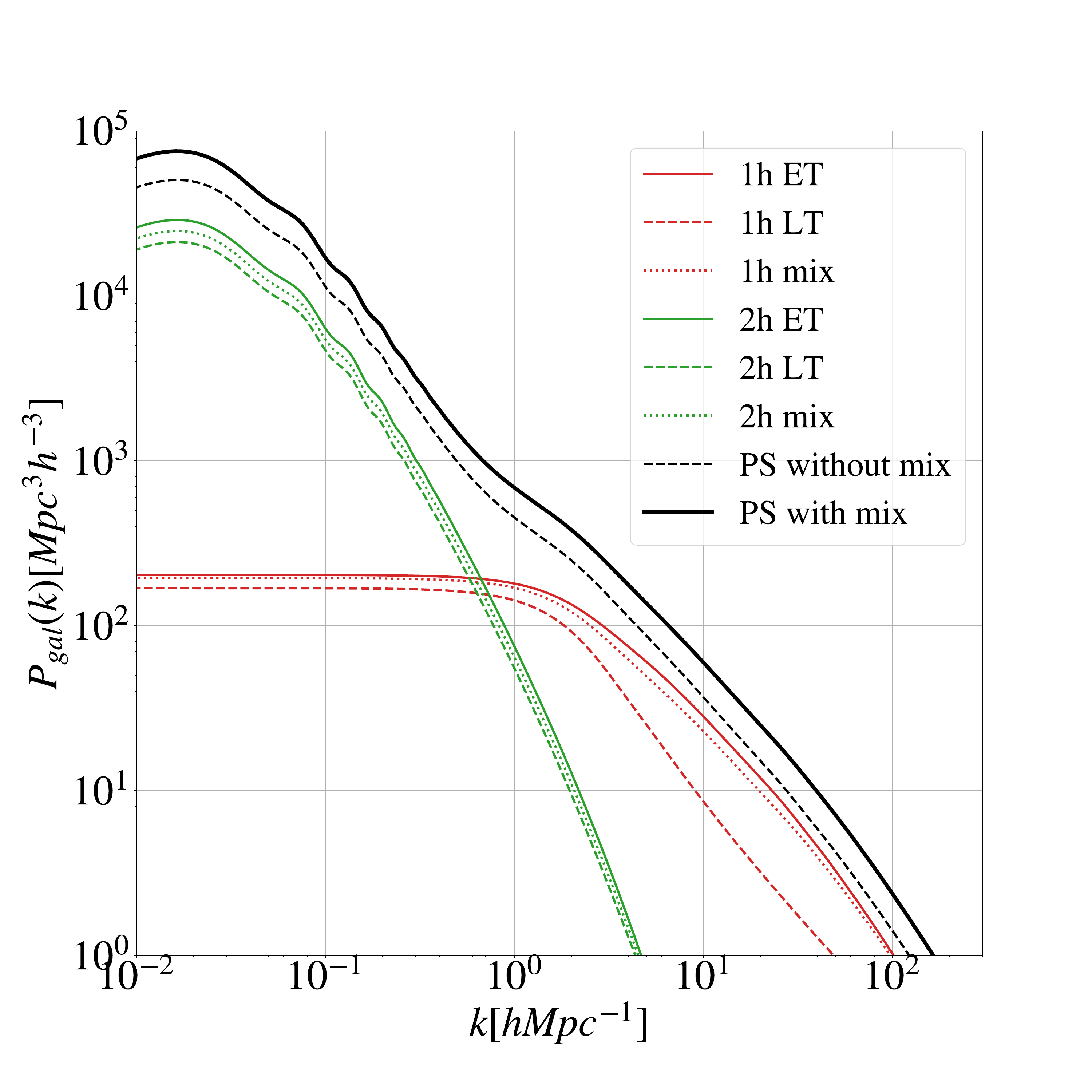}\\
\includegraphics[width=.5\textwidth]{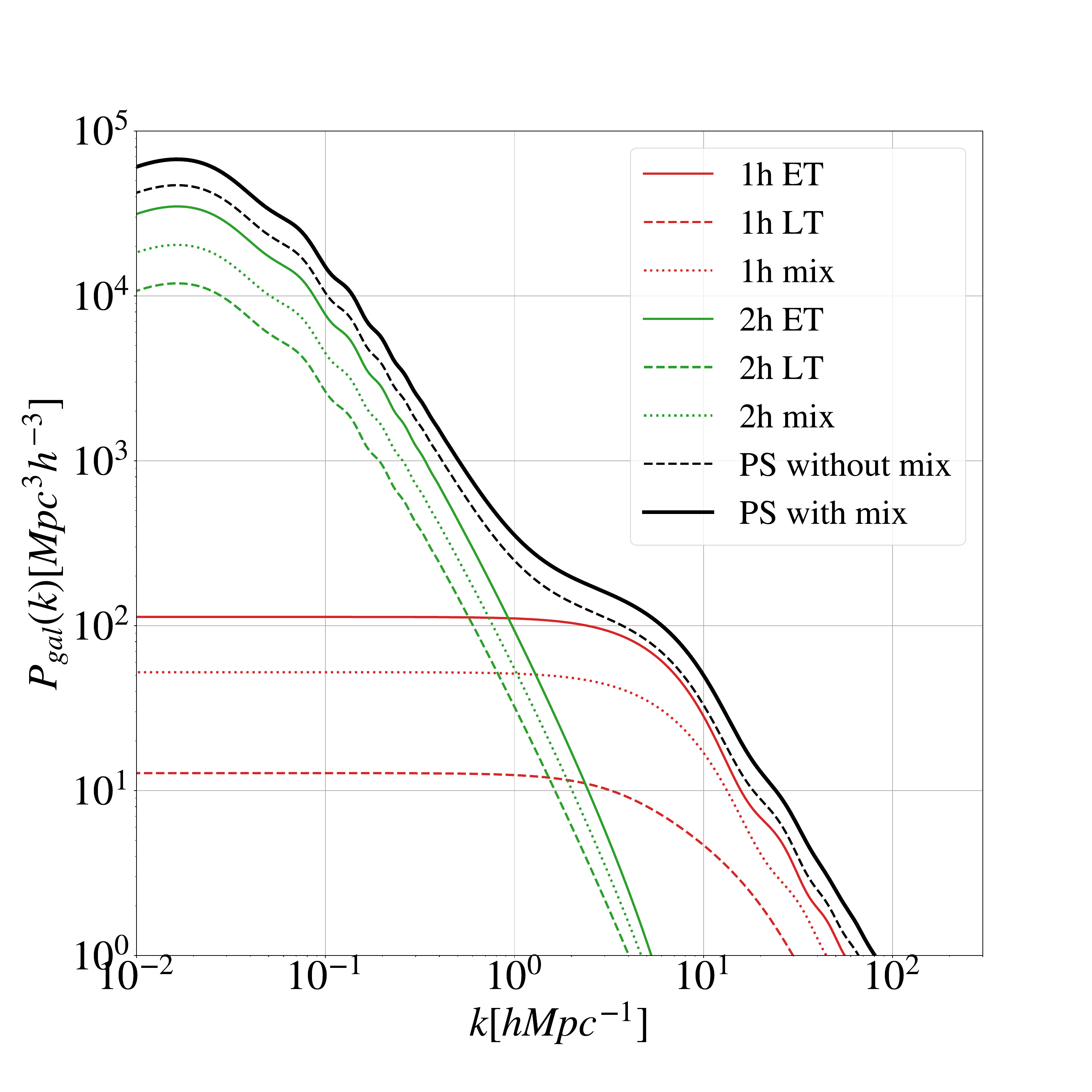}
\caption{\label{fig:gal PS} Galaxy power spectrum, evaluated at redshifts $z=0.001$ (upper panel) and $z=2.0$ (lower panel), broken down into the contributions from one-halo (1h) and two-halo (2h) terms, with (solid) and without (dashed) including the mixing terms between early-type (ET) and late-type (LT) galaxy populations. \underline{Red curves}: 1h terms from ET (\textit{solid}), LT (\textit{dashed}) galaxies, and the mix (1h mix, \textit{dotted}) between the two. \underline{Green curves}: 2h terms from ET (\textit{solid}), LT (\textit{dashed}) galaxies and the mixing (2h mix, \textit{dotted}) between the two. \underline{Black dashed curve}: galaxy power spectrum without mixing terms between the two galaxy populations (PS without mix). \underline{Black solid curve}: galaxy power spectrum with mixing terms (PS with mix).}
\end{figure}

In the upper and lower panel of Fig.~\ref{fig:gal PS} we show the different contributions to the galaxy power spectrum at redshifts $z=0.001$ and $z=2.0$, respectively, and highlight the difference between including (\textit{black solid curve}) and not including (\textit{black dashed curve}) the mixing terms in its computation. The 1h terms (\textit{red curves}) dominate at smaller angular scales, while the 2h terms (\textit{green curves}) are the leading contributions at large angular scales. Focusing on the different contributions to the 1h and the 2h terms coming from ET galaxies (\textit{solid}), LT galaxies (\textit{dashed}) and the mixing between the two galaxy populations (\textit{dotted}), the analysis of the two panels reveals that ET galaxies are always the dominant contribution. Specifically, even though at redshift $z=0.001$ the two contributions are comparable, going to higher redshifts, the contribution of the ET galaxies becomes a factor of $\sim10$ larger than the one of LT. Furthermore, our analysis shows that the difference between the galaxy power spectrum evaluated with and without the mixing terms is more pronounced at smaller redshifts. To our knowledge this is the first time that the galaxy power spectrum is explicitly broken down into two galaxy populations and with each of them including mixing terms. 

\section{The Cosmic Infrared Background emission}\label{sec:CIB}
The halo model provides information about the spatial distribution of galaxies but says nothing about their emission. 
In this section, we take the predictions of the halo model on how galaxies are distributed within dark matter halos and add modeling of the galaxies emission to present a physical description of the \textit{Cosmic Infrared Background} (CIB) power spectrum.  
The key quantity that encapsulates the information on how galaxies emit as a function of redshift is the emissivity function, describing the redshift distribution of the cumulative flux density of sources below the detection limit, $S_{\nu}^{ \text{lim}}$, of the instrument and defined as:
\begin{equation}\label{emissivity}
    j_\nu(z) = \int_0^{S_\nu^{\text{lim}}} \dfrac{\d ^3N(S_\nu,z)}{\d S_\nu \d z \d \Omega}S_\nu \d S_\nu .
\end{equation}
Here $\dfrac{\d ^3N(S_\nu,z)}{\d S_\nu \d z \d \Omega}$ is the surface density of sources per unit flux density and redshift interval: 
\begin{equation}\label{surface density}
    \dfrac{\d ^3N(S_\nu,z)}{\d S_\nu \d z \d \Omega} = \dfrac{\Phi(\log L_{\nu^\prime},z)}{L_{\nu^\prime} \ln{10}} \dfrac{\d L_{\nu^\prime}}{\d S_\nu}\dfrac{\d V}{\d z\d \Omega} ,
\end{equation}
where $\nu^\prime$ and $L_{\nu^\prime}$ are the observed frequency and luminosity, $\dfrac{\d V}{\d z\d \Omega}$ is the comoving volume per unit solid angle and $\Phi(\log L,z)$ is the luminosity function (LF), which quantifies the number of galaxies per unit volume with luminosity between $L$ and $L + \d L$. The observed dichotomy in the galaxy populations enters when modeling the luminosity functions. ET and LT galaxies not only show different clustering properties, but also different emission properties. For this reason, in this work we adopt the hybrid approach, firstly introduced by C13, to model the emissions of the two galaxy populations. 

From N-body simulations, ET galaxies are expected to populate more massive dark matter halos \citep{Wang:2010ik}, for which it is easier to model the evolution. In fact, it is possible to extract an analytic model for the LF of ET galaxies, obtained as a convolution of the halo formation rate, $\d n/\d t_{\text{vir}}$, with the galaxy luminosity distribution, $P(\log L, z)$ (see C13 for the complete overview of the model). To obtain the expression for the halo formation rate, we explicitly write $\d n$ from Eq.~\eqref{hmf} as:
\begin{equation}
    \d n = \dfrac{\overline{\rho}}{m^2}\nu f(\nu)\dfrac{\d\ln{\nu}}{\d \ln{m}} \d m ,
\end{equation}
and the halo formation rate, $\d n/\d t_{\text{vir}}$, has been evaluated as:
\begin{equation}
    \dfrac{\d n}{\d t_{\text{vir}}} = \d n \dfrac{\d \ln{(\nu f(\nu))}}{\d t_{\text{vir}}} .
\end{equation}
The galaxy luminosity distribution is evaluated assuming a log-normal distribution, so that:
\begin{equation}
    P(\log L|\log \overline{L})\d\log L = \dfrac{\exp\{-\log^2(L/\overline{L})/2\sigma^2\}}{\sqrt{2\pi\sigma^2}}\d\log L,
\end{equation}
where $\overline{L}$ and $\sigma$ are the mean luminosity and the dispersion of the distribution, respectively. The LF for the ET galaxies is then obtained as:
\begin{equation}
    \Phi(\log L,z) = \int_{M_{\text{vir}}^{\text{min}}}^{M_{\text{vir}}^{\text{max}}}\d M_{\text{vir}}\int_{z_{\text{vir}}^{\text{min}}}^{z_{\text{vir}}^{\text{max}}}\d z_{\text{vir}}\Biggl|\dfrac{\d t_{\text{vir}}}{\d z_{\text{vir}}} \Biggr|\dfrac{\d n}{\d t_{\text{vir}}}P(\log L,z) .
\end{equation}

Late-type galaxies are expected to populate less massive dark matter halos, which more likely merge with other low-mass dark matter halos. This makes the modeling more challenging \citep{Lapi:2010is}, and for this reason LT galaxies require an empirical parameterization of their LF (see \citealp{Saunders:1990kb}):
\begin{equation}
    \Phi(\log L,z) = \Phi^* \Biggl(\dfrac{L}{L^*}\Biggr)^{1-\alpha} \exp\Biggl\{-\dfrac{\log^2(1+L/L^*)}{2\sigma^2}\Biggr\} ,
\end{equation}
where $\Phi^*$ and $L^*$ are the characteristic density and luminosity, respectively, $\alpha$ is the slope in the low-luminosity regime and $\sigma$ is the dispersion of the galaxy population (see table 1 of C13 for the values of this parameters).

\subsection{CIB power spectrum}
The CIB emission is composed of two different components: a clustering term, describing the galaxy overdensities in the background tracing the dark matter distribution, and a shot noise term, describing the contribution of a diffuse background of unresolved sources.

To compute the clustering term of the CIB power spectrum we integrate the emissivity function, $j_\nu(z)$, describing how galaxies emit (Eq.~\eqref{emissivity}), together with the non-linear galaxy power spectrum from the halo model formalism (Eq.~\eqref{gg PS}):
\begin{equation}\label{CIB clust}
    C_{\ell,\nu\times\nu\prime}^{\text{clust}} = \int \dfrac{\d z}{\chi^2}\dfrac{\d z}{\d\chi}j_\nu(z)j_{\nu\prime}(z)P_{\text{gal}}(k=\ell/\chi,z) .
\end{equation}
As mentioned above the non-linear galaxy power spectrum that enters in Eq.~\eqref{CIB clust} includes three main contributions coming from ET galaxies, LT galaxies and the mixing among the two populations:
\begin{equation}
    P_{\text{gal}}(k) = P_{\text{gal}}^{\text{ET}}(k) + P_{\text{gal}}^{\text{LT}}(k) + P_{\text{gal}}^{\text{mix}}(k) .
\end{equation}
The emissivity functions in Eq.~\eqref{CIB clust} account for the contributions coming from both galaxy populations, \textit{i.e.}:
\begin{equation}
    j_\nu(z) = j_\nu^{\text{ET}}(z) + j_\nu^{\text{LT}}(z) . 
\end{equation}
When taking the product $j_\nu(z)j_{\nu\prime}(z)$ in Eq.~\eqref{CIB clust}, we find:
\begin{equation}
    \begin{aligned}
        j_\nu(z)j_{\nu\prime}(z) &= j_\nu^{\text{ET}}(z)j_{\nu\prime}^{\text{ET}}(z) + j_\nu^{\text{LT}}(z)j_{\nu\prime}^{\text{LT}}(z) + \\
        & \Bigl(j_\nu^{\text{ET}}(z)j_{\nu\prime}^{\text{LT}}(z) + j_\nu^{\text{LT}}(z)j_{\nu\prime}^{\text{ET}}(z)\Bigr) .
    \end{aligned} 
\end{equation}
In summary, we arrive to a clustering term of the CIB power spectrum composed of three terms:
\begin{equation}\label{CIB clust comp}
    C_{\ell,\nu\times\nu\prime}^{\text{clust}} = C_{\ell,\nu\times\nu\prime}^{\text{ET,clust}} + C_{\ell,\nu\times\nu\prime}^{\text{LT,clust}} + C_{\ell,\nu\times\nu\prime}^{\text{mix,clust}} .
\end{equation}

The shot noise (SN) component of the CIB power spectrum is obtained integrating the contribution of sources falling below the detection limit, $S_{\nu}^{\text{lim}}$, of a given experiment:
\begin{equation}
    C_{\ell,\nu}^{\shot} = \int_0^{S_\nu^{\text{lim}}} S_\nu^2 \dfrac{\d ^2N(S_\nu)}{\d S_\nu \d \Omega}\d S_\nu ,
\end{equation}
where $\d ^2N(S_\nu)/\d S_\nu \d \Omega$ is obtained integrating Eq.~\eqref{surface density} over redshift, and represents the differential number counts. The shot noise term is independent of the angular scale, thus representing a flat contribution to the CIB power spectrum when expressed in $C_\ell$s. In this study, the shot noise level for a specific frequency channel $\nu$, $\shot_{\nu}$, is modeled as:
\begin{equation}\label{CIB shot}
    C_{\ell,\nu\times\nu\prime}^{\shot} = \mathcal{C}_{\nu_1\times\nu_2}\sqrt{\shot_{\nu_1}\times \shot_{\nu_2}} ,
\end{equation}
where $\mathcal{C}_{\nu_1\times\nu_2}$ accounts for the possible correlation between emissions at different frequencies.

Finally, the CIB power spectrum is given by the sum of clustering and shot noise terms:
\begin{equation}\label{CIB PS}
    C_{\ell,\nu\times\nu\prime}^{\cib} = C_{\ell,\nu\times\nu\prime}^{\text{clust}} + C_{\ell,\nu\times\nu\prime}^{\shot} .
\end{equation} 

\begin{figure}[t]
\centering
\includegraphics[width=.45\textwidth]{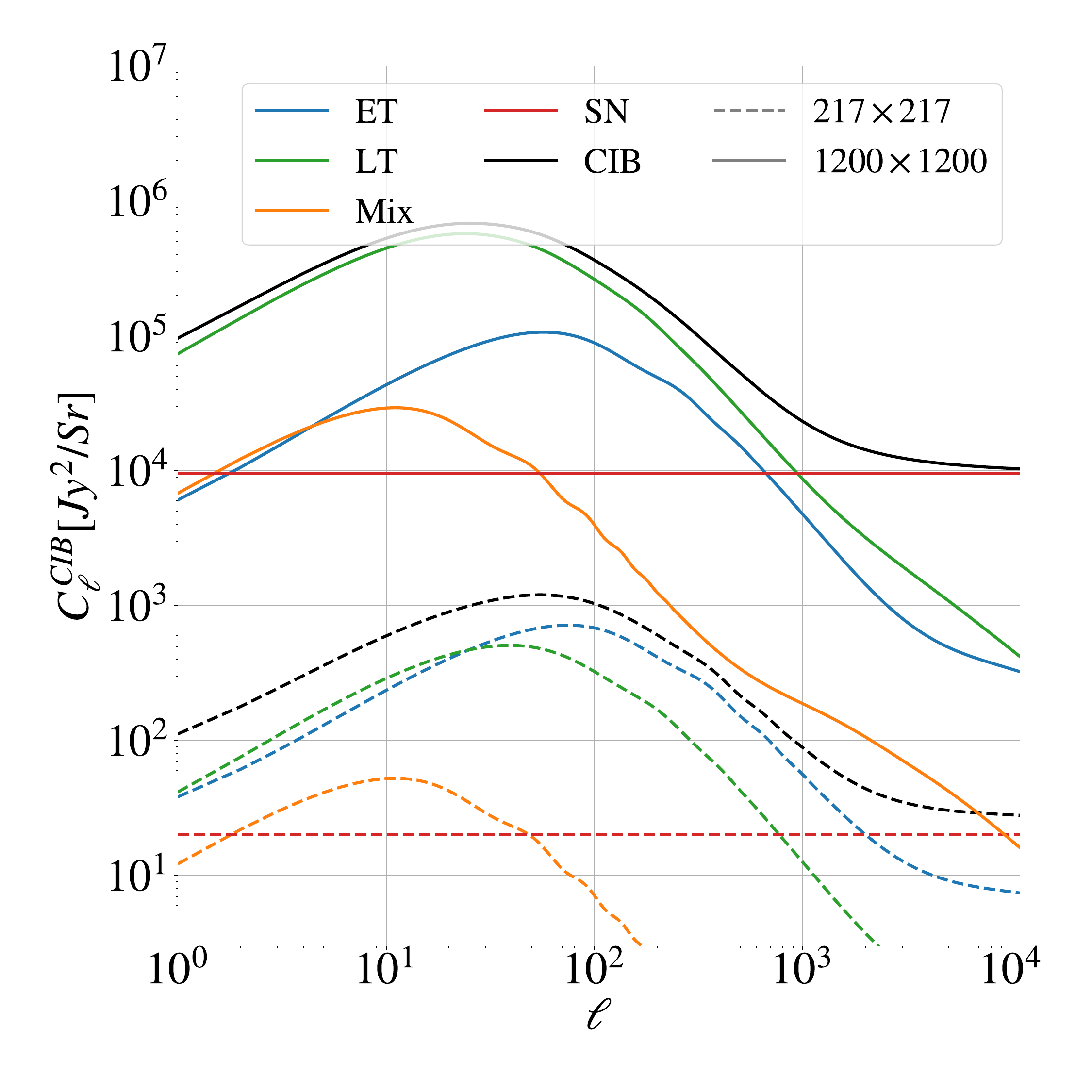}
\caption{\label{fig:CIB emission} Different contributions to the CIB power spectrum as predicted by our model, for two different frequency combinations: $217\times217 \ghz$ (\textit{dashed} curves), and $1200\times1200 \ghz$ (\textit{solid} curves). In both cases, the contribution from early-type (ET) galaxies is in \textit{blue}, for late-type (LT) galaxies in \textit{green} and the mixing between the two galaxy populations in \textit{orange}. The shot noise term is the flat contribution shown in \textit{red}. The total CIB power spectrum (Eq.~\eqref{CIB PS}) is shown in \textit{black}. The values of the model parameters used to produce these curves are:  $\log(M_{\text{min}}^{\text{EP}}/\Msun  h^{-1}) = 12.07$, $\log(M_{\text{min}}^{\text{LP}}/\Msun  h^{-1}) = 10.85$, $\alpha_{\text{EP}} = 1$, $\alpha_{\text{LP}} = 1$, ${\rm \shot}_{217} = 16$ and  ${\rm \shot}_{1200} = 9833$.}
\end{figure}

Figure \ref{fig:CIB emission} shows the model predictions for the CIB power spectrum (\textit{black curve}), highlighting the different contributions coming from ET (\textit{blue}), LT (\textit{green}), mixing terms (\textit{orange}) and shot noise (\textit{red}). This is plotted for two different frequency combinations: $217\times217~\ghz$ (dashed curves) and $1200\times1200~\ghz$ (solid curves), taking the two extremes of the datasets which we will use in Sec.~\ref{sec:results}.
The figure shows that the LT galaxies dominate at higher frequencies (comparing the solid and dashed green curves) and that large angular scales (low multipoles) are dominated by the clustering terms, while the contribution coming from the constant shot noise term becomes more and more relevant as we move to higher multipoles, \textit{i.e.} smaller angular scales.

\section{Dataset}\label{sec:dataset}
To constrain the parameters describing the clustering properties of the two galaxy populations, we fit our model to three datasets: the CIB power spectrum measurements from \cite{planck2013-pip56} (denoted as P14 in this work), the CIB power spectrum obtained from a re-analysis of \textit{Planck} data by \cite{Lenz:2019ugy} (hereafter L19), and the angular power spectra from \textit{Herschel}-SPIRE data presented by \cite{Viero:2018mdn} (hereafter V18).

\subsection{Planck data}\label{subsec:P14 data}
P14 presented measurements of the CIB power spectrum over five frequencies, spanning $143-857\ghz$. In this study, we use ten CIB power spectra from P14, selecting only four frequency channels (217, 353, 545 and 857 GHz). In analogy with P14, we do not include the 143 GHz measurements for model fits since this channel has significant CMB signal contaminating the CIB (the CMB power spectrum at $\ell=100$ is $\sim5000$ higher than that of the CIB). This can be overcome implementing large corrections in the analysis to remove the CMB, but neither P14 nor this work attempt to do this. 

The P14 dataset covers the multipole range from $\ell=150$ to $\ell=2500$, binned using a logarithmic binning, $\Delta\ell/\ell=0.3$. Across this range, P14 provides 8 data points for each CIB frequency auto- and cross-spectrum, resulting in a total of 80 data points in the dataset available for the model fit that will be described in Sec.~\ref{sec:results}. The errorbars associated to each point account for both cosmic variance and instrumental noise. Since P14 data come without a covariance matrix, we conduct our analysis building a diagonal covariance matrix from the spectra errorbars and therefore neglecting possible correlations between different multipoles.

Data from \planck\,have been calibrated in two different ways, depending on the frequency channel. Specifically, the calibrator for the 217 and 353 GHz channels is the CMB orbital dipole, which boasts superior accuracy, while the 545 and 857 GHz channels are calibrated using planetary measurements. The calibration method generates a frequency-dependent uncertainty, with all values reported in Table 6 of \cite{planck2014-a09} and used in later sections when fitting the spectra.
Additionally, updates to the calibrations were introduced between the first and the second \textit{Planck} Public Releases, \textit{i.e.} PR1 and PR2. 
The CIB spectra are from PR1 and to account for these calibration updates here we correct them with: 
\begin{equation}\label{eq:PR_recalibration}
    C_{\ell,\nu_1\times\nu_2}^{\cib} = \dfrac{C_{\ell,\nu_1\times\nu_2}^{\cib,PR1}}{corr_{\nu_1}\times corr_{\nu_2}} .
\end{equation}
$C_{\ell,\nu_1\times\nu_2}^{\cib}$ represents the calibration-corrected data fed in input to the fit, $C_{\ell,\nu_1\times\nu_2}^{\cib,PR1}$ are the original CIB power spectra presented in P14 and $corr_{\nu_i}$ are the calibration corrections for each frequency, as detailed in \cite{planck2014-a09}, and corresponding to 0.991, 0.997, 1.018, 1.033 for the 217, 353, 545 and 857 GHz frequency channels, respectively. 

As detailed in the next section, to compare the model predictions with the data we also need to apply color corrections. Specifically, P14 analysis computes the color corrections for each frequency channel using the CIB SED from \cite{Bethermin:2012ki}, ending up with color-correction factors of 1.119, 1.097, 1.068 and 0.995 for the 217, 353, 545 and 857 GHz frequency channels, respectively.

\subsection{Lenz data}\label{subsec:L19 data}
The second dataset explored in this study is L19 \citep{Lenz:2019ugy}, which provides six CIB power spectra using the 353, 545 and 857 GHz \planck\,maps -- L19 excludes both 143 and 217 GHz channels on the basis of CMB contamination. As described in their measurements paper, the authors of L19 employed a method based on neutral atomic hydrogen data to remove Galactic dust from the \planck\,intensity maps. This procedure allows to extract the CIB power spectra down to lower multipoles compared to P14, reaching $\ell\sim75$ (extending the P14 CIB power spectra starting from $\ell\sim150$). L19 CIB data are also presented with a different and less aggressive binning scheme than the one of P14. Specifically, the authors apply a linear binning, including 64 multipoles in each bin and resulting in a dataset with a total of 186 data points.

From the L19 dataset, we decide to exclude the first three bandpowers of each spectrum. We do this for two reasons: to better allow for a straightforward comparison with P14 data, and noting that the P14 data exhibit more power at lower multipoles compared to the L19 bandpowers (see Fig. 11 of L19). This could be due to a more aggressive dust cleaning procedure adopted in L19 but the consistency of the two datasets is beyond the scope of this work and not fully addressed in other literature for us to trust those additional bandpowers. We explore a few assumptions, including an investigation on the impact of the dust contamination in the L19 analysis in Appendix \ref{app:Lenz}.

As for P14, we apply to the L19 dataset color corrections and the calibration factors, using the same values quoted in the previous subsection. 

Similarly to P14, L19 spectra have been released without a covariance matrix. Once again, here we build a diagonal covariance between bandpowers, neglecting possible correlations between bands -- we explore more this assumption in Appendix \ref{app:Lenz}.

\subsection{SPIRE data}\label{subsec:V18 data}
The third dataset used in this work is the V18 CIB measurement from \textit{Herschel}-SPIRE observations. This covers higher frequencies and multipoles than those covered by the \planck\,data, providing complementary information on CIB physics. In particular, the higher angular resolution is expected to better probe the multipole region dominated by shot noise contribution, as opposed to the \planck\,scales which are more sensitive to the clustering term.

SPIRE observations approached the THz regime, covering 600, 857 and 1200 GHz. The CIB spectra span multipoles from $\ell = 600$ to $\ell = 11000$. Following the binning scheme of \cite{Amblard:2011gc}, the bandpowers are linearly binned for $\ell<1950$, and for higher multipoles a logarithmic binning with $\Delta\ell/\ell=720$ is applied. In total, this dataset consists of 132 data points.

The maps are calibrated with ununcertainty of $5\%$ for each frequency channel. The color corrections needed at 600, 857 and 1200 GHz are 0.9988, 0.9929 and 0.9957, respectively \citep{Lagache:2002xq}.

SPIRE spectra are accompanied by a covariance matrix which is, however, not well defined\footnote{We started our work using the full, published covariance matrix and encountered a number of problems. After performing some tests, we concluded that this matrix exhibits a pathological behaviour. We found it to be occasionally negative defined, thus not invertible. This was also noted by ~\cite{Maniyar:2020tzw}.}. In analogy with other studies (see e.g., \citealp{Maniyar:2020tzw}) to overcome issues with the available covariance, we revert to assuming uncorrelated errorbars also for V18 data, using a simple diagonal covariance matrix.

\section{Results}\label{sec:results}
Here we bring together the full model described in Sec.~\ref{sec:CIB} and the three datasets in Sec.~\ref{sec:dataset}. We explore the model against the multi-frequency, multi-scale observations from \planck\,and \textit{Herschel}-SPIRE individually, anticipating that, as also discussed in~\cite{Maniyar:2020tzw}, there is a tension among the results from P14 and V18 which prevents us to make a joint fit using both datasets.

To compare the predictions of the model to the observed data, we correct the theory to account for instrument-specific corrections as follows:
\begin{equation}\label{cib model data}
    C_{\ell,\nu_1\times\nu_2}^{\cib,\text{data}} = \mathcal{A}_{\nu_1x\nu_2}\times cc_{\nu_1} \times cc_{\nu_2} \times C_{\ell, \nu_1\times\nu_2}^{\cib,\text{model}} .
\end{equation}
Here, $C_{\ell, \nu_1\times\nu_2}^{\cib,\text{model}}$ is the CIB power spectrum as predicted by the model 
; $\mathcal{A}_{\nu_1\times\nu_2}=\sqrt{f_{\text{cal}}^{\nu_1}\times f_{\text{cal}}^{\nu_2}}$ accounts for the absolute calibration uncertainties, \textit{i.e.} $f_{\text{cal}}^{\nu_i}$ for each different frequency channels; $cc_{\nu_i}$ is the color correction per frequency channel.

To explore the parameters of the model, we perform a Monte Carlo Markov Chain (MCMC) analysis by making use of the {\tt emcee} ensemble sampler \citep{Foreman-Mackey:2012any}. The multi-dimensional parameter space is explored sampling a Gaussian likelihood:
\begin{equation}
    \log \mathcal{L}(C_{\ell}^{\text{data}}|C_{\ell}^{\text{model}}) \propto -\dfrac{1}{2}\displaystyle\sum_\ell \dfrac{\bigl(C_{\ell}^{\text{data}}-C_{\ell}^{\text{model}}\bigr)^2}{\sigma_\ell^2} ,
\end{equation}
where $\sigma_\ell^2$ represents the errorbars associated to the points of each dataset (i.e., representing a diagonal covariance matrix for all the three datasets). 

The analysis has taken into consideration different scenarios and the numbers of fixed or varied model parameters has been set accordingly in each case as we describe below and in the individual subsections.

We perform our analyses covering two scenarios for the shot noise term:
\begin{itemize}[noitemsep,topsep=0pt]
    \item In one case, we assume maximal correlation between the shot noise contributions, with the correlation coefficients in Eq.~\eqref{CIB shot} all set to 1.
    \item In another case, the correlation coefficients are treated as free parameters in the MCMC analysis.
\end{itemize}

The data considered in this work are able to constrain, with different degrees of sensitivity, the clustering parameters for both the early and the late galaxy populations, i.e., $M_{\text{min}}^{\text{EP}}$, $\alpha_{\text{EP}}$, $M_{\text{min}}^{\text{LP}}$ and $\alpha_{\text{LP}}$; the shot noise levels for the considered frequency channels, $\shot_{\nu_i}$; and the calibration factors, $f_{\text{cal}}^{\nu_i}$. However, the datasets covering \planck\,frequencies and multipole ranges are not able to constrain the $\alpha_{\text{EP}}$ clustering parameter (see Appendix~\ref{app:Planck_EP}) which we then fix to 1 for P14 and L19. On the other hand, as we will show later, all datasets considered in this work can constrain $\alpha_{\text{LP}}$ (either unconstrained or only weakly constrained in previous works) and this parameter is therefore always varied in our runs.

The choice of the prior distributions assigned to the free parameters, and the discussion and implications of the results from the individual model fits are discussed in the following sub-sections. 

\subsection{P14 data analysis}\label{subsec:P14 model fit}

For P14 we sample 10 and 16 free parameters for the two scenarios, respectively, with prior ranges listed in the second column of Table~\ref{tab: best fit planck}. As mentioned above, expecting no constraining power from \planck\,, in the baseline case we fix $\alpha_{\text{EP}}$=1 (and explore its variation in Appendix~\ref{app:Planck_EP}). We vary the calibration factors with a Gaussian prior centered on 1 and with errors taken as twice the uncertainty in Table 6 of \cite{planck2014-a09}, except for the calibration factor at 217 GHz which we fix to $f_\mathrm{cal}^{217}=1$\footnote{This choice is driven by the very small uncertainty of the \planck\, measurement at 217, and it has been verified extracting results treating $f_\mathrm{cal}^{217}$ as a free parameter and finding perfect agreement with the case in which it is fixed to unity.}. The clustering parameters and the shot noise levels are sampled from uniform distributions. When correlation coefficients are treated as free parameters, these are also sampled from uniform prior distributions.

\begin{table}[t!]
	\begingroup
	\nointerlineskip
	\vskip -3mm
	\tiny
	\setbox\tablebox=\vbox{
		\newdimen\digitwidth
		\setbox0=\hbox{\rm 0}
		\digitwidth=\wd0
		\catcode`*=\active
		\def*{\kern\digitwidth}
		\newdimen\signwidth
		\setbox0=\hbox{+}
		\signwidth=\wd0
		\catcode`!=\active
		\def!{\kern\signwidth}
		\halign{\hbox to 0.8in{#\leaderfil}\tabskip=1em&
			\hfil#\hfil\tabskip=10pt&
			\hfil#\hfil\tabskip=10pt&
			\hfil#\hfil\tabskip=10pt&
			\hfil#\hfil\tabskip=10pt&
			\hfil#\hfil\tabskip=10pt&
			\hfil#\hfil\tabskip=0pt\cr
			\noalign{\doubleline}
			\noalign{\vskip 3pt}
            \omit\hfil Parameter\hfil & Prior & Results & Results\cr
            \omit & \omit  & $\mathcal{C}_{\nu_1\times\nu_2}=1$ & $\mathcal{C}_{\nu_1\times\nu_2}$ open \\[1ex]\cr
			\noalign{\doubleline}
  $\log(M_{\text{min}}^{\text{EP}}/\Msun  h^{-1})$ & [10.7,12.8] & $11.45^{+0.16}_{-0.13}$ & $11.12\pm 0.19$\cr
  $\log(M_{\text{min}}^{\text{LP}}/\Msun  h^{-1})$ & [10.5,12.8] & $11.18^{+0.31}_{-0.27}$ & $11.56^{+0.21}_{-0.17}$\cr
  $\alpha_{\text{LP}}$ & [0.2,3.5] & $1.337^{+0.063}_{-0.072}$ & $1.436^{+0.075}_{-0.067}$\cr
			\noalign{\vskip 3pt\hrule\vskip 5pt}
  $\shot_{217}$ & [0,50] & $6.72\pm 0.76$ & $22\pm 4$\cr
  $\shot_{353}$ & [50,500] & $273\pm 16$ & $276\pm 20$\cr
  $\shot_{545}$ & [400,4000] & $1296\pm 120$ & $1247\pm 100$\cr
  $\shot_{857}$ & [200,8000] & $1827\pm 300$ & $1899^{+400}_{-600}$\cr
  $f^{353}_{\text{cal}}$ & $1\pm 0.0156$ & $0.999\pm 0.014$ & $1.010\pm 0.015$\cr
  $f^{545}_{\text{cal}}$ & $1\pm 0.122$ & $1.096\pm 0.032$ & $1.094\pm 0.032$\cr
  $f^{857}_{\text{cal}}$ & $1\pm 0.128$ & $1.289\pm 0.076$ & $1.200\pm 0.072$\cr
			\noalign{\vskip 3pt\hrule\vskip 5pt}
  $\mathcal{C}_{217 \times 353}$ & [-1, 1] & - & $0.648^{+0.068}_{-0.075}$\cr
  $\mathcal{C}_{217 \times 545}$ & [-1, 1] & - & $0.469\pm 0.054$\cr
  $\mathcal{C}_{217 \times 857}$ & [-1, 1] & - & $0.562\pm 0.070$\cr
  $\mathcal{C}_{353 \times 545}$ & [-1, 1] & - & $0.946^{+0.035}_{-0.031}$\cr
  $\mathcal{C}_{353 \times 857}$ & [-1, 1] & - & $0.937^{+0.056}_{-0.042}$\cr
  $\mathcal{C}_{545 \times 857}$ & [-1, 1] & - & $0.934^{+0.046}_{-0.039}$\cr
	\noalign{\vskip 5pt\hrule\vskip 3pt}}}
 \endPlancktable
\vskip 3pt
\caption{\textit{First column}; Model parameters sampled for the fit to P14 data. \textit{Second column}; priors and ranges of variation imposed on the model parameters. We adopt uniform priors for all parameters (ranges are within square brackets), except for the calibration factors for the 353, 545 and 857 GHz frequency channels which are varied with a Gaussian priors centered on 1. \textit{Third and fourth columns}; mean and standard deviation of the model parameters for the case in which the correlations for the shot noise are fixed to one and the case in which they are free to vary, respectively. In the former case, we obtain a $\chi^2 = 81$ with 80 points and 10 free parameters. The latter case has the same number of points, 16 free parameters and $\chi^2 = 45$.}\label{tab: best fit planck}
 \endgroup
\end{table}

\begin{figure*}[ht]
\centering
\includegraphics[width=1.\textwidth]{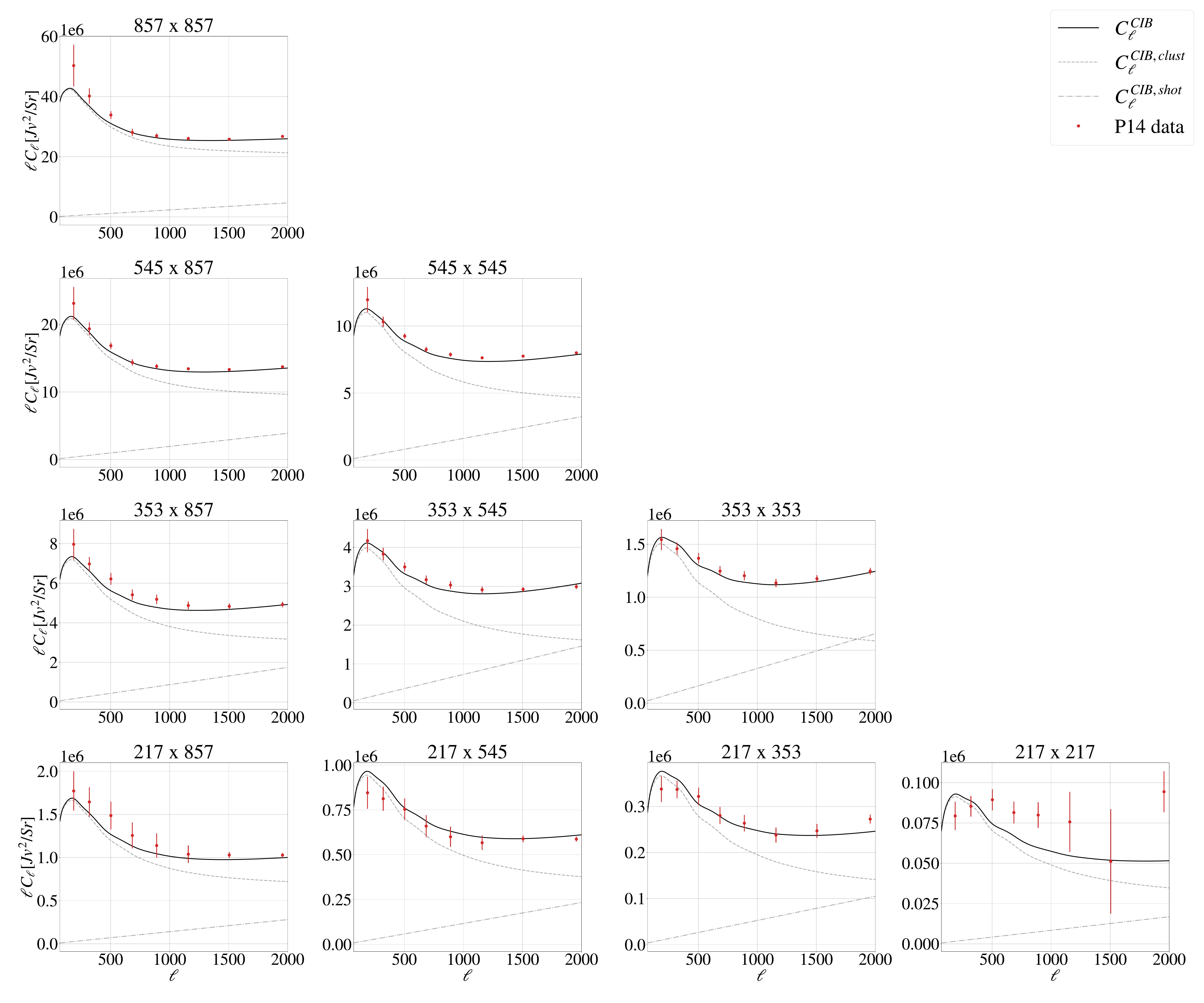}
\caption{\label{fig:planck model data}Comparison between P14 data and the model predictions obtained with the best-fit parameters in Table \ref{tab: best fit planck}. The CIB power spectrum, the clustering term and the shot noise term predicted by the model are represented with the \textit{solid black}, the \textit{dashed gray} and the \textit{dot-dashed gray} curves, respectively. P14 data with their error bars are represented in \textit{red}.}
\end{figure*}

\begin{figure}[t]
\centering
\includegraphics[width=.5\textwidth]{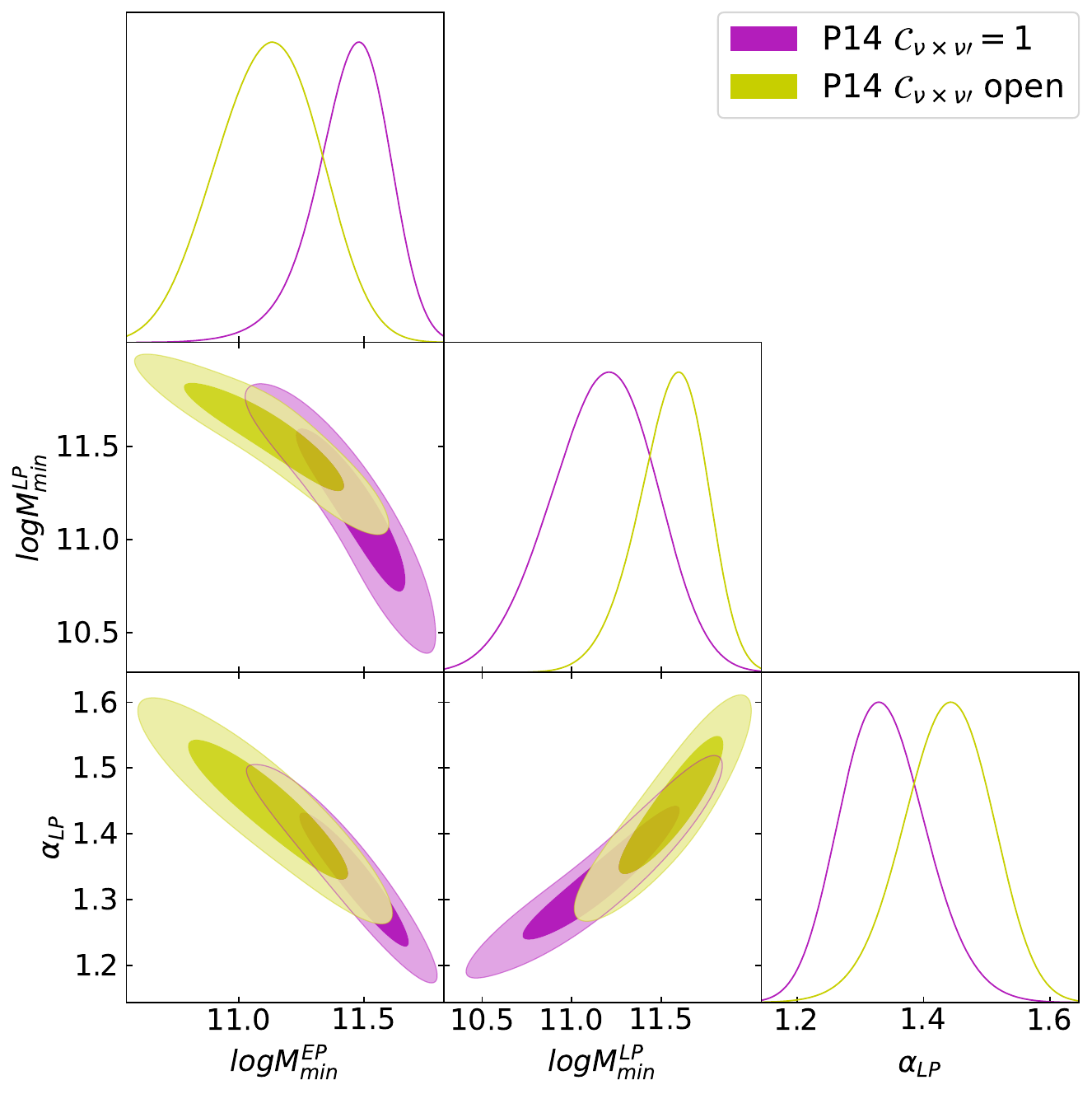}
\caption{\label{fig:planck clust best fit} Posterior probability distributions of the clustering parameters for the two scenarios where the correlations for the shot noise are fixed to unity (in \textit{purple}) and where they are free parameters in the fit (in \textit{green}). The posterior distributions for the two cases are in agreement, exploring the same regions of the parameter space. We fix $\alpha_{\text{EP}}=1$.}
\end{figure}

Results ($1\sigma$ confidence level) for the two model scenarios with fixed/varied correlations are reported in the third and the fourth columns of Table \ref{tab: best fit planck}. We obtain $\chi^2=81$ with 70 degrees of freedom when fixing the correlation coefficients, and $\chi^2=45$ when the correlation coefficients are free to vary (when the number of degrees of freedom is 64), with Probability To Exceed (PTE) of $17.9\%$ and $96.9\%$ for the two cases, respectively. We show how the best-fit model compares with the ten P14 CIB spectra in Fig.~\ref{fig:planck model data}. 

\begin{figure}[ht]
\centering
\includegraphics[width=.5\textwidth]{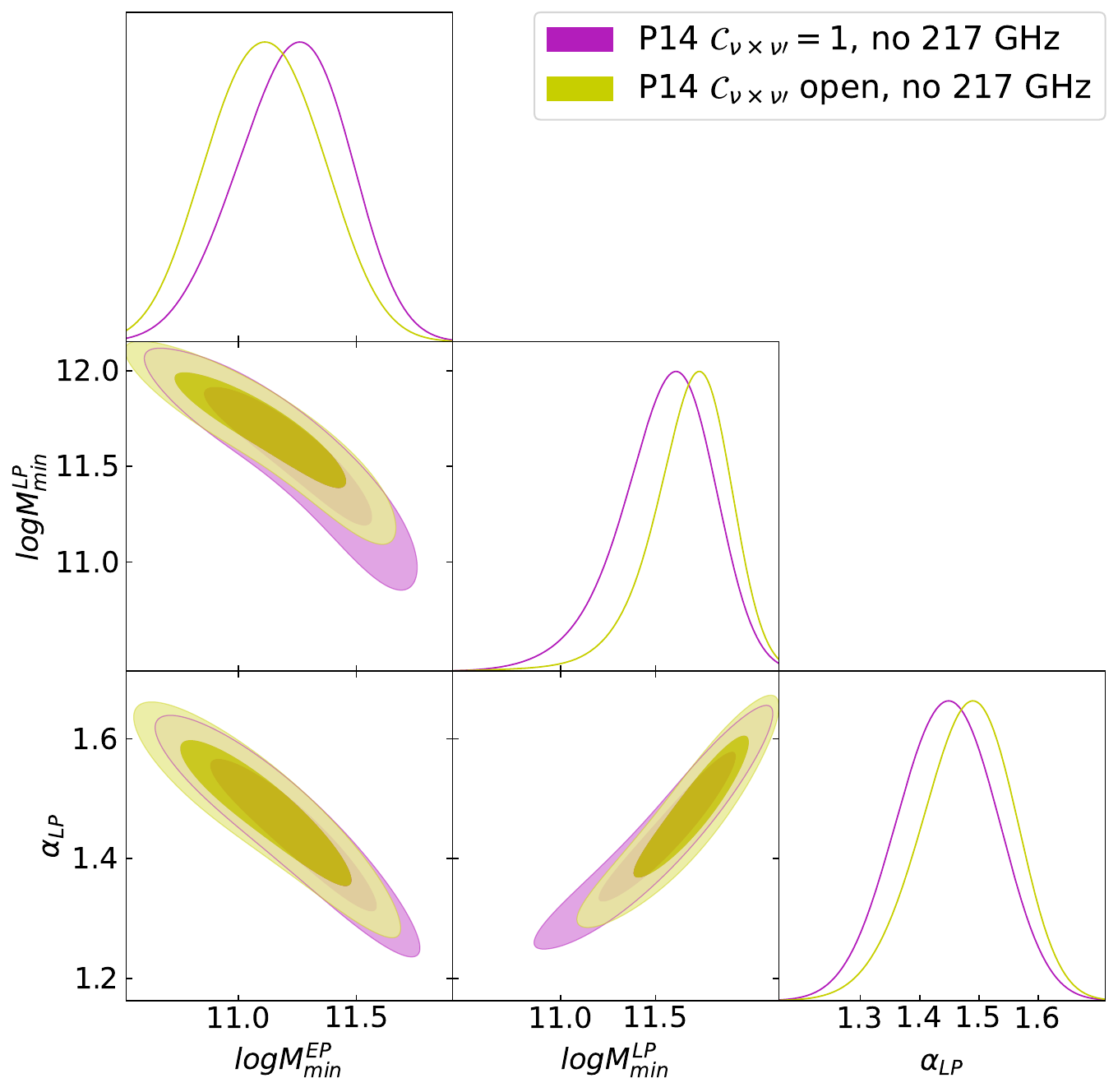}
\caption{\label{fig:planck WO 217}Same as Fig.~\ref{fig:planck clust best fit} excluding the 217 GHz frequency channel.}
\end{figure}

\begin{figure*}[ht]
\centering
\includegraphics[width=1.\textwidth]{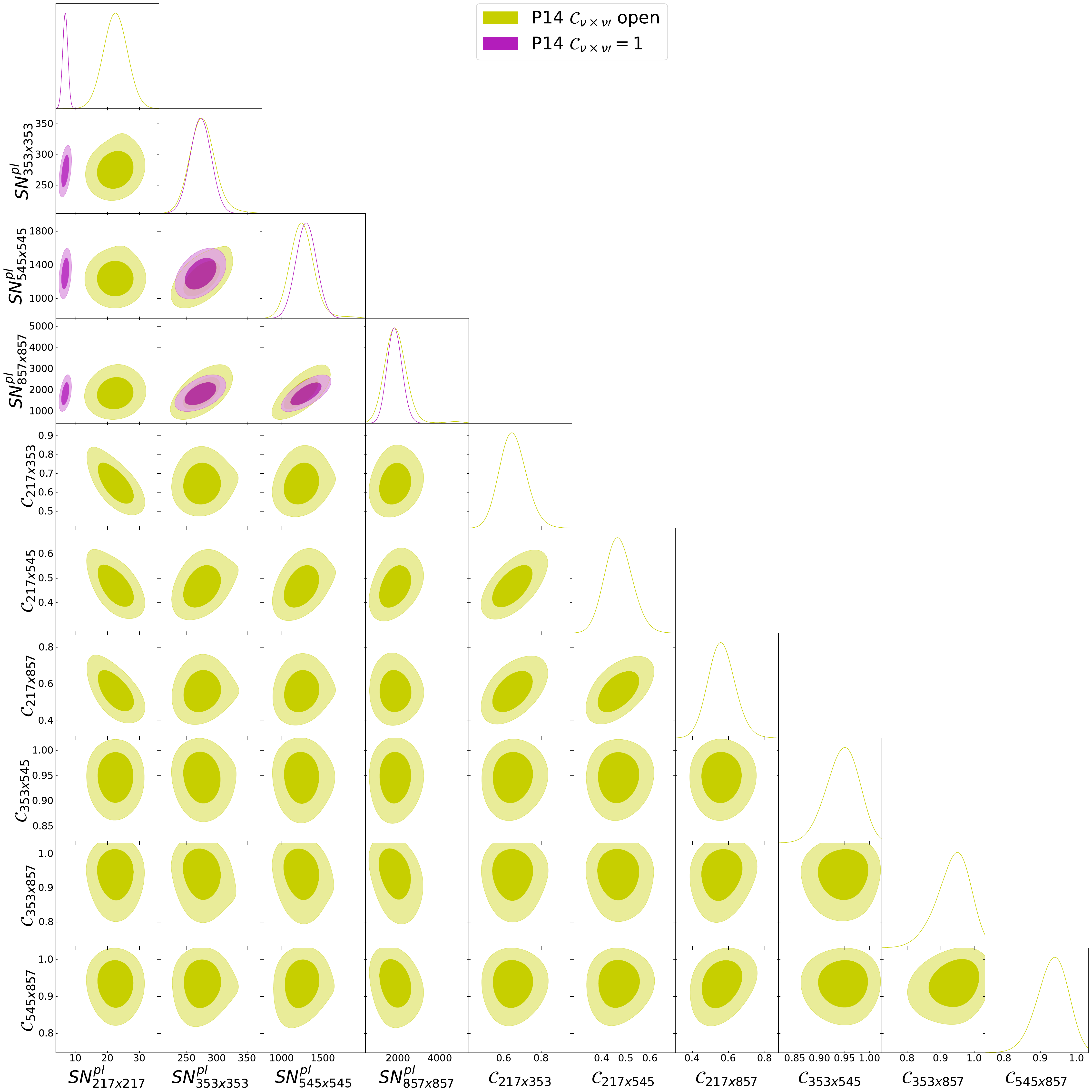}
\caption{\label{fig:shot corr}Posterior probability distributions of the shot noise levels for the scenario where the correlation coefficients are fixed to unity (in \textit{purple}), and posterior probability distributions for the shot noise levels and the correlation coefficients for the scenario where the latter are free to vary (in \textit{green}). This figure highlights the movement in the 217 GHz shot noise and shows the degeneracies with parameters characterising the spectra including at least one leg of the 217 GHz channel.}
\end{figure*}

In Fig.~\ref{fig:planck clust best fit}, we show the 1-dimensional posterior distributions and the 2-dimensional correlations (at 68 and 95\% confidence) of the clustering parameters for the two scenarios considered in this work. The two cases are in good agreement, and as expected we recover larger constraints for some parameters due to the higher number of degrees of freedom in the MCMC when leaving correlations free to vary.
The relative impact seen here on the best-fit values for the minimum mass of the ET and LT galaxies is likely due to the fact that, within the multipole range probed by P14 data, the contribution to the CIB power spectrum from clustering dominates, while the shot noise is expected to dominate at smaller angular scales. Therefore, the different treatment of the shot noise levels does not have a significant impact on the clustering parameters. 
Motivated by the results found on the L19 dataset (see next section), we re-run the analysis removing the spectra involving 217 GHz and note much closer agreements between the two scenarios as shown in Fig.~\ref{fig:planck WO 217}. To understand this we need to look at the full parameter space explored by the fit. We find a tension in the two scenarios for the value of the shot noise level of the 217 GHz frequency channel, shown in Fig.~\ref{fig:shot corr}. Specifically, we obtain a value for the shot noise level which is higher in the case with free correlations than in the case with fixed correlations at $\sim 4\sigma$ level. We also see that the correlation coefficients involving the 217 GHz frequency channel, i.e., $\mathcal{C}_{217\times i}$ with $i = 353, 545, 857$ (see fourth column of Table~\ref{tab: best fit planck}), are significantly lower than unity. We explain this behaviour by noting the degeneracy between the shot noise level and the correlation coefficients at 217 GHz (see Fig.~\ref{fig:shot corr}), which P14 data are not able to break. Specifically, they are anti-correlated, meaning that a shift of the shot noise level towards a lower value, closer to the one obtained in the first scenario, leads to higher values of the $\mathcal{C}_{217\times i}$.

When comparing our results with other literature findings, we note significant deviations between the values for the minimum masses of both galaxy populations extracted here with those found in the literature, more evident for the ET galaxies. Specifically, we find the minimum mass, $\log (M_{\text{min}}^{\text{EP}}/M_\odot h^{-1})$, to be lower than previously reported values (see Table~\ref{tab: literature values clust}). In particular, when comparing the value we obtain when fixing the correlation coefficients for the shot noise to unity, $\log(M_{\text{min}}^{\text{EP}}/\Msun  h^{-1}) = 11.45^{+0.16}_{-0.13}$, with the ones of C13, $\log(M_{\text{min}}^{\text{EP}}/\Msun  h^{-1}) = 12.00^{+0.04}_{-0.04}$, and X12, $\log(M_{\text{min}}^{\text{EP}}/\Msun  h^{-1}) = 12.09^{+0.06}_{-0.06}$, we find a tension at the level of $\sim 4\sigma$. In contrast, our analysis shows a higher value of the minimum mass for LT galaxies than the expected value found in C13 of $\log(M_{\text{min}}^{\text{LP}}/\Msun  h^{-1}) = 10.85\pm0.06$, but it is still compatible with the current constraint within $\sim 1\sigma$. This discrepancy raises a critical point as far as the model is concerned: our results suggest that the minimum mass for ET galaxies is not higher than that for LT, contrary to prior assumptions. We speculate that this inconsistency may stem from the limited frequency and multipole ranges covered by P14 data. These ranges may not be sufficiently wide to effectively distinguish between the contributions of the two galaxy populations to the CIB spectrum, and, consequently, brake the degeneracies among all the three clustering parameters (see Fig.~\ref{fig:planck clust best fit}). Concerning $\alpha_{\text{LP}}$, as already mentioned, previous analyses only considered it fixed to unity. Here, $\alpha_{\text{LP}}$ is a free parameter of the model fit and is constrained at the best fit values of $\alpha_{\text{LP}}=1.337^{+0.063}_{-0.072}$ and $\alpha_{\text{LP}}=1.436^{+0.075}_{-0.067}$ for the case where the correlation coefficients are fixed and free to vary, respectively.

We now comment on the shot-noise correlation coefficients, in the scenario where they are free to vary (fourth column of Table \ref{tab: best fit planck}). To ease the comparison with previous \planck\, analyses, we also report in Table \ref{tab: corr shot Planck} the values of the correlation coefficients extrapolated from Table 6 of P14, as:
\begin{equation}\label{corr coeff}
    \mathcal{C}_{\nu_1\times\nu_2} = \dfrac{C_{\ell,\nu_1\times\nu_2}^{\text{Planck}}}{\sqrt{C_{\ell,\nu_1\times\nu_1}^{\text{Planck}}C_{\ell,\nu_2\times\nu_2}^{\text{Planck}}}} ,\text{with}\, \nu_1\neq\nu_2 ,
\end{equation}
where $C_{\ell,\nu_1\times\nu_2}^{\text{Planck}}$ represent the entries of Table 6 of P14. We observe that the correlation coefficients involving the 217 GHz frequency channel, $\mathcal{C}_{217 \times i}$ with $i = 353, 545, 857$, are always lower than the values reported in Table \ref{tab: corr shot Planck}, even though they are still compatible within $\sim 2\sigma$. On the other hand, the remaining correlation coefficients are compatible with the values in Table \ref{tab: corr shot Planck} within $\sim1\sigma$. The observed low correlation coefficients are likely due to non-trivial degeneracies between the free parameters of the model, which the specific data are not able to efficiently break. In other words, the number of free parameters for this second scenario is too high and the model overfits the data, and this is confirmed by the low $\chi^2$ obtained in this scenario ($\chi^2 = 45$).

\begin{table}[t!]
	\begingroup
	\nointerlineskip
	\vskip -3mm
	\footnotesize
	\setbox\tablebox=\vbox{
		\newdimen\digitwidth
		\setbox0=\hbox{\rm 0}
		\digitwidth=\wd0
		\catcode`*=\active
		\def*{\kern\digitwidth}
		\newdimen\signwidth
		\setbox0=\hbox{+}
		\signwidth=\wd0
		\catcode`!=\active
		\def!{\kern\signwidth}
		\halign{\hbox to 0.8in{#\leaderfil}\tabskip=1em&
			\hfil#\hfil\tabskip=10pt&
			\hfil#\hfil\tabskip=10pt&
			\hfil#\hfil\tabskip=10pt&
			\hfil#\hfil\tabskip=10pt&
			\hfil#\hfil\tabskip=10pt&
			\hfil#\hfil\tabskip=0pt\cr
			\noalign{\doubleline}
			\noalign{\vskip 3pt}
            \omit  & 217 & 353 & 545 & 857\\[1ex]\cr
			\noalign{\doubleline}
217 & 1 & $0.98\pm0.24$ & $0.88\pm0.21$ & $0.72\pm0.17$ \cr
 353 & - & 1 & $0.95\pm0.23$ &  $0.81\pm0.20$ \cr
 545 & - & - & 1 & $0.93\pm0.23$\cr
 857 & - & - & - & 1 \cr
	\noalign{\vskip 5pt\hrule\vskip 3pt}}}
 \endPlancktable
\vskip 3pt
\caption{Correlation coefficients for the shot noise cross spectra evaluated from Eq. \eqref{corr coeff} using the shot noise levels presented in Table 6 of P14.}
\label{tab: corr shot Planck}
 \endgroup
\end{table}
 
\subsection{L19 data analysis}\label{subsec:L19 model fit}
In this section, we present the analysis of the L19 dataset. We note that, as explained in more details later, our model fits to L19 for both scenarios of correlations result in a poor $\chi^2$. We, therefore, do not attempt to interpret the goodness of the fit to this dataset and present results in this section only as a qualitative outcome. 

The choice of the free parameters in the model fit and their prior distributions mirror what was described in the previous section and are listed in the first and second column of Table~\ref{tab: best fit lenz}, respectively. As done for the analysis of P14 data, $\alpha_{\text{EP}}$ is fixed to unity, and the priors for the calibration factors of the 353, 545, and 857 GHz channels are based on \planck\, uncertainties.

\begin{figure}[t]
\centering
\includegraphics[width=.5\textwidth]{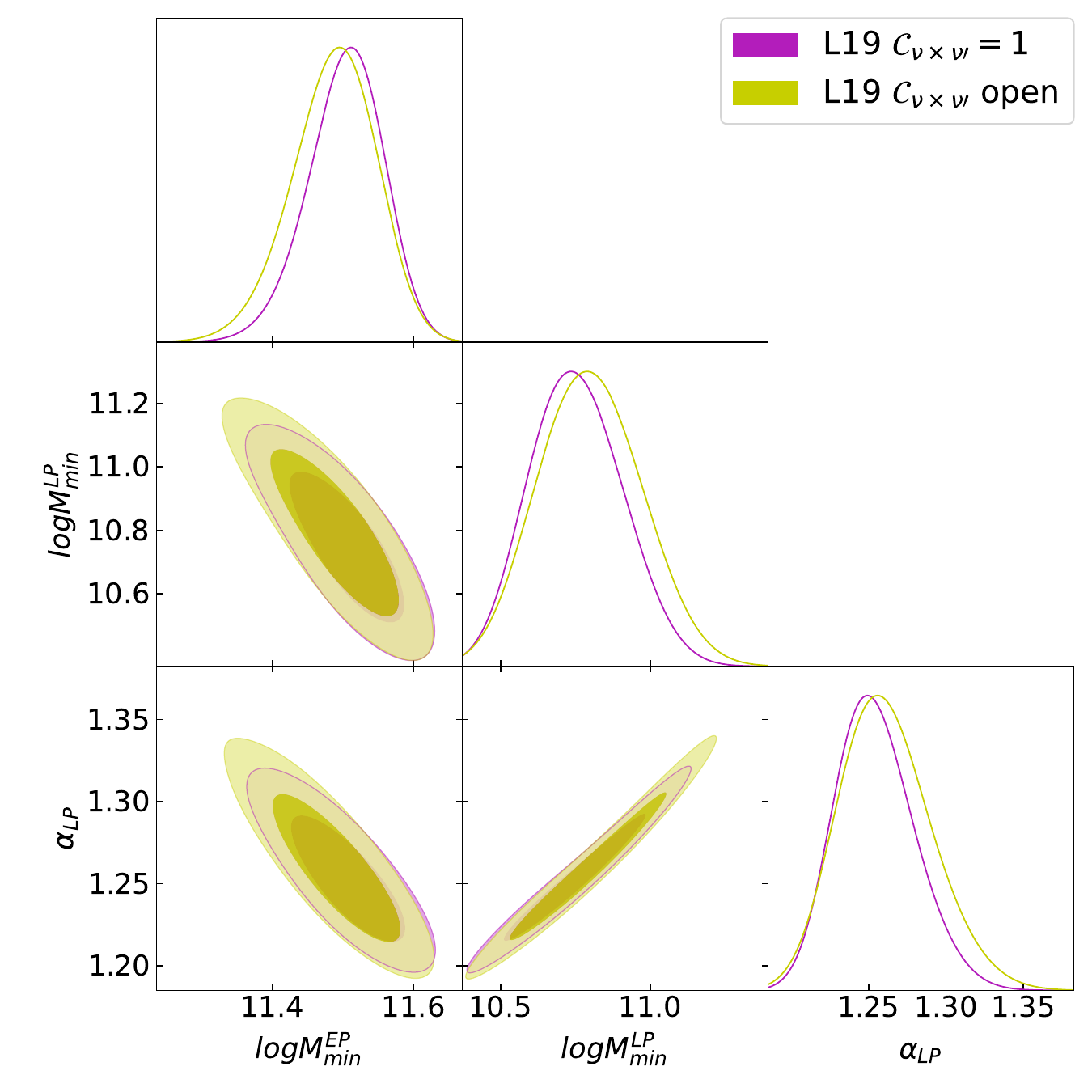}
\caption{\label{fig:lenz clust best fit}Same as Fig.~\ref{fig:planck clust best fit} for the L19 CIB dataset.}
\end{figure}

\begin{table}[t!]
	\begingroup
	\nointerlineskip
	\vskip -3mm
	\tiny
	\setbox\tablebox=\vbox{
		\newdimen\digitwidth
		\setbox0=\hbox{\rm 0}
		\digitwidth=\wd0
		\catcode`*=\active
		\def*{\kern\digitwidth}
		\newdimen\signwidth
		\setbox0=\hbox{+}
		\signwidth=\wd0
		\catcode`!=\active
		\def!{\kern\signwidth}
		\halign{\hbox to 0.8in{#\leaderfil}\tabskip=1em&
			\hfil#\hfil\tabskip=10pt&
			\hfil#\hfil\tabskip=10pt&
			\hfil#\hfil\tabskip=10pt&
			\hfil#\hfil\tabskip=10pt&
			\hfil#\hfil\tabskip=10pt&
			\hfil#\hfil\tabskip=0pt\cr
			\noalign{\doubleline}
			\noalign{\vskip 3pt}
            \omit\hfil Parameter\hfil & Prior & Results & Results\cr
            \omit & \omit  & $\mathcal{C}_{\nu_1\times\nu_2}=1$ & $\mathcal{C}_{\nu_1\times\nu_2}$ open \\[1ex]\cr
			\noalign{\doubleline}
  $\log(M_{\text{min}}^{\text{EP}}/\Msun  h^{-1})$ & [10.7,12.8] & $11.505^{+0.057}_{-0.049}$ & $11.488^{+0.063}_{-0.055}$\cr
  $\log(M_{\text{min}}^{\text{LP}}/\Msun  h^{-1})$ & [10.5,12.8] & $10.75^{+0.15}_{-0.17}$ & $10.80\pm 0.16$\cr
  $\alpha_{\text{LP}}$ & [0.2,3.5] & $1.254^{+0.023}_{-0.028}$ & $1.261^{+0.027}_{-0.032}$\cr
\noalign{\vskip 3pt\hrule\vskip 5pt}
  $\shot_{353}$ & [50,500] & $261.8\pm 6.9$ & $264.1\pm 7.1$\cr
  $\shot_{545}$ & [400,4000] & $1396\pm 45$ & $1396\pm 46$\cr
  $\shot_{857}$ & [200,8000] & $2391\pm 120$ & $2425\pm 130$\cr
  $f^{353}_{\text{cal}}$ & $1\pm 0.0156$ & $0.965\pm 0.015$ & $0.966\pm 0.015$\cr
  $f^{545}_{\text{cal}}$ & $1\pm 0.122$ & $1.174\pm 0.020$ & $1.175\pm 0.021$\cr
  $f^{857}_{\text{cal}}$ & $1\pm 0.128$ & $1.468^{+0.042}_{-0.038}$ & $1.460^{+0.048}_{-0.043}$\cr
\noalign{\vskip 3pt\hrule\vskip 5pt}
  $\mathcal{C}_{353 \times 545}$ & [-1, 1] & - & $0.9838\pm 0.0076$\cr
  $\mathcal{C}_{353 \times 857}$ & [-1, 1] & - & $0.9970^{+0.0036}_{-0.0022}$\cr
  $\mathcal{C}_{545 \times 857}$ & [-1, 1] & - & $0.9851^{+0.0093}_{-0.0082}$\cr
	\noalign{\vskip 5pt\hrule\vskip 3pt}}}
 \endPlancktable
\vskip 3pt
\caption{Same as Table~\ref{tab: best fit planck} for the L19 dataset.}
\label{tab: best fit lenz}
 \endgroup
\end{table}

In Table \ref{tab: best fit lenz}, we also report the $1\sigma$ confidence level results in both the scenarios of fixed (third column) and variable (fourth column) shot noise correlation coefficients. In Fig.~\ref{fig:lenz clust best fit}, we show the posterior distributions of the clustering parameters for the two scenarios of shot noise correlation coefficients fixed to unity (in \textit{purple}), and free-to-vary correlation (in \textit{green}). The results show perfect agreement between the two scenarios, with only a minor broadening of the distributions in the case of increased degrees of freedom in the fit. Specifically, we note that the posterior distributions deviate much less than those of P14 seen in Fig. \ref{fig:planck clust best fit}, confirming that the exclusion of the 217 GHz frequency channel -- not included at all in the L19 dataset -- makes the results of the fits more stable among the two scenarios.

The model fits to the L19 data result in poor goodness-of-the-fit metrics, in particular in exceedingly high $\chi^2$ values (specifically we obtained a $\chi^2$ of 655 and 650 for 159 and 156 degrees of freedom, respectively). However, we cannot definitely conclude that this is caused by deficits in the model. Appendix~\ref{app:Lenz} is devoted to a detailed and closer examination of the L19 dataset; we show there how a smooth theory cannot meet the bandpowers as published, and that changing assumptions in the dataset is also not enough to improve the $\chi^2$ and that the qualitative results presented here remain valid.

\begin{figure}[t]
\centering
\includegraphics[width=.5\textwidth]{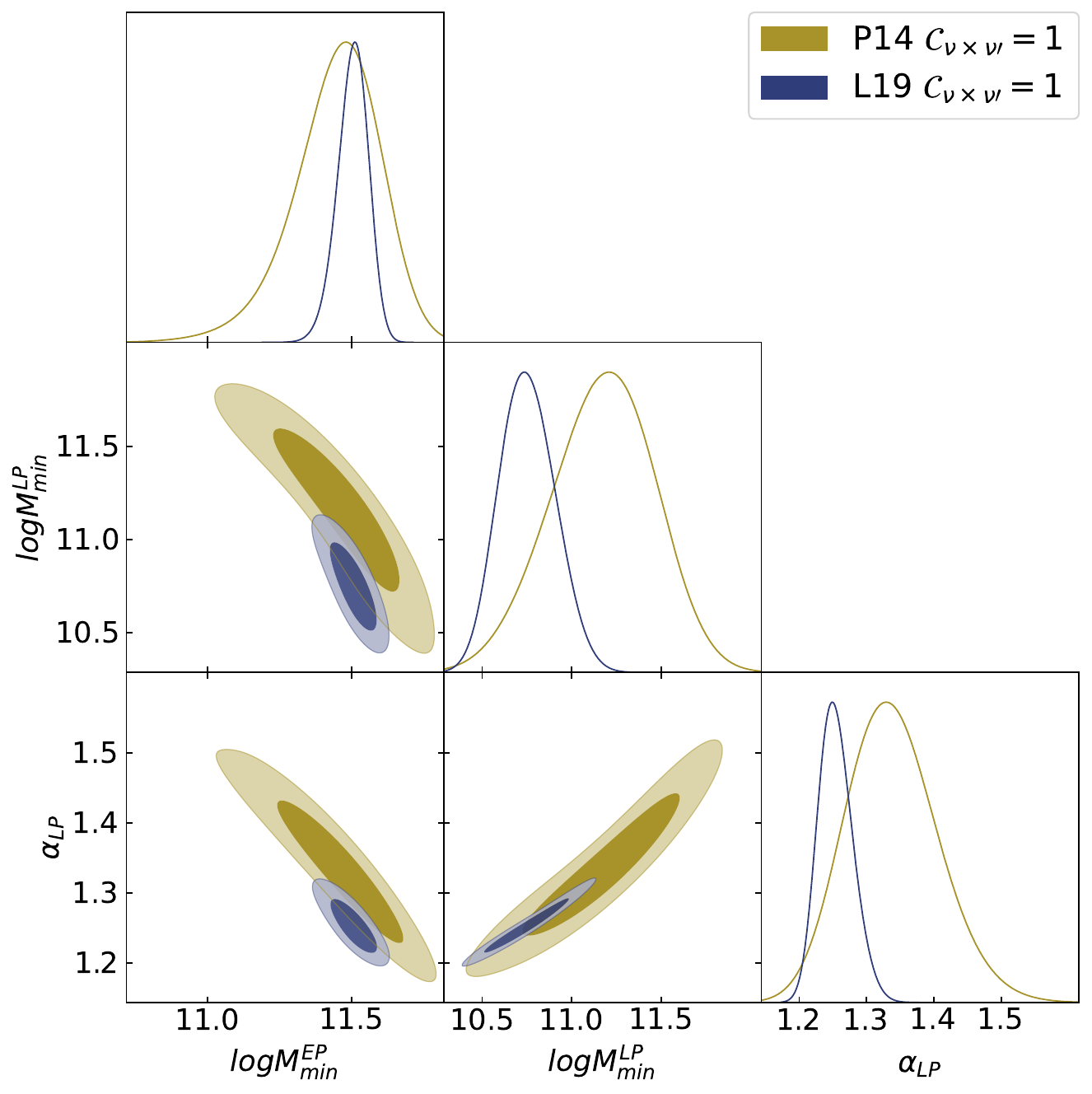}
\caption{\label{fig:lenz Planck comp} Comparison of the posterior distributions of the clustering parameters for P14 (\textit{dark yellow}) and L19 (\textit{dark purple}) data in the case in which the correlation coefficients of the shot noise are fixed to unity.}
\end{figure}

\begin{figure}[t]
\centering
\includegraphics[width=.5\textwidth]{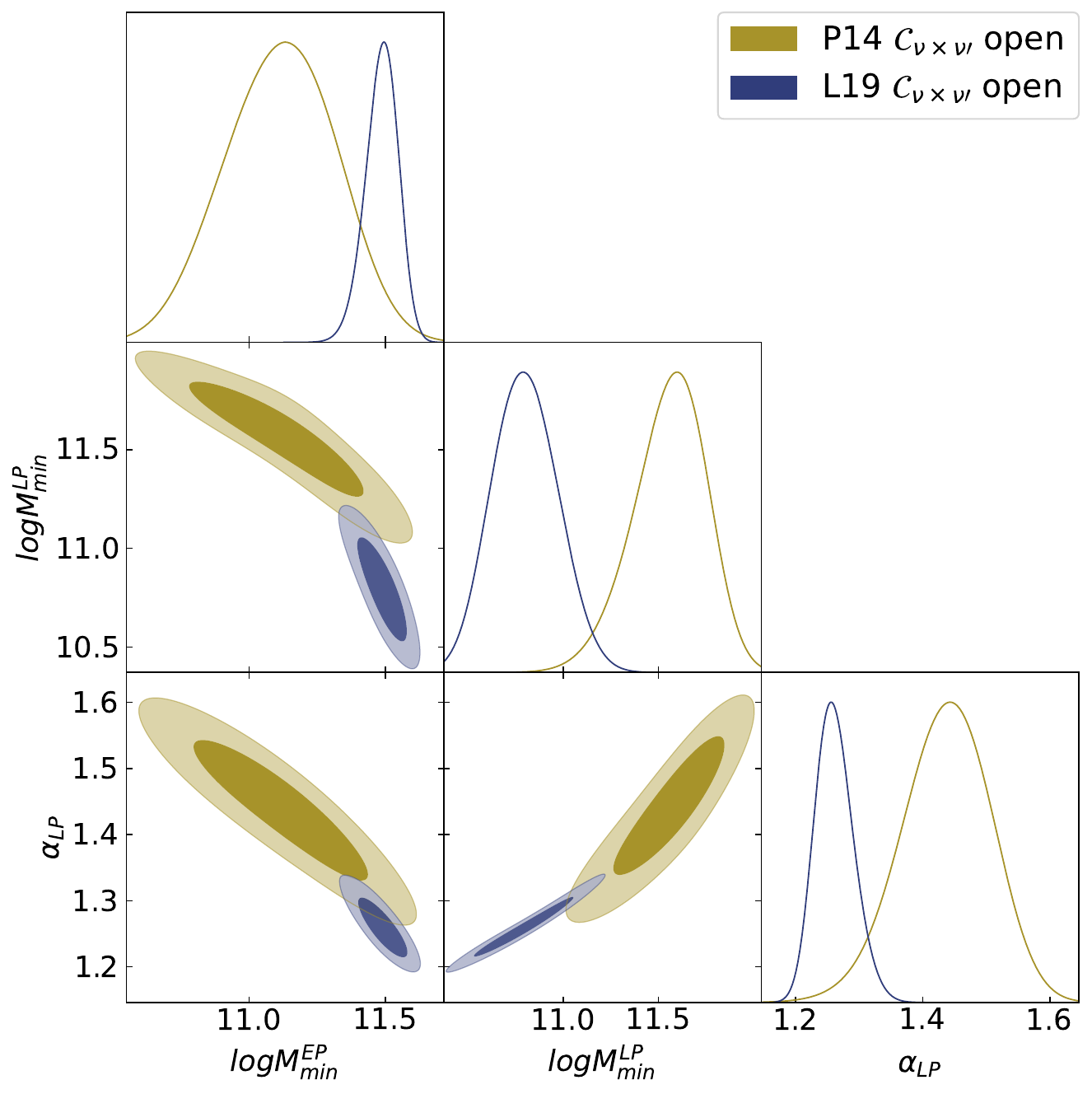}
\caption{\label{fig:lenz Planck comp open} Same as Fig.~\ref{fig:lenz Planck comp} for the case in which the correlation coefficients of the shot noise are free to vary.}
\end{figure}

Without focusing on the exact results from L19, in the following we do a qualitative exploration of how the posteriors compare to those of P14. The clustering parameters for the two scenarios explored in this study are shown in Fig.~\ref{fig:lenz Planck comp} and \ref{fig:lenz Planck comp open}. The overlap between the two datasets is good (the largest shift is noted in the 1D posterior distribution of the minimum masses for the LT galaxies in Fig.~\ref{fig:lenz Planck comp open}, approaching a $\sim 3\sigma$ difference). 


\subsection{\textit{Herschel}-SPIRE data analysis}\label{subsec:V18 model fit}
We now move to discuss the results from the V18 SPIRE dataset.

The model parameters and the prior distributions adopted in the analysis are reported in the first and second column of Table~\ref{tab: best fit SPIRE}, respectively. Unlike the previous datasets, the SPIRE data allows to constrain the $\alpha_{\text{EP}}$ clustering parameter. We therefore let it free to vary along with the other clustering parameters. The calibration factors for the three frequency channels, $f_{\text{cal}}^i$ with $i=600,857,1200$, are varied with a Gaussian prior centered on 1 and a $1\sigma$ width determined as twice the calibration accuracy reported in \cite{Valtchanov:2017}. As done for the P14 and L19 analyses, we consider two scenarios of fixed and free-to-vary shot-noise correlation coefficients.

\begin{figure*}[ht]
\centering
\includegraphics[width=1.\textwidth]{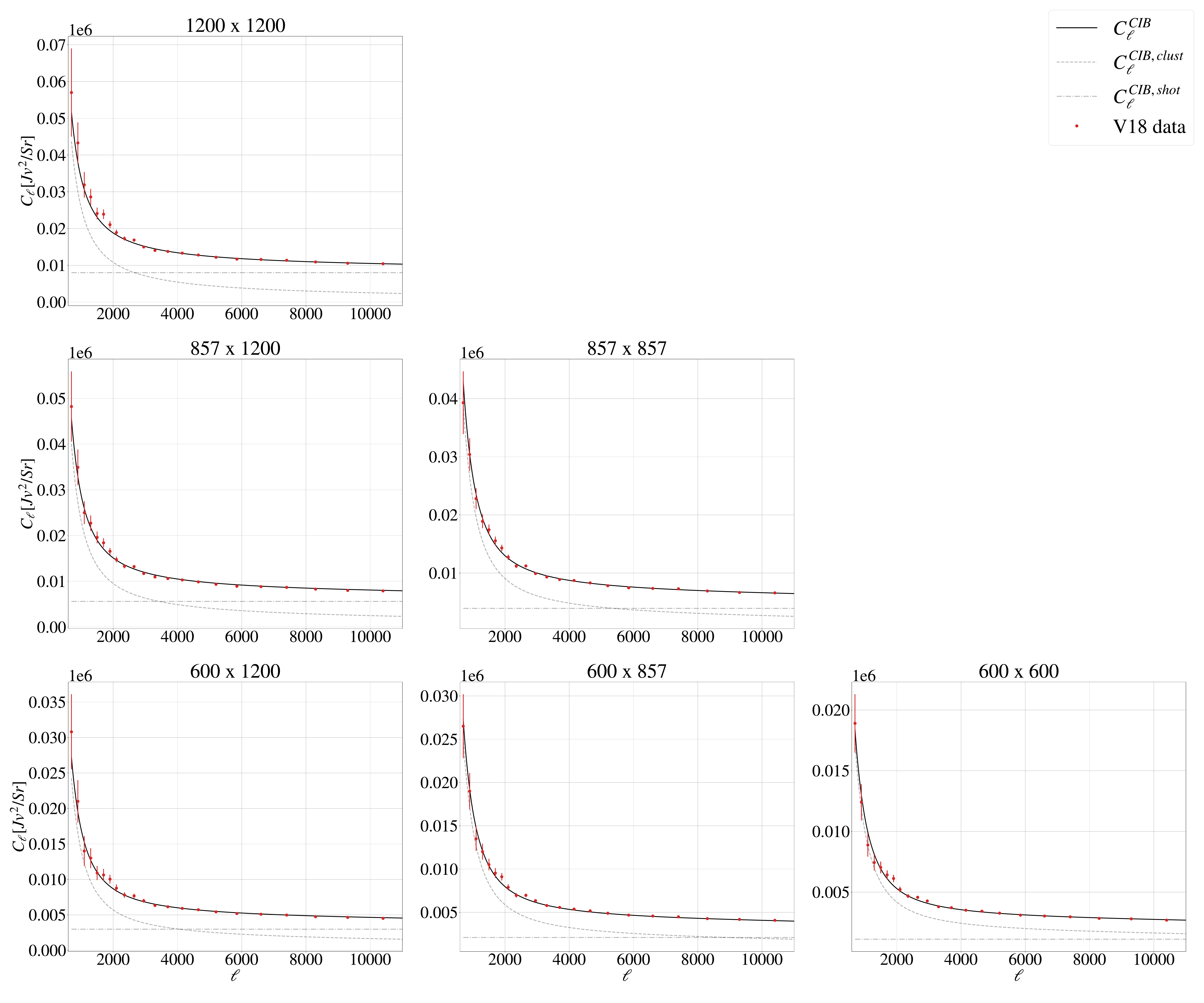}
\caption{\label{fig:SPIRE model data}Same as Fig.~\ref{fig:planck model data} for the V18 dataset and using the best-fit parameters from Table \ref{tab: best fit SPIRE}.}
\end{figure*}

The results of the model fit ($68\%$ confidence level) for the two cases under study are reported in the third and the fourth columns of Table~\ref{tab: best fit SPIRE}. We obtain $\chi^2 = 101$ for 122 degrees of freedom, when fixing the correlation coefficients for the shot noise, and $\chi^2 = 80$ for the scenario where the correlation coefficients are free to vary and the number of degrees of freedom is 119. The spectra-model comparison is shown in Fig.~\ref{fig:SPIRE model data}, for the case where the correlation coefficients for the shot noise are fixed to unity across all the V18 spectra. Specifically, V18 data starts to be sensitive not only to the information coming from the clustering part, but also to the shot noise contribution. We find good agreement both in the multipole region where the clustering dominates and in the region when the shot noise dominates, at all frequencies.

\begin{figure}[t]
\centering
\includegraphics[width=.5\textwidth]{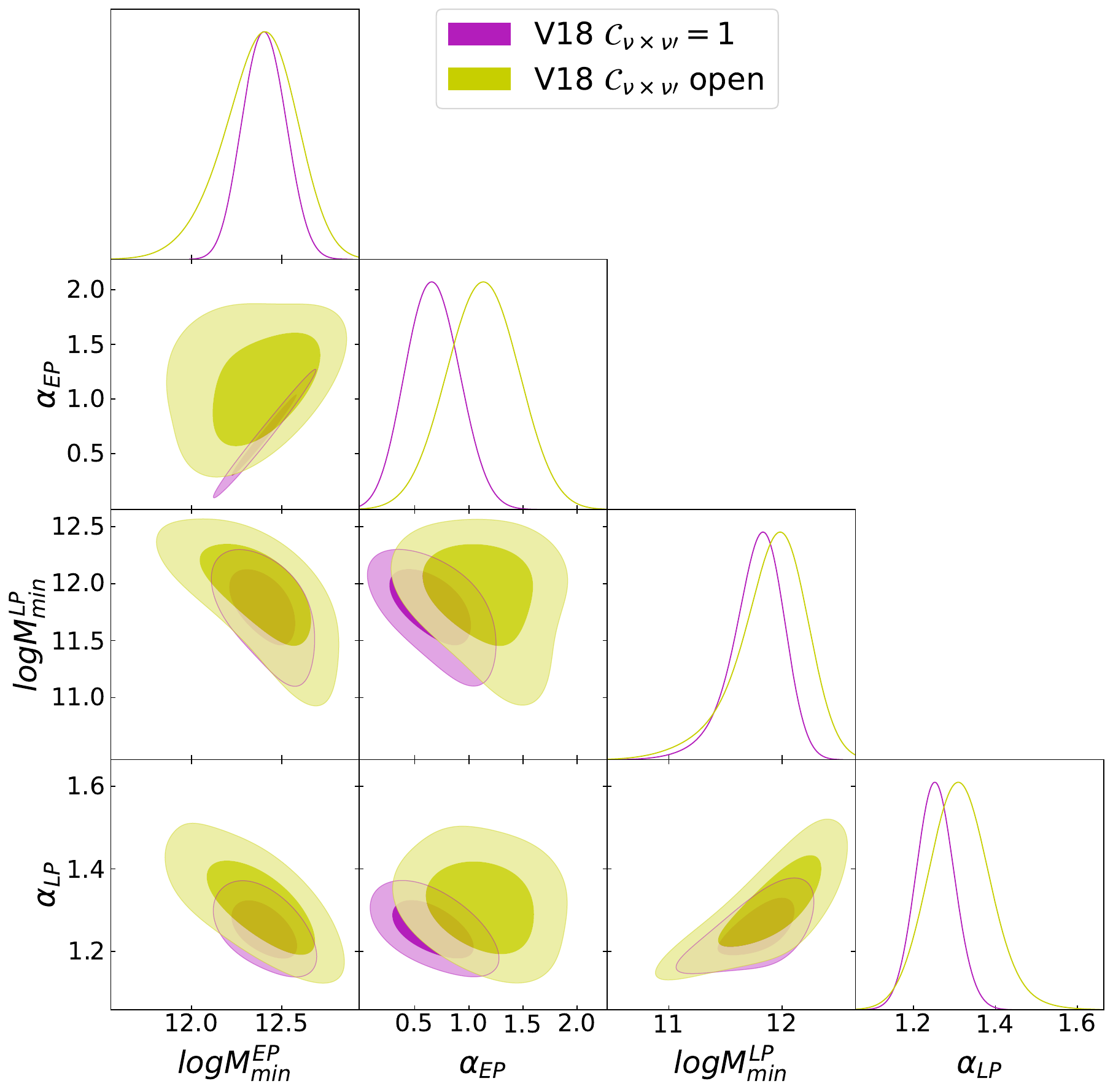}
\caption{\label{fig:SPIRE clust best fit}Same as Fig.~\ref{fig:planck clust best fit} for the V18 CIB dataset.}
\end{figure}

Figure~\ref{fig:SPIRE clust best fit} shows the posterior distributions of the clustering parameters when fixing the correlation coefficients for the shot noise to unity (\textit{purple}) and when letting them free to vary (\textit{green}).

\begin{table}[t!]
	\begingroup
	\nointerlineskip
	\vskip -3mm
	\tiny
	\setbox\tablebox=\vbox{
		\newdimen\digitwidth
		\setbox0=\hbox{\rm 0}
		\digitwidth=\wd0
		\catcode`*=\active
		\def*{\kern\digitwidth}
		\newdimen\signwidth
		\setbox0=\hbox{+}
		\signwidth=\wd0
		\catcode`!=\active
		\def!{\kern\signwidth}
		\halign{\hbox to 0.8in{#\leaderfil}\tabskip=1em&
			\hfil#\hfil\tabskip=10pt&
			\hfil#\hfil\tabskip=10pt&
			\hfil#\hfil\tabskip=10pt&
			\hfil#\hfil\tabskip=10pt&
			\hfil#\hfil\tabskip=10pt&
			\hfil#\hfil\tabskip=0pt\cr
			\noalign{\doubleline}
			\noalign{\vskip 3pt}
            \omit\hfil Parameter\hfil & Prior & Results & Results\cr
            \omit & \omit  & $\mathcal{C}_{\nu_1\times\nu_2}=1$ & $\mathcal{C}_{\nu_1\times\nu_2}$ open \\[1ex]\cr
			\noalign{\doubleline}
  $\log(M_{\text{min}}^{\text{EP}}/\Msun  h^{-1})$ & [10.7,12.8] & $12.40\pm 0.11$ & $12.38^{+0.22}_{-0.18}$\cr
  $\alpha_{\text{EP}}$ & [0.2,3.5] & $0.67\pm 0.23$ & $1.12\pm 0.31$\cr
  $\log(M_{\text{min}}^{\text{LP}}/\Msun  h^{-1})$ & [10.5,12.8] & $11.79^{+0.25}_{-0.19}$ & $11.91^{+0.34}_{-0.25}$\cr
  $\alpha_{\text{LP}}$ & [0.2,3.5] & $1.253\pm 0.044$ & $1.311\pm 0.071$\cr
\noalign{\vskip 3pt\hrule\vskip 5pt}
  ${\rm \shot}_{600}$ & [300,3000] & $1117^{+120}_{-140}$ & $1788\pm 300$\cr
  ${\rm \shot}_{857}$ & [2000,7000] & $3739\pm 340$ & $5153\pm 530$\cr
  ${\rm \shot}_{1200}$ & [5000,13000] & $8506\pm 710$ & $9688\pm 900$\cr
  $f^{600}_{\text{cal}}$ & $1\pm 0.11$ & $1.055\pm 0.071$ & $1.046\pm 0.084$\cr
  $f^{857}_{\text{cal}}$ & $1\pm 0.11$ & $1.072\pm 0.064$ & $0.973\pm 0.066$\cr
  $f^{1200}_{\text{cal}}$ & $1\pm 0.11$ & $0.962\pm 0.066$ & $0.907\pm 0.069$\cr
\noalign{\vskip 3pt\hrule\vskip 5pt}
  $\mathcal{C}_{600 \times 857}$ & [-1, 1] & - & $0.983^{+0.014}_{-0.012}$\cr
  $\mathcal{C}_{600 \times 1200}$ & [-1, 1] & - & $0.885\pm 0.032$\cr
  $\mathcal{C}_{857 \times 1200}$ & [-1, 1] & - & $0.9558\pm 0.0092$\cr
	\noalign{\vskip 5pt\hrule\vskip 3pt}}}
 \endPlancktable
\vskip 3pt
\caption{Same as Table~\ref{tab: best fit planck} for the V18 dataset. When the correlation coefficients are fixed to unity, we obtain $\chi^2 = 101$ with 132 points and 10 free parameters. In the case where the correlation coefficients are free to vary, we have 13 free parameters and obtain $\chi^2 = 80$.}
\label{tab: best fit SPIRE}
 \endgroup
\end{table}

The best-fit values which we obtain for the minimum mass of the ET galaxies are higher than those obtained in the analysis conducted on P14 data. They are also higher than the ones previously found in C13 and X12 (and reported in Table~\ref{tab: literature values clust}). Similarly, the minimum mass for LT galaxies in both scenarios explored in this work appears to be higher that the value found in C13 ($\log(M_{\text{min}}^{\text{LP}}/\Msun  h^{-1}) = 10.85\pm0.06$), totalling a $\sim 4\sigma$ tension between our work and previous literature. This difference could be explained by the fact that we are exploring an extended model compared to C13 who fixed $\alpha_{\text{LP}} = 1$. In fact, from Fig.~\ref{fig:SPIRE clust best fit} we note that, due to the positive correlation between $\log(M_{\text{min}}^{\text{LP}})$ and $\alpha_{\text{LP}}$, values of $\alpha_{\text{LP}}$ closer to unity will pull $\log(M_{\text{min}}^{\text{LP}})$ towards lower values. 

Our results also showcase V18 data ability to distinguish among the two galaxy populations and therefore their sensitivity to $\alpha_{\text{EP}}$ which is unconstrained by \planck. The constraints on $\alpha_{\text{EP}}$ are in agreement with semi-analytical models \citep{Gao:2004au,Hansen:2007fy}, predicting $\alpha_{\text{EP}}\leq 1.0$. However, we note that these results are not in agreement with the estimated values of $\alpha_{\text{EP}} = 1.55 \pm 0.05$ and $\alpha_{\text{EP}} = 1.81 \pm 0.04$ found in C13 and X12, corresponding to a tension of $\sim 4\sigma$ and $5\sigma$ respectively. The discrepancy between our findings and those of C13 and X12 reduces when considering the correlation coefficients of the shot noise as free parameters. Specifically, the $\alpha_{\text{EP}}$ clustering parameter in that scenario is compatible with the values of C13 and X12 within $1\sigma$ and $2\sigma$, respectively. We can explain this better agreement looking at how the model parameters act on the spectra. The $\alpha_{\text{EP}}$ clustering parameter (as well as $\alpha_{\text{LP}}$) re-scale the overall power, similarly to calibration factors. From Table~\ref{tab: best fit SPIRE}, we see that in the case of free correlations, the shot noise values change and carry with them the correlation factors which go low and reduce power in the spectra. This is then compensated by power added back in by higher clustering parameters. Concerning the spectral index of LT galaxies, a previous study conducted by C13 shows that $\alpha_{\text{LP}}$ was only weakly constrained and therefore kept fixed at unity. Here we show that the V18 data are capable of constraining this parameter at $\alpha_{\text{LP}} = 1.253\pm 0.044$ and $\alpha_{\text{LP}}=1.311\pm 0.071$ for the two scenarios. 

Shot noise levels depend on the scenario under scrutiny. When assuming maximal correlation between frequencies (correlation coefficients fixed to unity), the shot noise values are a factor of $\sim 2$ smaller than the levels predicted by the model of \cite{Bethermin:2012ki} and reported in Table 4 of \cite{Lagache:2019xto}. Allowing shot noise correlation coefficients to vary, or in other words, giving more freedom to the model fit, leads the parameters to move towards values that are closer to the ones reported in the literature.

We conclude by commenting on the calibration factors. They are generally consistent with 1 within $1\sigma$ regardless of the treatment of the shot noise correlation coefficients. We only report a low value of the 1200GHz factor $f_{\text{cal}}^{1200}$, which deviates from 1 at $2\sigma$ in the case of free-to-vary correlations between shot noise levels. This may compensate for a lower value of the shot noise level at 1200GHz with respect to the value found in the case of fixed shot-noise correlations.

\subsection{P14 and V18 comparison}\label{subsec:P14-SPIRE}
Given the general agreement between the posterior distributions of P14 and L19 shown in Sec.~\ref{subsec:L19 model fit} and given the qualitative nature of our L19 results, in this section we compare the results between \planck\, and SPIRE data taking the P14 and V18 results.

\begin{figure}[t]
\centering
\includegraphics[width=.5\textwidth]{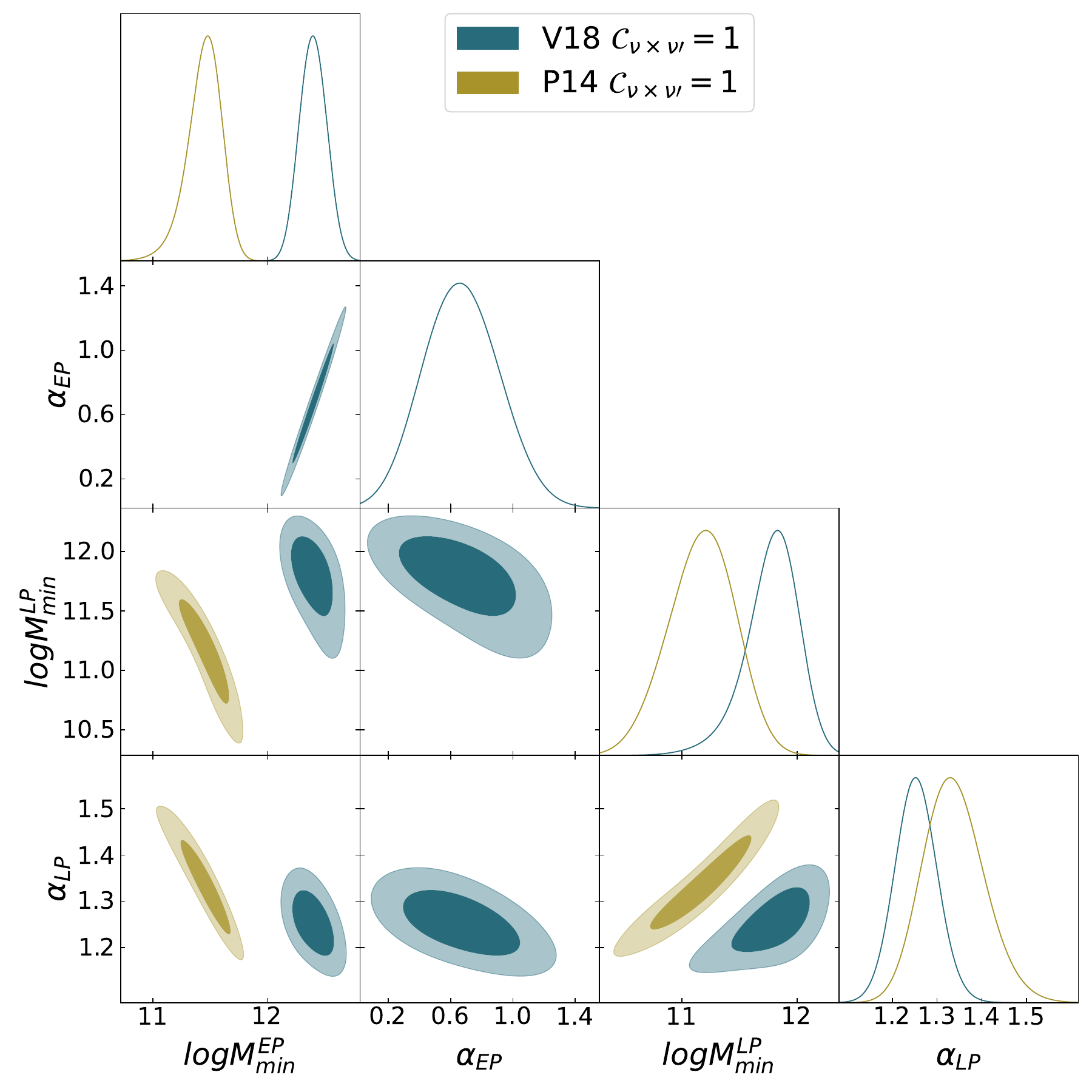}
\caption{\label{fig:comparison Planck SPIRE ones} Comparison of the posterior distributions of the clustering parameters from P14 (\textit{dark yellow}) and V18 (\textit{dark blue}) data for the case in which the correlation coefficients of the shot noise are fixed to unity. The two posterior distributions are in tension, not exploring the same regions of the parameter space.}
\end{figure}

\begin{figure}[t]
\centering
\includegraphics[width=.5\textwidth]{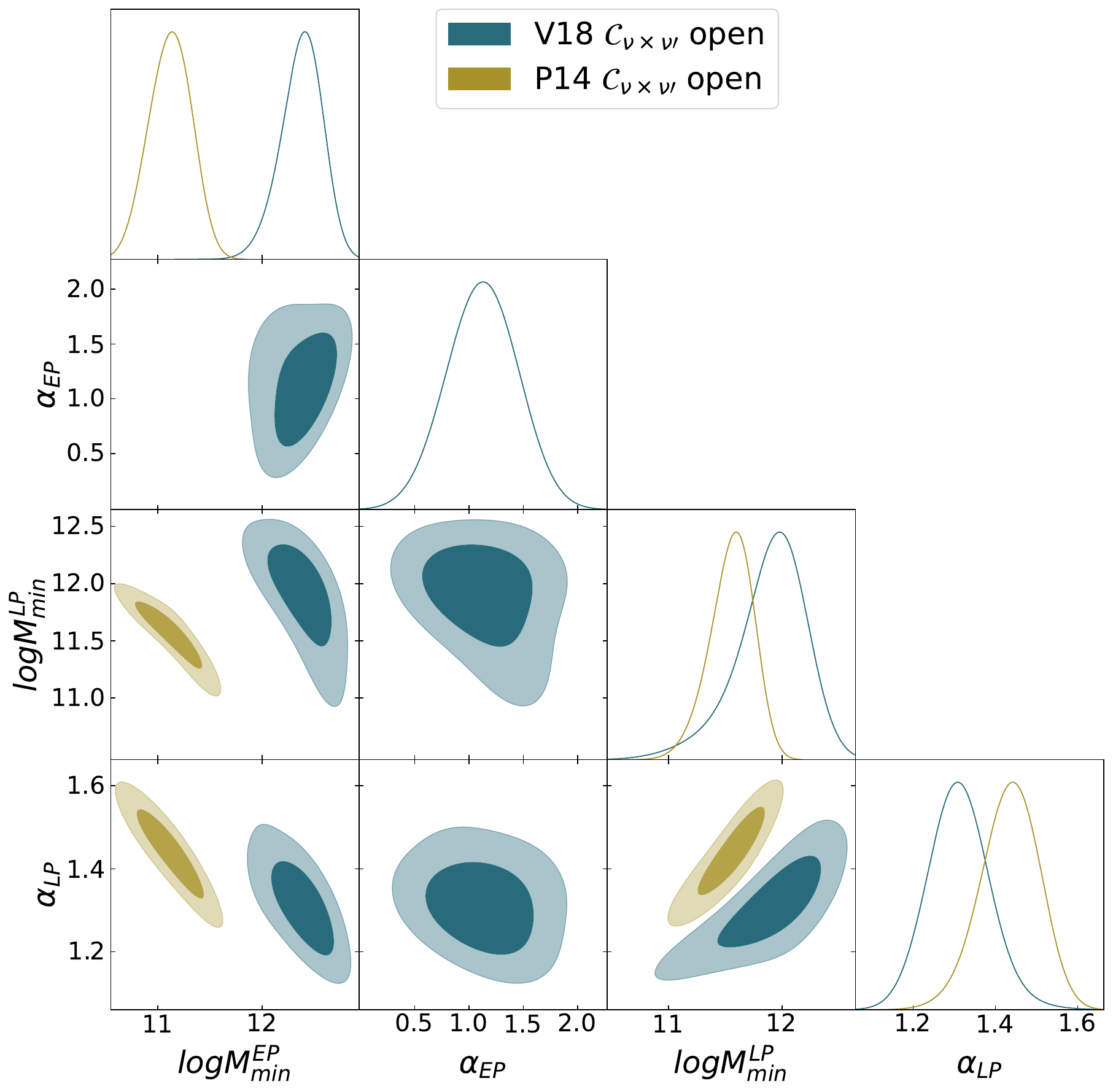}
\caption{\label{fig:comparison Planck SPIRE open} Same as Fig.~\ref{fig:comparison Planck SPIRE ones}, for the case in which the correlation coefficients of the shot noise are free to vary in the analysis. The two datasets are still in tension.}
\end{figure}

Figures \ref{fig:comparison Planck SPIRE ones} and \ref{fig:comparison Planck SPIRE open} show the posterior distributions of the clustering parameters for V18 (in \textit{dark blue}) and P14 (in \textit{dark yellow}) for the case in which the correlation coefficients of the shot noise are fixed to unity and the case in which they are free to vary, respectively. From the inspection of these figures, we note a tension between the two datasets in the best-fit values of the minimum mass of the ET galaxies, at the level of $\sim5\sigma$ for both scenarios. We find better agreement for the $\log(M_{\text{min}}^{\text{LP}}/\Msun  h^{-1})$ and $\alpha_{\text{LP}}$ clustering parameters. Specifically, the best-fit values for the minimum mass for LT galaxies from the two datasets are in agreement within $\sim2\sigma$ when fixing the correlation coefficients, and $\sim1\sigma$ for the case where the correlation coefficients are free to vary. Lastly, the best-fit values for the $\alpha_{\text{LP}}$ clustering parameter are in agreement within $\sim1\sigma$ for the two datasets and for both the scenarios explored in this study. We remind the reader that the P14 data do not allow to provide significant constraining power for $\alpha_\mathrm{EP}$, which is thus fixed to 1.  


We performed a number of tests to verify this comparison and to understand the impact of method and model assumptions on the agreement/disagreement. In particular, for P14 we tried excluding the 217 GHz frequency channel and fixing the $\alpha_{\text{EP}}$ clustering parameter to values different from 1 (\textit{e.g.} $\alpha_{\text{EP}} = 0.5$ and $\alpha_{\text{EP}} = 1.5$). For V18, we performed the MCMC fixing the $\alpha_{\text{EP}}$ clustering parameter in order to recover the same setup used for P14. None of these tests improved the agreement between the two datasets. Given the tension between the two datasets, we do not perform a joint analysis. The tests we have done and the limited availability to compare overlapping scales do not allow us to clearly identify the source of the discrepancy between the datasets. The tension may equally point at inherent differences in the datasets or a failure in the ability of our model to capture the complexity of the physics under scrutiny (for example, we are not including the luminosity dependence of the HOD, as done in \citealp{Viero:2012uq}). The analysis of independent data from other surveys may help shed light into this tension. 
 
\section{Conclusions and future perspectives}
In this work, we applied a halo model formalism to derive a physically motivated model of the CIB emission. The model includes contributions to the CIB emission generated by two different types of galaxy populations, namely early-type (ET) and late-type (LT) galaxies, which show different halo distribution and emission characteristics. For the first time, the mixing terms between these populations has also been fully integrated into the calculations. In the model, we distinguish between a contribution from the clustering of matter to the CIB power spectrum (dominating at larger angular scales) and a shot noise contribution (dominating at smaller scales). The model has been compared against three state-of-the-art datasets — data from the official \planck\, release (P14), a reanalysis of \planck\, data by Lenz (L19), and data from \textit{Herschel}-SPIRE (V18). The datasets have been chosen so to cover two observational regimes: lower frequency and lower multipole ranges (\planck); high-frequency, high-multipoles ranges (SPIRE). The two regimes map onto different sensitivities to the model parameters and contributions from galaxy populations. 

We found that P14 data provide a better handle to the clustering parameters. However, the frequency range probed with P14 is not wide enough to effectively distinguish between contributions from the two galaxy populations. Qualitatively, L19 data say the same story. The same does not apply to V18. In this case, the analysis allows to constrain the clustering parameters of both ET and LT galaxies. 

When comparing the results from P14 and V18, we report -- in agreement with previous analyses (see e.g., \citealp{Maniyar:2020tzw}) -- a tension between the two datasets. This discrepancy is severe enough to prevent us from performing a joint fit. At this stage, we are unable to clearly pinpoint the source of this discrepancy, which might be due to either the data themselves or to model assumptions (or both).

Our work also highlights that the results are highly dependent on the choices made in the analysis and on which priors and external information are imposed on the parameters. We provide a long list of tests in this paper but found limitations in the constraining power of the datasets at hand, we need new observations to be able to lock in further the model. 

The tension observed between the low- and high-frequency/multipole range experiments suggests that surveys spanning wider frequency/angular scale coverage could help shed light on the current discrepancies, as well as advance our understanding of the physics of CIB emission.

The approach used here for the CIB can be extended to provide a physical description of the Sunyaev Zeldovich (SZ) effect. Furthermore, being CIB and SZ tracers of the same underlying matter distribution, the extension of the halo model approach to the SZ effect will also allow properly predict the correlation between the two tracers (building on other existing literature). 

\section*{Acknowledgements}
We thank Laura Salvati and Ian Harrison for useful discussions during the preparation of this work. We thank Abhishek Maniyar for valuable information about the \texttt{halomodel\_cib\_tsz\_cibxtsz} code \citep{Maniyar:2020tzw}. We acknowledge the use of \texttt{numpy} \citep{harris2020array}, \texttt{matplotlib} \citep{Hunter:2007} and \texttt{getdist} \citep{Lewis:2019xzd} software packages. We acknowledge the CINECA award under the ISCRA initiative, for the availability of high performance computing resources. EC acknowledges support from the European Research Council (ERC) under the European Union's Horizon 2020 research and innovation programme (Grant agreement No. 849169). GZ and MG are funded by the European Union (ERC, RELiCS, project number 101116027). Views and opinions expressed are however those of the author(s) only and do not necessarily reflect those of the European Union or the European Research Council Executive Agency. Neither the European Union nor the granting authority can be held responsible for them. MG, GZ, CC, MG and LP~acknowledge the financial support from the COSMOS network (www.cosmosnet.it) through the ASI (Italian Space Agency) Grants 2016-24-H.0 and 2016-24-H.1-2018. 

\appendix
\section{Extended analyses of the P14 dataset}\label{app:Planck}
\subsection{Adding a freely varying $\alpha_{\text{EP}}$}\label{app:Planck_EP}
\begin{figure}[ht]
\centering
\includegraphics[width=.5\textwidth]{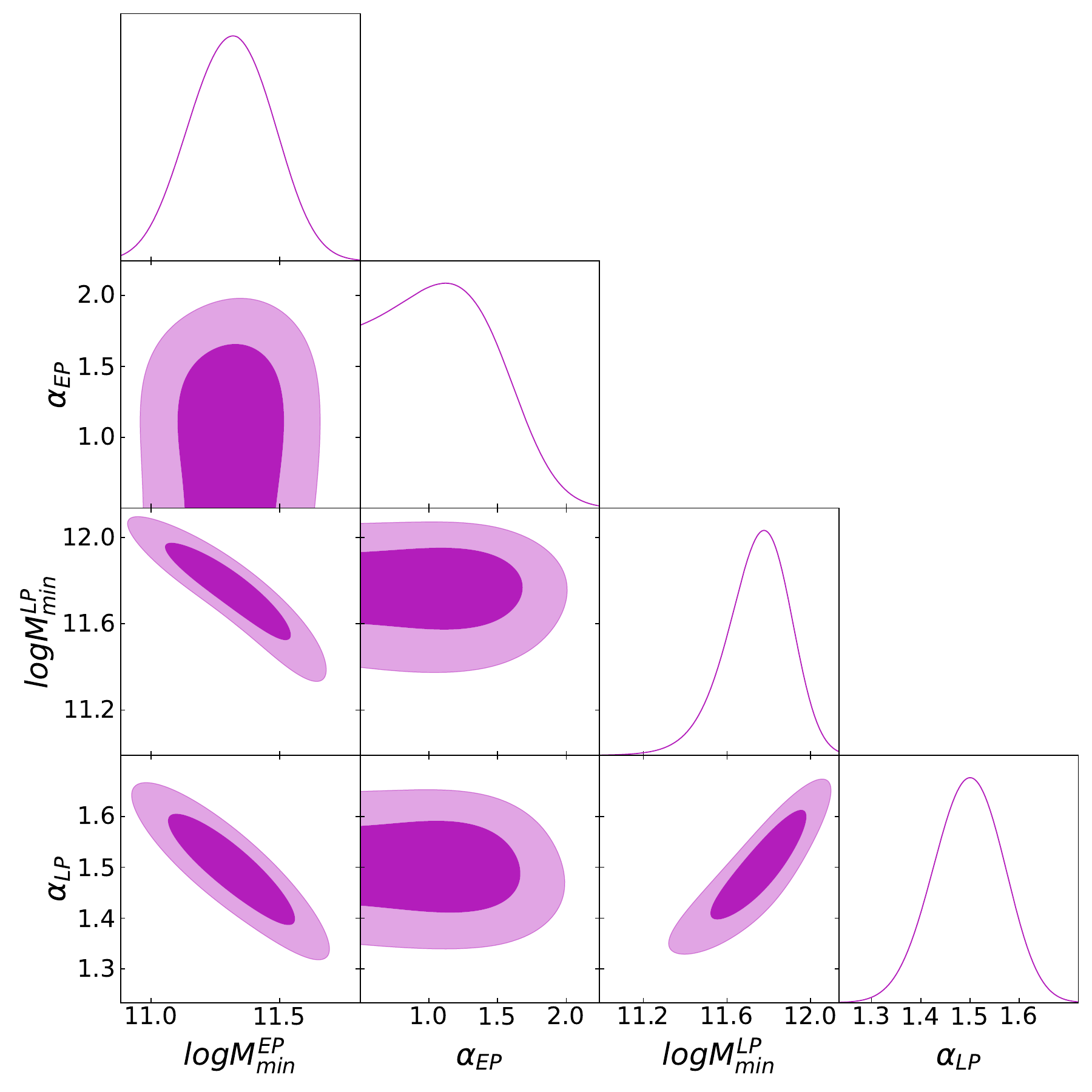}
\caption{\label{fig:planck alpha open} Posterior distributions of the clustering parameters from the analysis of the P14 dataset when allowing $\alpha_{\text{EP}}$ to vary freely.}
\end{figure}

We report here the results of the fit to P14 in an extended model where $\alpha_{\text{EP}}$ is free to vary. We sample $\alpha_{\text{EP}}$ from a uniform distribution in the range [0.2, 3.5]. The priors on the remaining parameters are the same as those reported in Table~\ref{tab: best fit planck}.
Figure \ref{fig:planck alpha open} shows the posterior distributions of the clustering parameters. 

P14 does not fully constrain $\alpha_{\text{EP}}$ beyond setting an upper bound of $\alpha_{\text{EP}}<1.14$ at 95\% CL. In addition, this parameter does not show significant correlation with the other clustering parameters. These considerations justify our choice of fixing $\alpha_{\text{EP}}=1$ in the baseline results presented in the main text.

\subsection{Full fit results for P14} \label{app:Planck_triangle}
\begin{figure*}[ht]
\centering
\includegraphics[width=1.\textwidth]{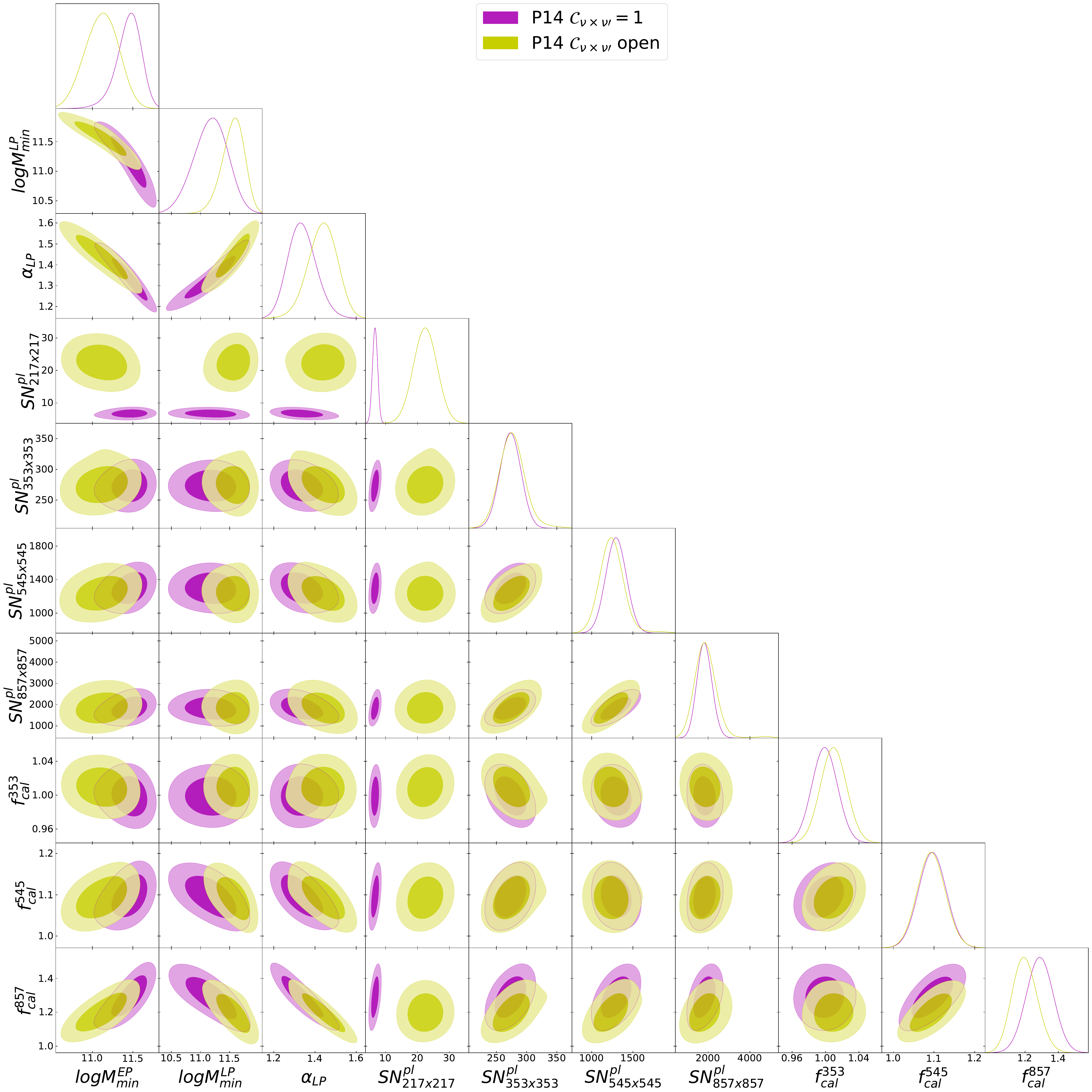}
\caption{\label{fig:planck tot comp}Same as Fig.~\ref{fig:planck clust best fit}, for all the parameters sampled in the MCMC analysis of P14 data.}
\end{figure*}

In Fig.~\ref{fig:planck tot comp}, we report the full triangle plot showing the the 1-dimensional posterior distributions and the 2-dimensional correlations (at 68 and 95\% confidence) of all the parameters sampled in the MCMC analysis discussed in the main text (i.e., the extended version of Fig.~\ref{fig:planck clust best fit} in Sec.\ref{subsec:P14 model fit}).

As mentioned above, we observe good agreement between the case where the shot noise correlation coefficients are fixed to unity and the case where they are free to vary, with the only exception concerning the shot noise level for the 217 GHz frequency channel. We refer the reader to the discussion done in Sec.~\ref{subsec:P14 model fit}.

\section{Extended analyses and tests of the L19 dataset}\label{app:Lenz}
\subsection{Full fit results for L19}
In this section, we show the qualitative comparison between L19 bandpowers and the theoretical predictions for the CIB spectra obtained using the best-fit values reported in Table~\ref{tab: best fit lenz} for the scenario where the shot noise correlation coefficients are fixed to unity.

\begin{figure*}[ht]
\centering
\includegraphics[width=1.\textwidth]{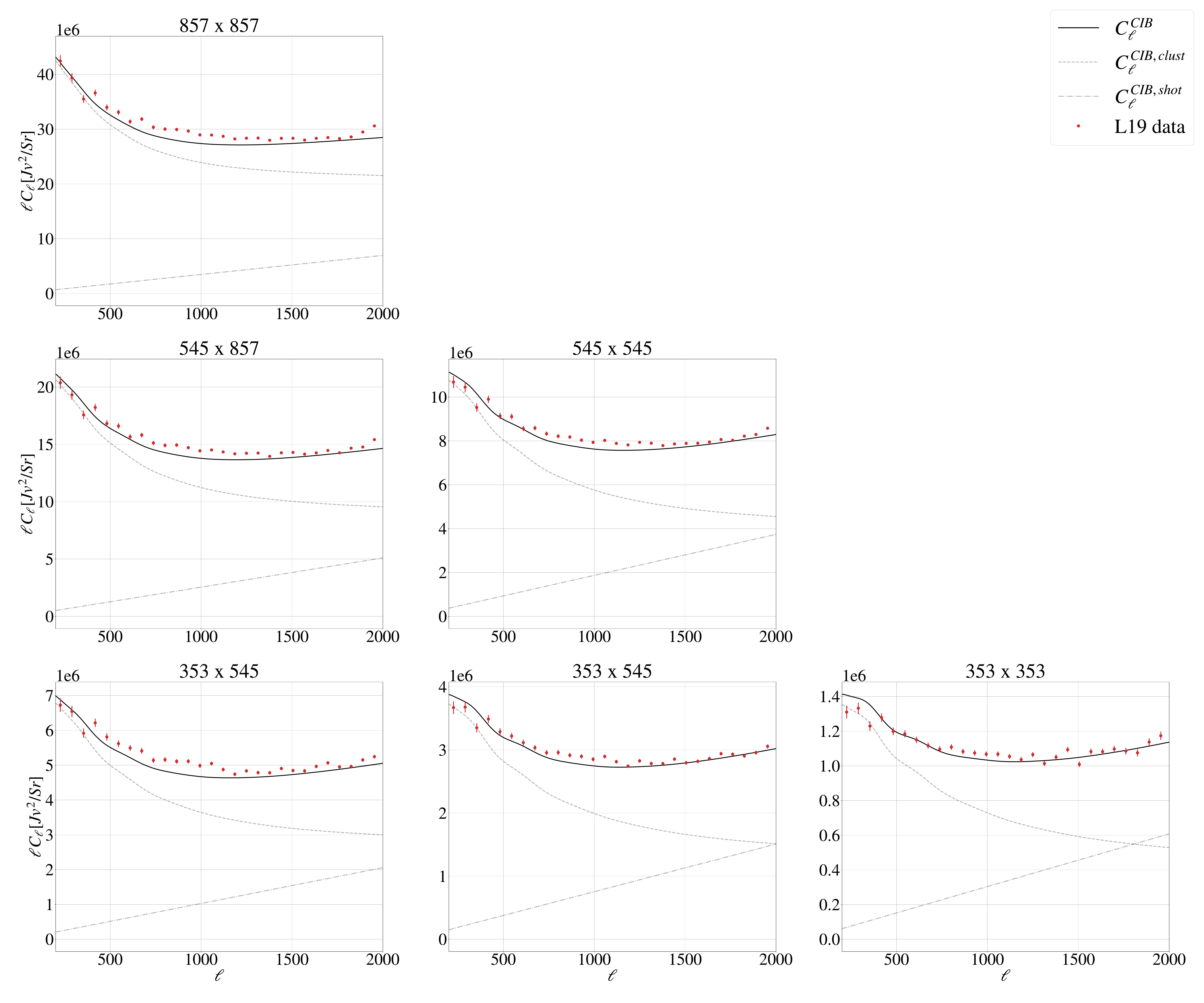}
\caption{\label{fig:Lenz model data}Same as Fig.~\ref{fig:planck model data} for the L19 dataset (red points) and using the best-fit parameters from Table~\ref{tab: best fit lenz}.}
\end{figure*}

From Fig.~\ref{fig:Lenz model data} we can immediately infer that, in the absence of strong correlations between points, the fit will not have a good $\chi^2$ (as noted in the main text in Sec.~\ref{subsec:L19 model fit}). A striking feature of this figure is the oscillating pattern of the L19 dataset which cannot be easily captured by the smooth predictions of our model.

\begin{figure*}[ht]
\centering
\includegraphics[width=1.\textwidth]{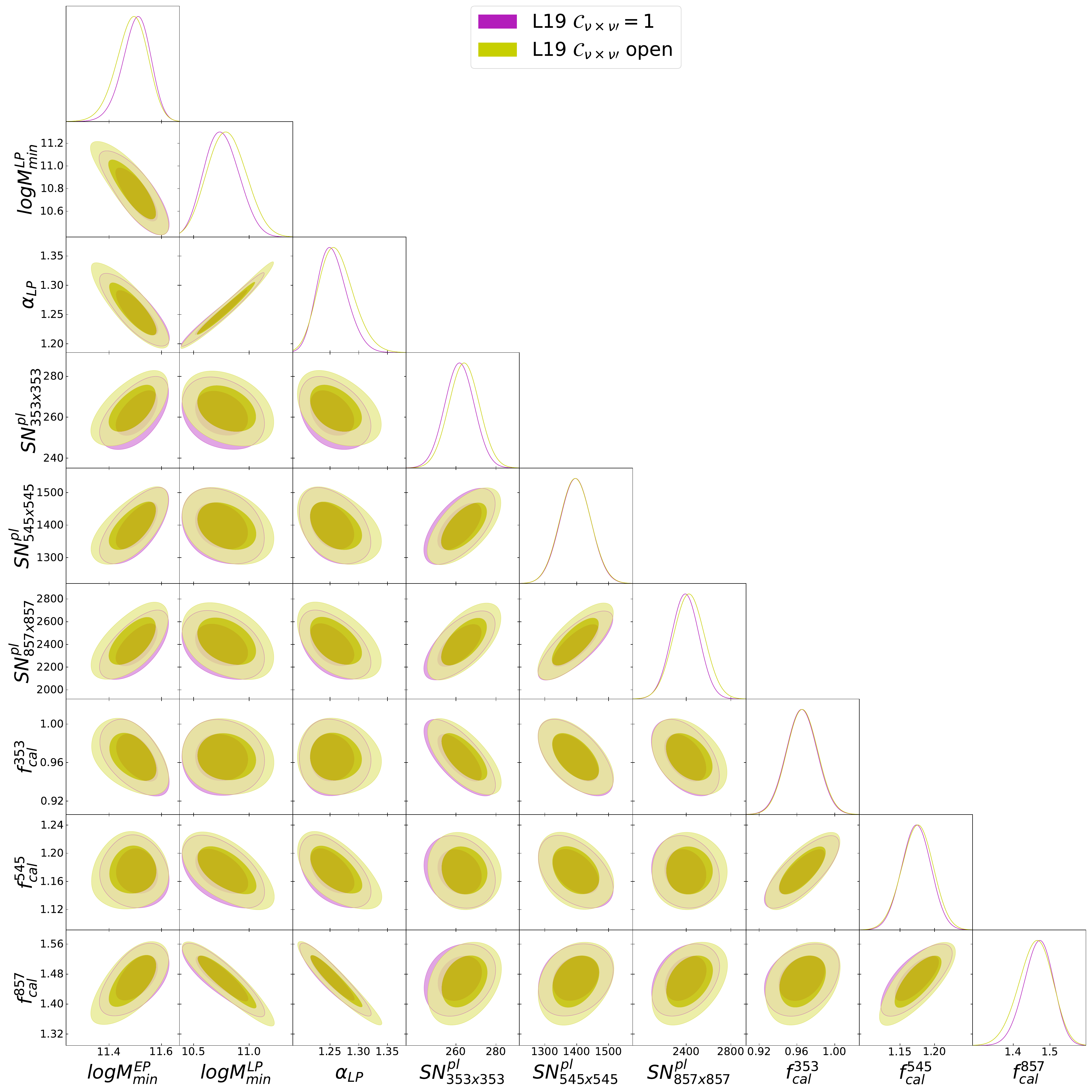}
\caption{\label{fig:Lenz tot comp}Same as Fig.~\ref{fig:planck tot comp} for L19 data.
}
\end{figure*}

In Fig.~\ref{fig:Lenz tot comp}, we report the full triangle plot of all the parameters for the two scenarios detailed in this study.

\subsection{Tests of the L19 CIB spectra}
To understand the L19 results further, we ran various tests on the L19 dataset taking the CIB intensity maps that they use and deriving from scratch CIB spectra. 

\begin{figure}
    \centering
    \includegraphics[width=.49\textwidth]{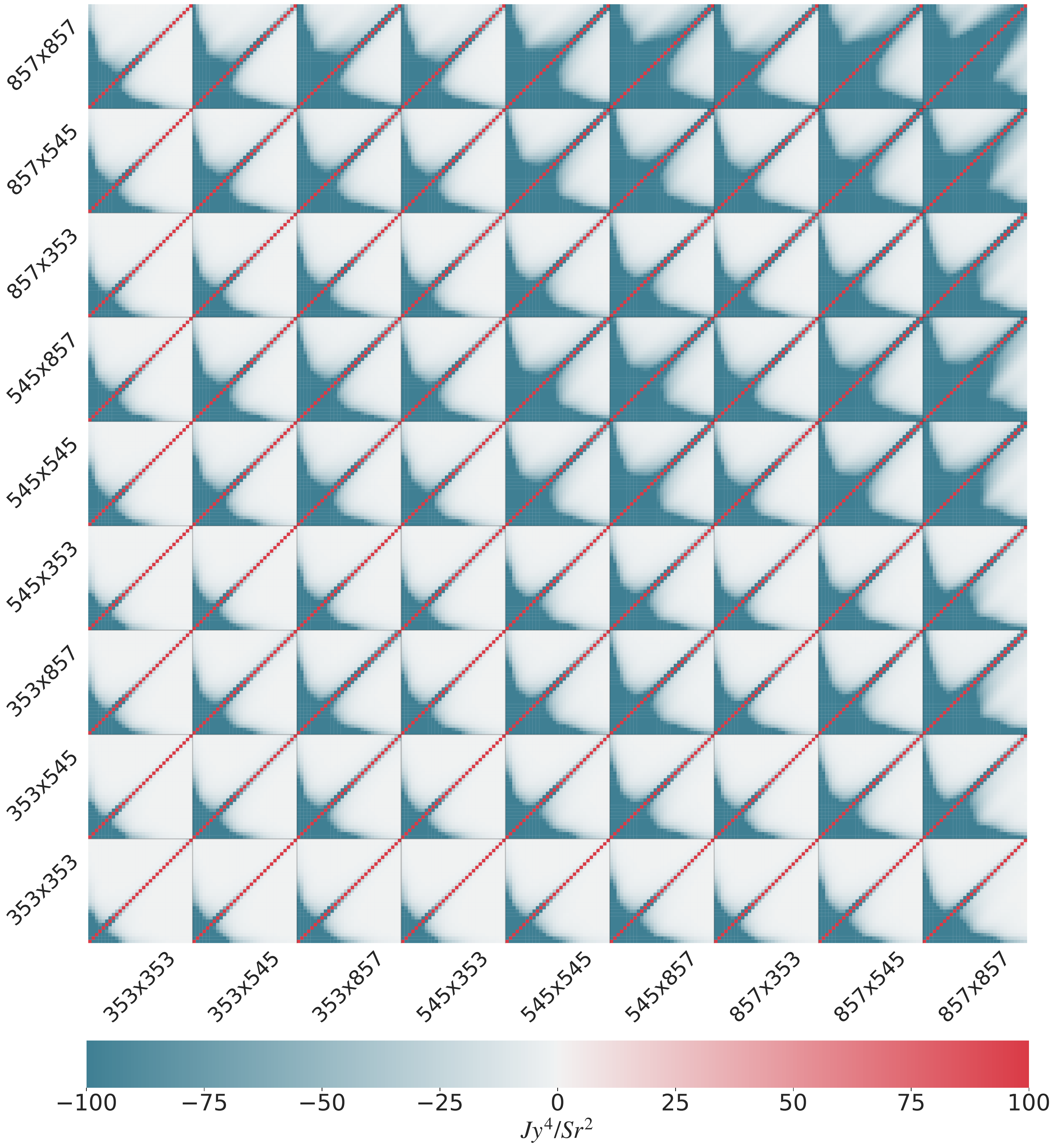}
    \caption{Covariance matrix of the L19 CIB measurements.}
    \label{fig:cov_mat}
\end{figure}

\begin{figure}
    \centering
    \includegraphics[width=.49\textwidth]{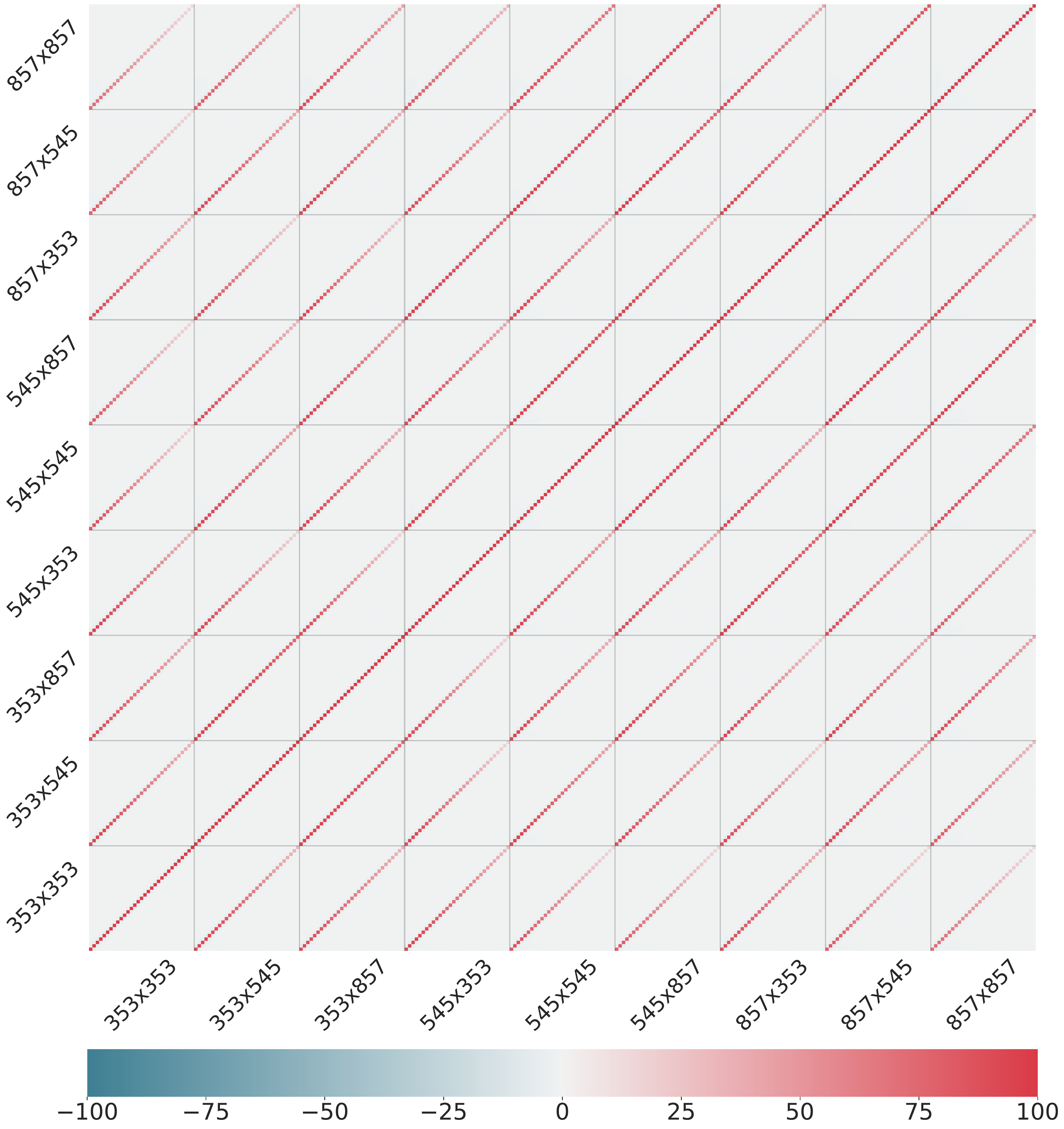}
    \caption{Correlation matrix of the L19 CIB measurements.}
    \label{fig:corr_mat}
\end{figure}

As initial step, we implement L19's pipeline to compute the CIB power spectra and extend it to build a Gaussian covariance matrix. We do this to explore the impact of possible bin-to-bin correlations in the fit and in the $\chi^2$. The covariance matrix that we obtain is shown in Fig.~\ref{fig:cov_mat}, while Fig.~\ref{fig:corr_mat} reports the correlation matrix. 
From these, we note that there is no significant correlation amongst different multipoles, but there is very significant correlation between the same bin at different frequencies. However, as we detail below, re-running the analysis including this new matrix showed no impact on the $\chi^2$ of the fit.


\begin{figure*}
    \centering
    \includegraphics[width=\textwidth]{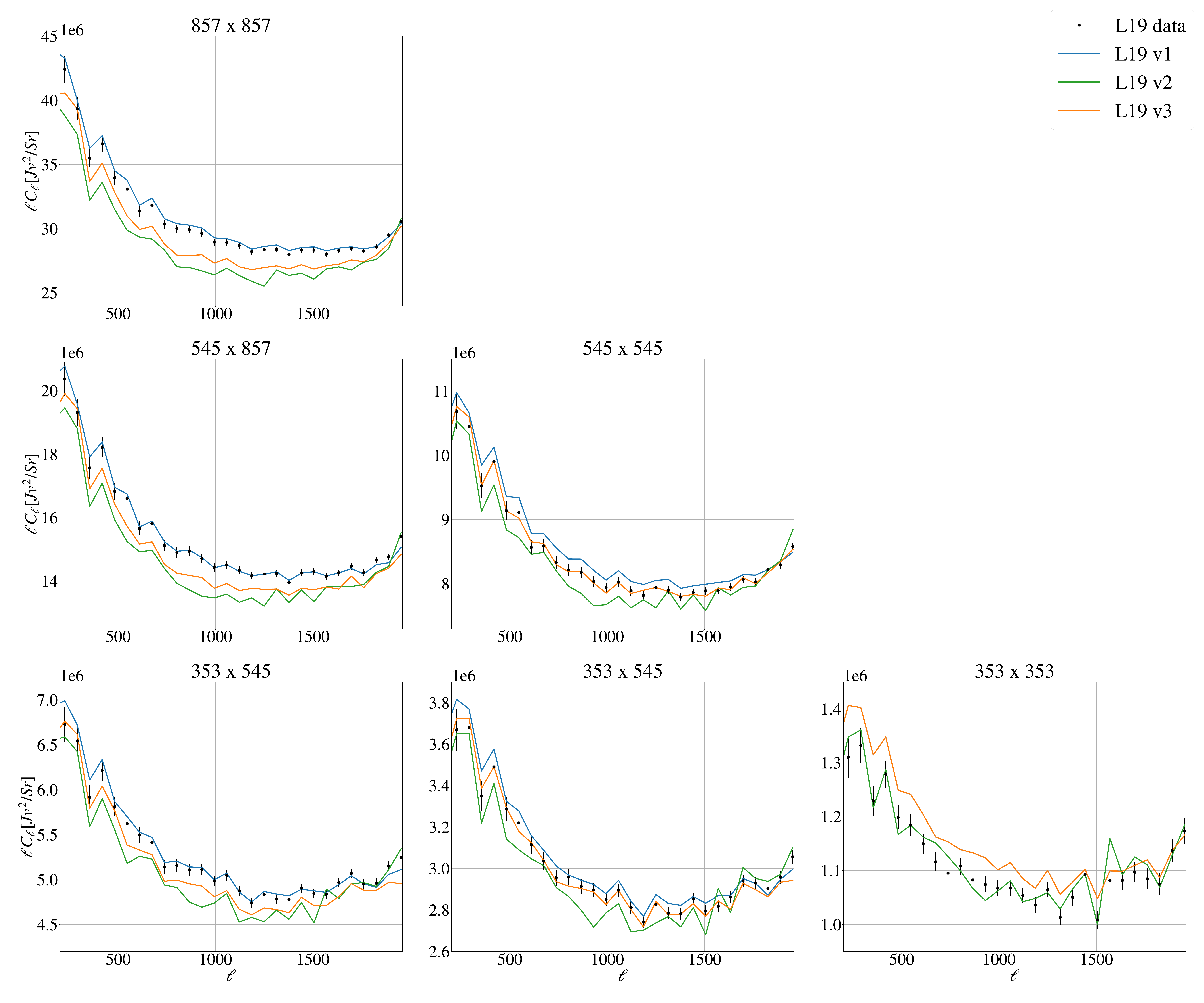}
    \caption{CIB power spectra from L19 (\textit{black} points) compared to the spectra obtain after a re-analysis of the maps and applying different neutral hydrogen column density thresholds. We represent in blue the dataset derived after the application of a threshold of $2.5 \times 10^{20} \text{cm}^{-2}$ across all three frequency channels, in \textit{green} the threshold applied to all frequency channels is of $1.5 \times 10^{20} \text{cm}^{-2}$, and in \textit{orange} we show the dataset obtained applying a different threshold depending on the frequency channel and following the suggestion of L19 ($N_{HI} = 2.5 \times 10^{20} \text{cm}^{-2}, 2.0 \times 10^{20} \text{cm}^{-2}$ and $1.8 \times 10^{20} \text{cm}^{-2}$ for 353, 545 and 857 GHz respectively.)}
    \label{fig:dataset comparison}
\end{figure*}

This covariance matrix accompanied three different configurations of the spectra:
\begin{itemize}[noitemsep,topsep=0pt]
    \item The first version of the dataset was obtained applying the same neutral hydrogen column density ($N_{HI}$) threshold of $2.5 \times 10^{20} \text{cm}^{-2}$ across all three frequency channels (blue curve in Fig.~\ref{fig:dataset comparison}).
    \item The second dataset applied a lower $N_{HI}$ threshold of $1.5 \times 10^{20} \text{cm}^{-2}$ across all channels to explore the impact of dust and in particular of dust residuals in the analysis (green curve in Fig.~\ref{fig:dataset comparison}).
    \item The third dataset has been obtained following the prescription of L19 and imposing a different threshold for each frequency channel. Specifically we set $N_{HI} = 2.5 \times 10^{20} \text{cm}^{-2}, ~2.0 \times 10^{20} \text{cm}^{-2}$ and $1.8 \times 10^{20} \text{cm}^{-2}$ for 353, 545 and 857 GHz respectively (orange curve in Fig.~\ref{fig:dataset comparison}).
\end{itemize}
Figure \ref{fig:dataset comparison} compares these three datasets with the publicly available L19 CIB power spectra (shown as black data points with errorbars).

We start our tests by using the first dataset to investigate the impact of different covariance choices on the goodness of the fit. To do this we go through:
\begin{enumerate}[noitemsep,topsep=0pt]
    \item Running a MCMC with the full covariance matrix;
    \item Using the full covariance matrix but limiting the analysis to the information from the frequency auto-spectra only;
    \item Employing only the diagonal of the covariance matrix;
    \item Exploring only individual frequency.
\end{enumerate}
The first run concluded that even when using the full covariance matrix, the model does not attain a good $\chi^2$, continuing to find values of $\sim 700$. Furthermore, we also observed very different behaviors in the posteriors and best-fit values compared to the fiducial run presented in Sec.~\ref{subsec:L19 model fit}. Specifically, the run using the full covariance matrix yielded very similar values for the minimum masses of ET and LT galaxies, compared to differences of $\sim 3 \sigma$ in the nominal case. Comparing the parameters posteriors with those obtained in the fiducial case presented in the main text (see Fig.~\ref{fig:lenz clust best fit} for the clustering parameters only and Fig.~\ref{fig:Lenz tot comp} for the full triangle plot), we noted that the minimum mass for the ET galaxies is no longer degenerate with the other clustering parameters. In contrast, the minimum mass for the LT galaxies and the $\alpha_{\text{LP}}$, and all the remaining free parameters of the fit, show the same degeneracies of the fiducial case. The runs that employed the full covariance with the frequency auto-spectra only and that using only the diagonal of the covariance matrix produced results that aligned more closely with the original L19 dataset findings, both in terms of best-fit values and parameter space degeneracies. This better alignment with our main findings, especially when considering the run employing only the diagonal of the covariance matrix, was expected as it closely resembles the scenario presented in Sec.~\ref{subsec:L19 model fit}. However, none of these tests led to better values of the $\chi^2$.

\begin{figure}
    \centering
    \includegraphics[width=.5\textwidth]{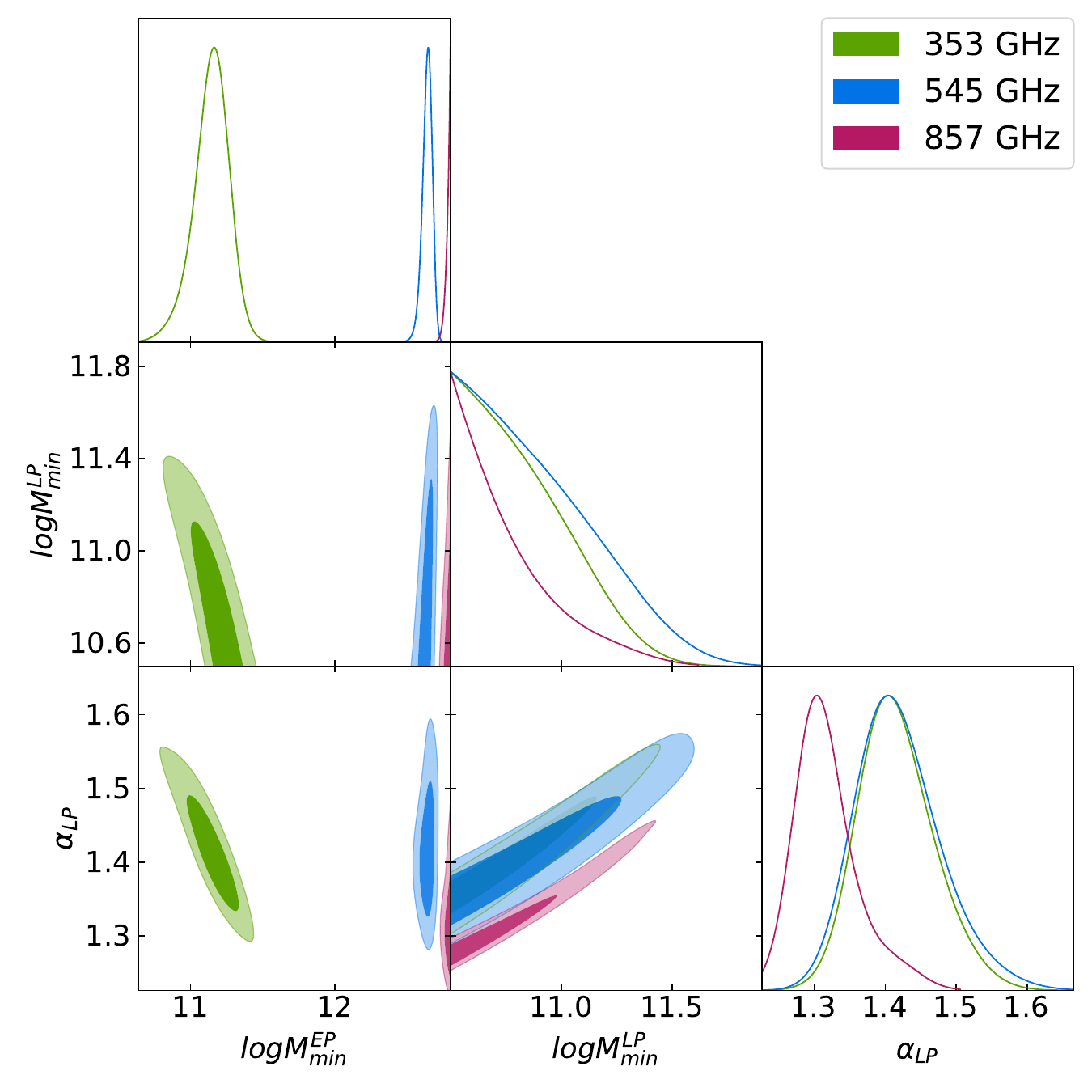}
    \caption{Comparison of the posterior distributions of clustering parameters from the single frequency MCMC using the first L19 dataset. 353, 545 and 857 GHz frequency channels are represented in green, blue and pink, respectively.}
    \label{fig:single_freq_ds1}
\end{figure}

When using the full covariance matrix but limiting the analysis to the information from the frequency auto-spectra only and employing only the diagonal of the covariance matrix, we observe that the agreement with the fiducial case presented in Sec.~\ref{subsec:L19 model fit} is stronger for the clustering parameters and slightly worse for the shot noise levels and the calibration factors, especially for the highest frequency channel. This led us to explore a possible dust residual which would be more important at higher frequencies. For this reason, we further investigated the model fit using single frequency data. None of the single frequency MCMC analyses could effectively constrain the minimum mass of the LT galaxies or the shot noise levels. Fig.~\ref{fig:single_freq_ds1} compares the recovered clustering parameters and shows a strong tension in the minimum mass of the ET galaxies between the 353 GHz frequency channel and the 545 and 857 GHz frequency channels. Specifically, the two higher frequency channels result in much higher best-fit values for the minimum mass of ET galaxies, both when compared to the best-fit value for the single frequency run involving the 353 GHz frequency channel, and when compared to the best-fit value of the minimum mass found in our fiducial case and reported in Table~\ref{tab: best fit lenz}. Recalling that this clustering parameter acts as a re-scaling of the power spectrum, the high values could possibly be hinting at an excess of power coming from dust residuals. 

\begin{figure}
    \centering
    \includegraphics[width=.5\textwidth]{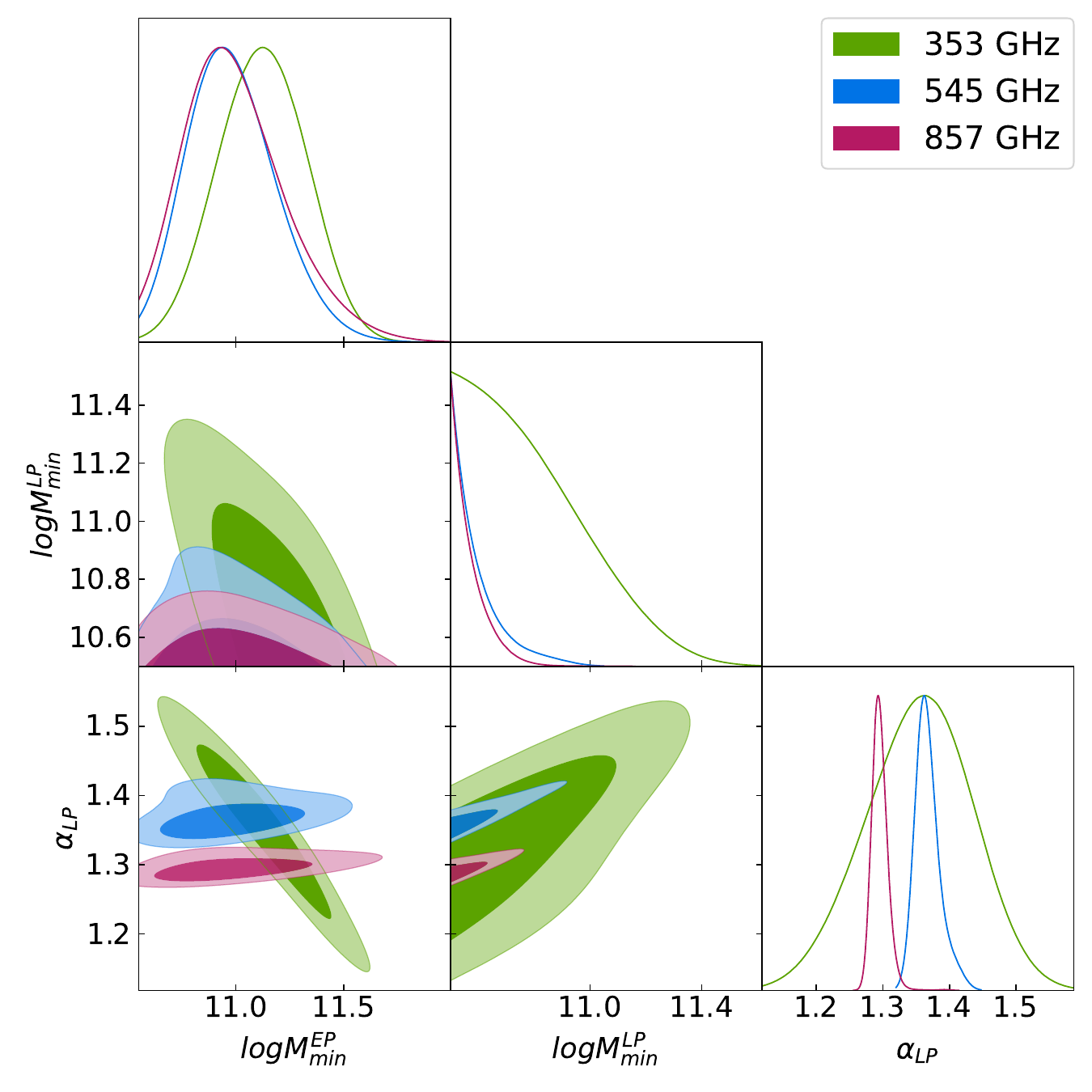}
    \caption{Same as Fig.~\ref{fig:single_freq_ds1} for the second L19 dataset.}
    \label{fig:single_freq_ds2}
\end{figure}

To better explore this hypothesis, we evaluate the model fit to single frequency data from the second version of the dataset (see Fig.~\ref{fig:single_freq_ds2}) which employs a more aggressive mask and therefore should have smaller contamination. We do not find a difference in the constraining power for the minimum mass of LT galaxies and for the shot noise levels. However, the behaviour of the posterior distributions of the clustering parameters changes significantly. In particular, there is no tension between the minimum masses of ET galaxies, supporting the hypothesis of a dust residual impacting the results from higher frequencies in the previous dataset. 

\begin{figure}
    \centering
    \includegraphics[width=.5\textwidth]{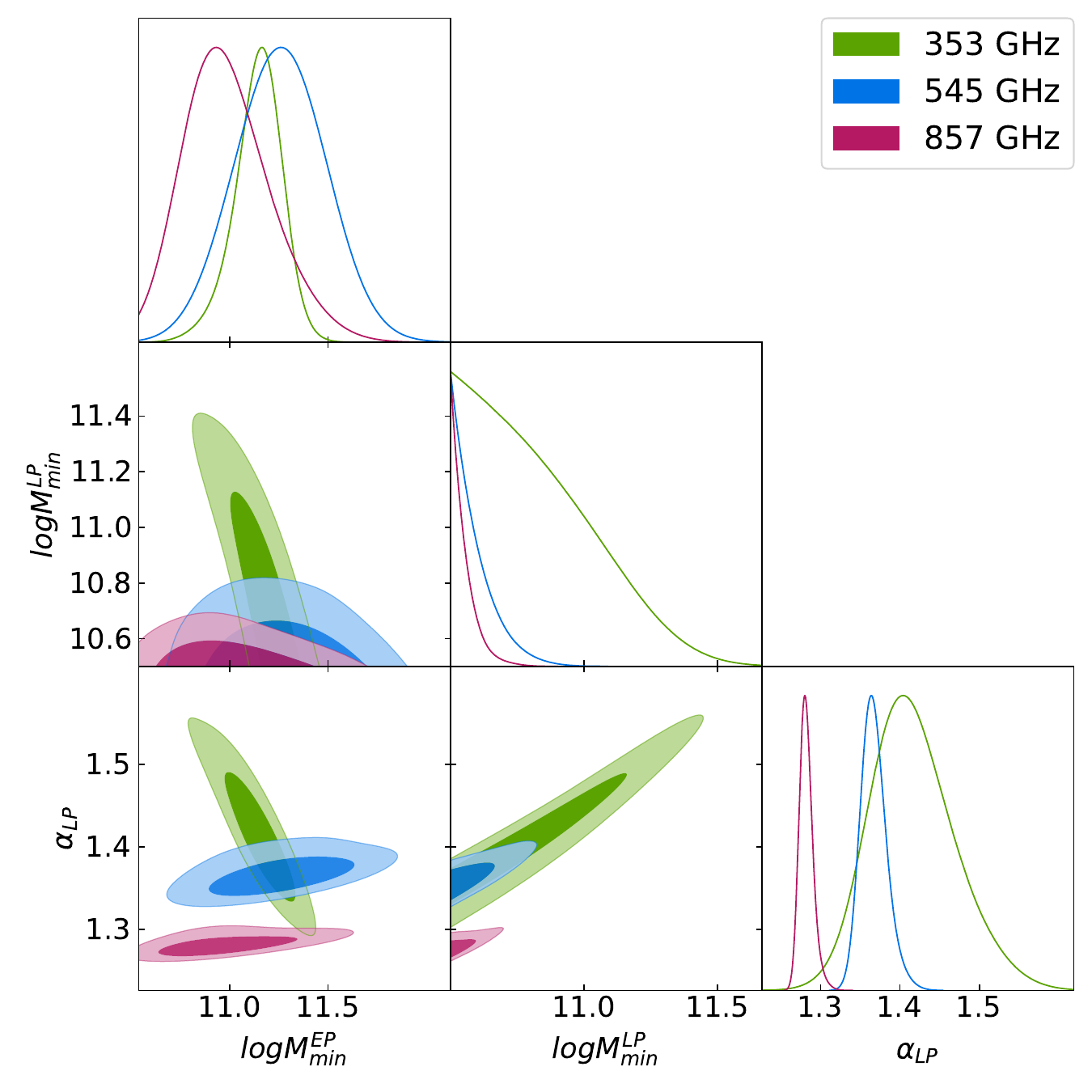}
    \caption{Same as Fig.~\ref{fig:single_freq_ds1} for the third L19 dataset.}
    \label{fig:single_freq_ds3}
\end{figure}

Finally, the third dataset, which used different thresholds for the three frequency channels, showed no significant differences from the second dataset (see Fig.~\ref{fig:single_freq_ds3}), except for the 353 GHz frequency channel which provided more constraining power due to more data retained.

This exploration highlights that the performance of our model in the multipole region where clustering terms dominate is highly sensitive to dust presence. Ensuring a dataset free from foreground contamination is crucial for testing our model and achieving reliable predictions.

\section{Extended analyses of the V18 dataset}\label{app:Spire}
\begin{figure*}[ht]
\centering
\includegraphics[width=1.\textwidth]{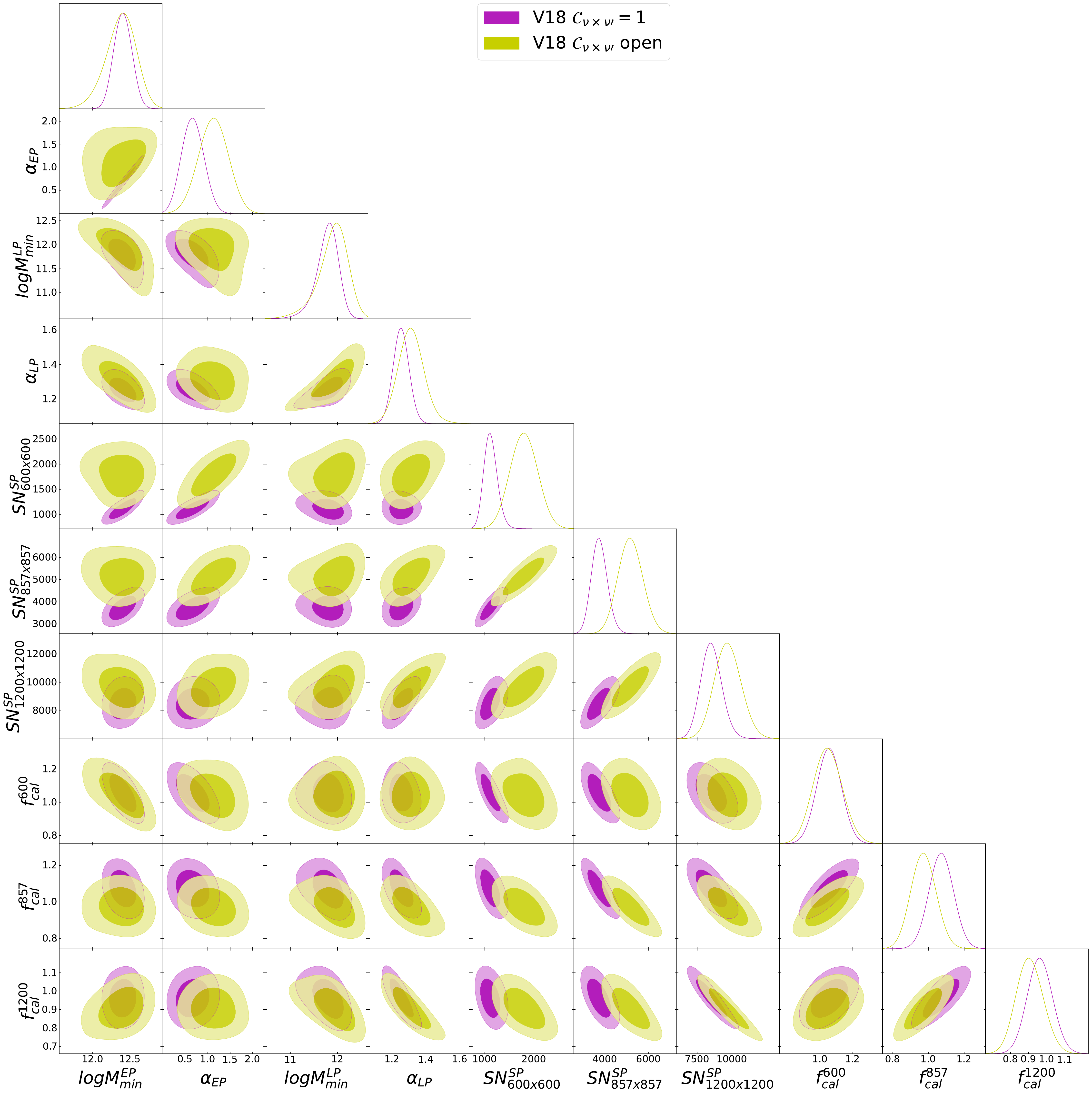}
\caption{\label{fig:SPIRE tot comp}Same as Fig.~\ref{fig:SPIRE clust best fit} for all the parameters sampled in the MCMC analysis of the V18 dataset.}
\end{figure*}

In Figure \ref{fig:SPIRE tot comp} we report the full triangle plot from the analysis of the V18 dataset described in Sec.~\ref{subsec:V18 model fit}. 

\section{Structure of the code, settings and parameters}\label{app:code}
%

In this Appendix we describe the structure and the features of the code used for this work\footnote{The code is made available at \url{https://github.com/giorgiazagatti/CIB_halomodel.git}}.
The code is composed of seven python modules:
\begin{itemize}[noitemsep,topsep=0pt]
    \item \texttt{cosmology.py} defines a class for the quantities needed to compute or read the matter power spectrum. If the user does not want to use the tabulated linear matter power spectrum, the module \texttt{matterPS.py} computes the linear matter power spectrum using \camb\ with the cosmology provided by the user.
    \item \texttt{utils.py} is the class which computes all the ingredients of the halo model, i.e., the halo mass function, the halo bias and the halo density profile.
    \item \texttt{HODS\_mod.py} defines a class for the computation of the number of central and satellite galaxies and the average galaxy number density for both ET and LT galaxies, employing the formalism of the halo occupation distribution described in this work.
    \item \texttt{spectraldependence.py} computes the emissivity functions for each frequency channel needed for the computation of the CIB power spectrum.
    \item \texttt{power\_spectrum.py} computes the CIB power spectra, both the clustering and the shot noise terms.
    \item \texttt{theory\_CIB.py} is the main module which has to be run for the computation of the CIB power spectrum.
\end{itemize}

The settings and the parameters needed for the computation of the CIB power spectrum are listed in the parameter file, which is organized in three main blocks:
\begin{itemize}
    \item the first block for the general settings;
    \item the second block specifies the experimental properties (\planck\,or SPIRE for the specific case of this study);
    \item the third block is devoted to the parameter settings (both cosmological and model parameters).
\end{itemize}

The first block of general settings allows to define the linear matter power spectrum and to choose the normalization of the power spectra ($C_\ell$'s or $D_\ell$'s), the multipole range for its computation, and the redshifts. Specifically, for the linear matter power spectrum there are three options:
\begin{enumerate}[noitemsep,topsep=0pt]
    \item use the linear matter power spectrum pre-computed using \camb\ and utilized within this study. In this case the code only reads the default linear matter power spectrum;
    \item use the linear matter power spectrum pre-computed by the user. In this case the user has to provide the path to their file;
    \item compute the linear matter power spectrum using a different cosmology with respect to the one employed in this work. In this case, the code computes the linear matter power spectrum using \camb.
\end{enumerate}

The second block defines the specific quantities of each experiment. The default parameter files contain \planck\,and SPIRE frequency channels, units of measure of the data, effective frequency for the CIB emission and the color corrections for each frequency channel.

The third block lists the parameters employed by the code and contains cosmological parameters, fixed parameters (such as the CMB temperature and the coefficients required for the conversion factors depending if the final power spectra are in CMB or flux units), clustering parameters related to the HOD and power spectra parameters like the shot noise levels, the correlation coefficients and the calibration factors for the different frequency channels.

\bibliographystyle{aat}
\interlinepenalty=10000
\bibliography{bibliography,Planck_bib}

\def\eprinttmppp@#1arXiv:@{#1}
\providecommand{\arxivlink[1]}{\href{http://arxiv.org/abs/#1}{arXiv:#1}}
\def\eprinttmp@#1arXiv:#2 [#3]#4@{\ifthenelse{\equal{#3}{x}}{\ifthenelse{
\equal{#1}{}}{\arxivlink{\eprinttmppp@#2@}}{\arxivlink{#1}}}{\arxivlink{#2}
  [#3]}}
\providecommand{\eprintlink}[1]{\eprinttmp@#1arXiv: [x]@}
\providecommand{\eprint}[1]{\eprintlink{#1}}
\providecommand{\adsurl}[1]{\href{#1}{ADS}}
\begin{thebibliography}{75}
\expandafter\ifx\csname natexlab\endcsname\relax\def\natexlab#1{#1}\fi

\bibitem[{Amblard {et~al.}(2011)}]{Amblard:2011gc}
Amblard, A. {et~al.}, {Sub-millimetre galaxies reside in dark matter halos with
  masses greater than $3\times 10^{11}$ solar masses}. 2011, Nature, 470, 510,
  \eprint{1101.1080}

\bibitem[{Asgari {et~al.}(2023)Asgari, Mead, \& Heymans}]{Asgari:2023mej}
Asgari, M., Mead, A.~J., \& Heymans, C., {The halo model for cosmology: a
  pedagogical review}. 2023, \eprint{2303.08752}

\bibitem[{Berlind {et~al.}(2003)Berlind, Weinberg, Benson, Baugh, Cole, Dave,
  Frenk, Jenkins, Katz, \& Lacey}]{Berlind:2002rn}
Berlind, A.~A., Weinberg, D.~H., Benson, A.~J., {et~al.}, {The Halo occupation
  distribution and the physics of galaxy formation}. 2003, Astrophys. J., 593,
  1, \eprint{astro-ph/0212357}

\bibitem[{Bethermin {et~al.}(2012)}]{Bethermin:2012ki}
Bethermin, M. {et~al.}, {A unified empirical model for infrared galaxy counts
  based on observed physical evolution of distant galaxies}. 2012, Astrophys.
  J. Lett., 757, L23, \eprint{1208.6512}

\bibitem[{Bhattacharya {et~al.}(2011)Bhattacharya, Heitmann, White, Lukic,
  Wagner, \& Habib}]{Bhattacharya:2010wy}
Bhattacharya, S., Heitmann, K., White, M., {et~al.}, {Mass Function Predictions
  Beyond LCDM}. 2011, Astrophys. J., 732, 122, \eprint{1005.2239}

\bibitem[{Bocquet {et~al.}(2020)Bocquet, Heitmann, Habib, Lawrence, Uram,
  Frontiere, Pope, \& Finkel}]{Bocquet:2020tes}
Bocquet, S., Heitmann, K., Habib, S., {et~al.}, {The Mira-Titan Universe. III.
  Emulation of the Halo Mass Function}. 2020, Astrophys. J., 901, 5,
  \eprint{2003.12116}

\bibitem[{Bond {et~al.}(1991)Bond, Cole, Efstathiou, \& Kaiser}]{Bond:1990iw}
Bond, J.~R., Cole, S., Efstathiou, G., \& Kaiser, N., {Excursion set mass
  functions for hierarchical Gaussian fluctuations}. 1991, Astrophys. J., 379,
  440

\bibitem[{Bullock {et~al.}(2001)Bullock, Kolatt, Sigad, Somerville, Kravtsov,
  Klypin, Primack, \& Dekel}]{Bullock:1999he}
Bullock, J.~S., Kolatt, T.~S., Sigad, Y., {et~al.}, {Profiles of dark haloes.
  Evolution, scatter, and environment}. 2001, Mon. Not. Roy. Astron. Soc., 321,
  559, \eprint{astro-ph/9908159}

\bibitem[{Cai {et~al.}(2013)Cai, Lapi, Xia, De~Zotti, Negrello, Gruppioni,
  Rigby, Castex, Delabrouille, \& Danese}]{Cai:2013wna}
Cai, Z.-Y., Lapi, A., Xia, J.-Q., {et~al.}, {A hybrid model for the evolution
  of galaxies and Active Galactic Nuclei in the infrared}. 2013, Astrophys. J.,
  768, 21, \eprint{1303.2335}

\bibitem[{Child {et~al.}(2018)Child, Habib, Heitmann, Frontiere, Finkel, Pope,
  \& Morozov}]{Child:2018skq}
Child, H.~L., Habib, S., Heitmann, K., {et~al.}, {Halo Profiles and the
  Concentration\textendash{}Mass Relation for a $\Lambda$CDM Universe}. 2018,
  Astrophys. J., 859, 55, \eprint{1804.10199}

\bibitem[{Cooray \& Sheth(2002)}]{Cooray:2002dia}
Cooray, A. \& Sheth, R.~K., {Halo Models of Large Scale Structure}. 2002, Phys.
  Rept., 372, 1, \eprint{astro-ph/0206508}

\bibitem[{Correa {et~al.}(2015)Correa, Wyithe, Schaye, \&
  Duffy}]{Correa:2015dva}
Correa, C.~A., Wyithe, J. S.~B., Schaye, J., \& Duffy, A.~R., {The accretion
  history of dark matter haloes \textendash{} III. A physical model for the
  concentration\textendash{}mass relation}. 2015, Mon. Not. Roy. Astron. Soc.,
  452, 1217, \eprint{1502.00391}

\bibitem[{Courtin {et~al.}(2011)Courtin, Rasera, Alimi, Corasaniti, Boucher, \&
  Fuzfa}]{Courtin:2010gx}
Courtin, J., Rasera, Y., Alimi, J.~M., {et~al.}, {Imprints of dark energy on
  cosmic structure formation: II) Non-Universality of the halo mass function}.
  2011, Mon. Not. Roy. Astron. Soc., 410, 1911, \eprint{1001.3425}

\bibitem[{Crocce {et~al.}(2010)Crocce, Fosalba, Castander, \&
  Gaztanaga}]{Crocce:2009mg}
Crocce, M., Fosalba, P., Castander, F.~J., \& Gaztanaga, E., {Simulating the
  Universe with MICE: The abundance of massive clusters}. 2010, Mon. Not. Roy.
  Astron. Soc., 403, 1353, \eprint{0907.0019}

\bibitem[{Despali {et~al.}(2016)Despali, Giocoli, Angulo, Tormen, Sheth, Baso,
  \& Moscardini}]{Despali:2015yla}
Despali, G., Giocoli, C., Angulo, R.~E., {et~al.}, {The universality of the
  virial halo mass function and models for non-universality of other halo
  definitions}. 2016, Mon. Not. Roy. Astron. Soc., 456, 2486,
  \eprint{1507.05627}

\bibitem[{Diemer \& Joyce(2019)}]{Diemer:2018vmz}
Diemer, B. \& Joyce, M., {An accurate physical model for halo concentrations}.
  2019, Astrophys. J., 871, 168, \eprint{1809.07326}

\bibitem[{Diemer \& Kravtsov(2015)}]{Diemer:2014gba}
Diemer, B. \& Kravtsov, A.~V., {A universal model for halo concentrations}.
  2015, Astrophys. J., 799, 108, \eprint{1407.4730}

\bibitem[{Duffy {et~al.}(2008)Duffy, Schaye, Kay, \&
  Dalla~Vecchia}]{Duffy:2008pz}
Duffy, A.~R., Schaye, J., Kay, S.~T., \& Dalla~Vecchia, C., {Dark matter halo
  concentrations in the Wilkinson Microwave Anisotropy Probe year 5 cosmology}.
  2008, Mon. Not. Roy. Astron. Soc., 390, L64, [Erratum:
  Mon.Not.Roy.Astron.Soc. 415, L85 (2011)], \eprint{0804.2486}

\bibitem[{Eales {et~al.}(1999)Eales, Lilly, Gear, Dunne, Bond, Hammer,
  Le~Fevre, \& Crampton}]{Eales:1998fn}
Eales, S., Lilly, S., Gear, W., {et~al.}, {The canada-uk deep submillimetre
  survey: first submillimetre images, the source counts, and resolution of the
  background}. 1999, Astrophys. J., 515, 518, \eprint{astro-ph/9808040}

\bibitem[{Eke {et~al.}(2001)Eke, Navarro, \& Steinmetz}]{Eke:2000av}
Eke, V.~R., Navarro, J.~F., \& Steinmetz, M., {The Power spectrum dependence of
  dark matter halo concentrations}. 2001, Astrophys. J., 554, 114,
  \eprint{astro-ph/0012337}

\bibitem[{Fixsen {et~al.}(1996)Fixsen, Cheng, Gales, Mather, Shafer, \&
  Wright}]{Fixsen:1996nj}
Fixsen, D.~J., Cheng, E.~S., Gales, J.~M., {et~al.}, {The Cosmic Microwave
  Background spectrum from the full COBE FIRAS data set}. 1996, Astrophys. J.,
  473, 576, \eprint{astro-ph/9605054}

\bibitem[{Foreman-Mackey {et~al.}(2013)Foreman-Mackey, Hogg, Lang, \&
  Goodman}]{Foreman-Mackey:2012any}
Foreman-Mackey, D., Hogg, D.~W., Lang, D., \& Goodman, J., {emcee: The MCMC
  Hammer}. 2013, Publ. Astron. Soc. Pac., 125, 306, \eprint{1202.3665}

\bibitem[{Gao {et~al.}(2004)Gao, White, Jenkins, Stoehr, \&
  Springel}]{Gao:2004au}
Gao, L., White, S. D.~M., Jenkins, A., Stoehr, F., \& Springel, V., {The
  Subhalo populations of lambda-CDM dark halos}. 2004, Mon. Not. Roy. Astron.
  Soc., 355, 819, \eprint{astro-ph/0404589}

\bibitem[{Hall {et~al.}(2010)}]{Hall:2009rv}
Hall, N.~R. {et~al.}, {Angular Power Spectra of the Millimeter Wavelength
  Background Light from Dusty Star-forming Galaxies with the South Pole
  Telescope}. 2010, Astrophys. J., 718, 632, \eprint{0912.4315}

\bibitem[{Hansen {et~al.}(2009)Hansen, Sheldon, Wechsler, \&
  Koester}]{Hansen:2007fy}
Hansen, S.~M., Sheldon, E.~S., Wechsler, R.~H., \& Koester, B.~P., {The Galaxy
  Content of SDSS Clusters and Groups}. 2009, Astrophys. J., 699, 1333,
  \eprint{0710.3780}

\bibitem[{Harris {et~al.}(2020)Harris, Millman, van~der Walt, Gommers,
  Virtanen, Cournapeau, Wieser, Taylor, Berg, Smith, Kern, Picus, Hoyer, van
  Kerkwijk, Brett, Haldane, del R{'{\i}}o, Wiebe, Peterson,
  G{'{e}}rard-Marchant, Sheppard, Reddy, Weckesser, Abbasi, Gohlke, \&
  Oliphant}]{harris2020array}
Harris, C.~R., Millman, K.~J., van~der Walt, S.~J., {et~al.}, Array programming
  with {NumPy}. 2020, Nature, 585, 357

\bibitem[{Howlett {et~al.}(2012)Howlett, Lewis, Hall, \&
  Challinor}]{Howlett:2012mh}
Howlett, C., Lewis, A., Hall, A., \& Challinor, A., {CMB power spectrum
  parameter degeneracies in the era of precision cosmology}. 2012, \jcap, 1204,
  027, \eprint{1201.3654}

\bibitem[{Hughes {et~al.}(1998)}]{Hughes:1998sx}
Hughes, D. {et~al.}, {High - redshift star formation in the Hubble Deep Field
  revealed by a submillimeter - wavelength survey}. 1998, Nature, 394, 241,
  \eprint{astro-ph/9806297}

\bibitem[{Hunter(2007)}]{Hunter:2007}
Hunter, J.~D., Matplotlib: A 2D graphics environment. 2007, Computing in
  Science \& Engineering, 9, 90

\bibitem[{Jenkins {et~al.}(2001)Jenkins, Frenk, White, Colberg, Cole, Evrard,
  Couchman, \& Yoshida}]{Jenkins:2000bv}
Jenkins, A., Frenk, C.~S., White, S. D.~M., {et~al.}, {The Mass function of
  dark matter halos}. 2001, Mon. Not. Roy. Astron. Soc., 321, 372,
  \eprint{astro-ph/0005260}

\bibitem[{Kwan {et~al.}(2013)Kwan, Bhattacharya, Heitmann, \&
  Habib}]{Kwan:2012nd}
Kwan, J., Bhattacharya, S., Heitmann, K., \& Habib, S., {Cosmic Emulation: The
  Concentration-Mass Relation for wCDM Universes}. 2013, Astrophys. J., 768,
  123, \eprint{1210.1576}

\bibitem[{Lagache {et~al.}(2020)Lagache, B\'ethermin, Montier, Serra, \&
  Tucci}]{Lagache:2019xto}
Lagache, G., B\'ethermin, M., Montier, L., Serra, P., \& Tucci, M., {Impact of
  polarised extragalactic sources on the measurement of CMB B-mode
  anisotropies}. 2020, Astron. Astrophys., 642, A232, \eprint{1911.09466}

\bibitem[{Lagache {et~al.}(2003)Lagache, Dole, \& Puget}]{Lagache:2002xq}
Lagache, G., Dole, H., \& Puget, J.~L., {Modeling the infrared galaxy evolution
  using a phenomenological approach}. 2003, Mon. Not. Roy. Astron. Soc., 338,
  555, \eprint{astro-ph/0209115}

\bibitem[{Lapi \& Cavaliere(2011)}]{Lapi:2010is}
Lapi, A. \& Cavaliere, A., {Dark Matter Halos: The Dynamical Basis of Effective
  Empirical Models}. 2011, Adv. Astron., 2011, 903429, \eprint{1010.2602}

\bibitem[{Lenz {et~al.}(2019)Lenz, Dor\'e, \& Lagache}]{Lenz:2019ugy}
Lenz, D., Dor\'e, O., \& Lagache, G., {Large-scale Maps of the Cosmic Infrared
  Background from Planck}. 2019, Astrophys. J., 883, 75, \eprint{1905.00426}

\bibitem[{Lewis(2019)}]{Lewis:2019xzd}
Lewis, A., {GetDist: a Python package for analysing Monte Carlo samples}. 2019,
  \eprint{1910.13970}

\bibitem[{Lewis {et~al.}(2000)Lewis, Challinor, \& Lasenby}]{Lewis:1999bs}
Lewis, A., Challinor, A., \& Lasenby, A., {Efficient computation of CMB
  anisotropies in closed FRW models}. 2000, \apj, 538, 473,
  \eprint{astro-ph/9911177}

\bibitem[{Ludlow {et~al.}(2016)Ludlow, Bose, Angulo, Wang, Hellwing, Navarro,
  Cole, \& Frenk}]{Ludlow:2016ifl}
Ludlow, A.~D., Bose, S., Angulo, R.~E., {et~al.}, {The
  mass\textendash{}concentration\textendash{}redshift relation of cold and warm
  dark matter haloes}. 2016, Mon. Not. Roy. Astron. Soc., 460, 1214,
  \eprint{1601.02624}

\bibitem[{Ludlow {et~al.}(2014)Ludlow, Navarro, Angulo, Boylan-Kolchin,
  Springel, Frenk, \& White}]{Ludlow:2013vxa}
Ludlow, A.~D., Navarro, J.~F., Angulo, R.~E., {et~al.}, {The
  mass\textendash{}concentration\textendash{}redshift relation of cold dark
  matter haloes}. 2014, Mon. Not. Roy. Astron. Soc., 441, 378,
  \eprint{1312.0945}

\bibitem[{Maccio' {et~al.}(2008)Maccio', Dutton, \& Bosch}]{Maccio:2008pcd}
Maccio', A.~V., Dutton, A.~A., \& Bosch, F. C. v.~d., {Concentration, Spin and
  Shape of Dark Matter Haloes as a Function of the Cosmological Model: WMAP1,
  WMAP3 and WMAP5 results}. 2008, Mon. Not. Roy. Astron. Soc., 391, 1940,
  \eprint{0805.1926}

\bibitem[{Mak {et~al.}(2017)Mak, Challinor, Efstathiou, Lagache, \&
  Lagache}]{Mak:2016ykk}
Mak, D. S.~Y., Challinor, A., Efstathiou, G., Lagache, G., \& Lagache, G.,
  {Measurement of CIB power spectra over large sky areas from Planck HFI maps}.
  2017, Mon. Not. Roy. Astron. Soc., 466, 286, \eprint{1609.08942}

\bibitem[{Maniyar {et~al.}(2018)Maniyar, B\'ethermin, \&
  Lagache}]{Maniyar:2018xfk}
Maniyar, A.~S., B\'ethermin, M., \& Lagache, G., {Star formation history from
  the cosmic infrared background anisotropies}. 2018, Astron. Astrophys., 614,
  A39, \eprint{1801.10146}

\bibitem[{Maniyar {et~al.}(2021)Maniyar, B\'ethermin, \&
  Lagache}]{Maniyar:2020tzw}
Maniyar, A.~S., B\'ethermin, M., \& Lagache, G., {Simple halo model formalism
  for the cosmic infrared background and its correlation with the thermal
  Sunyaev-Zel\textquoteright{}dovich effect}. 2021, Astron. Astrophys., 645,
  A40, \eprint{2006.16329}

\bibitem[{McClintock {et~al.}(2019)McClintock, Rozo, Becker, DeRose, Mao,
  McLaughlin, Tinker, Wechsler, \& Zhai}]{McClintock:2018uyf}
McClintock, T., Rozo, E., Becker, M.~R., {et~al.}, {The Aemulus Project II:
  Emulating the Halo Mass Function}. 2019, Astrophys. J., 872, 53,
  \eprint{1804.05866}

\bibitem[{Mo \& White(1996)}]{Mo:1995cs}
Mo, H.~J. \& White, S. D.~M., {An Analytic model for the spatial clustering of
  dark matter halos}. 1996, Mon. Not. Roy. Astron. Soc., 282, 347,
  \eprint{astro-ph/9512127}

\bibitem[{Murray {et~al.}(2021)Murray, Diemer, Chen, Neuhold, Schnapp, Peruzzi,
  Blevins, \& Engelman}]{Murray:2020dcd}
Murray, S.~G., Diemer, B., Chen, Z., {et~al.}, {TheHaloMod: An online
  calculator for the halo model}. 2021, Astron. Comput., 36, 100487,
  \eprint{2009.14066}

\bibitem[{Navarro {et~al.}(1997)Navarro, Frenk, \& White}]{Navarro:1996gj}
Navarro, J.~F., Frenk, C.~S., \& White, S. D.~M., {A Universal density profile
  from hierarchical clustering}. 1997, Astrophys. J., 490, 493,
  \eprint{astro-ph/9611107}

\bibitem[{Neto {et~al.}(2007)Neto, Gao, Bett, Cole, Navarro, Frenk, White,
  Springel, \& Jenkins}]{Neto:2007vq}
Neto, A.~F., Gao, L., Bett, P., {et~al.}, {The statistics of lambda CDM Halo
  Concentrations}. 2007, Mon. Not. Roy. Astron. Soc., 381, 1450,
  \eprint{0706.2919}

\bibitem[{Okoli \& Afshordi(2016)}]{Okoli:2015dta}
Okoli, C. \& Afshordi, N., {Concentration, Ellipsoidal Collapse, and the
  Densest Dark Matter haloes}. 2016, Mon. Not. Roy. Astron. Soc., 456, 3068,
  \eprint{1510.03868}

\bibitem[{Peacock(2007)}]{Peacock:2007cw}
Peacock, J.~A., {Testing anthropic predictions for Lambda and the CMB
  temperature}. 2007, Mon. Not. Roy. Astron. Soc., 379, 1067,
  \eprint{0705.0898}

\bibitem[{Peacock \& Smith(2000)}]{Peacock:2000qk}
Peacock, J.~A. \& Smith, R.~E., {Halo occupation numbers and galaxy bias}.
  2000, Mon. Not. Roy. Astron. Soc., 318, 1144, \eprint{astro-ph/0005010}

\bibitem[{{\sorthelp{Planck Collaboration 2011R}}{Planck Collaboration
  XVIII}(2011)}]{planck2011-6.6}
{\sorthelp{Planck Collaboration 2011R}}{Planck Collaboration XVIII},
  {\textit{Planck} early results. XVIII. The power spectrum of cosmic infrared
  background anisotropies}. 2011, \aap, 536, A18, \eprint{1101.2028}

\bibitem[{{\sorthelp{Planck Collaboration 2014ZE}}{Planck Collaboration
  XXX}(2014)}]{planck2013-pip56}
{\sorthelp{Planck Collaboration 2014ZE}}{Planck Collaboration XXX},
  {\textit{Planck} 2013 results. XXX. Cosmic infrared background measurements
  and implications for star formation}. 2014, \aap, 571, A30,
  \eprint{1309.0382}

\bibitem[{{\sorthelp{Planck Collaboration 2015H}}{Planck Collaboration
  VIII}(2016)}]{planck2014-a09}
{\sorthelp{Planck Collaboration 2015H}}{Planck Collaboration VIII},
  {\textit{Planck} 2015 results. VIII. High Frequency Instrument data
  processing: Calibration and maps}. 2016, \aap, 594, A8, \eprint{1502.01587}

\bibitem[{Prada {et~al.}(2012)Prada, Klypin, Cuesta, Betancort-Rijo, \&
  Primack}]{Prada:2011jf}
Prada, F., Klypin, A.~A., Cuesta, A.~J., Betancort-Rijo, J.~E., \& Primack, J.,
  {Halo concentrations in the standard LCDM cosmology}. 2012, Mon. Not. Roy.
  Astron. Soc., 423, 3018, \eprint{1104.5130}

\bibitem[{Press \& Schechter(1974)}]{Press:1973iz}
Press, W.~H. \& Schechter, P., {Formation of galaxies and clusters of galaxies
  by selfsimilar gravitational condensation}. 1974, Astrophys. J., 187, 425

\bibitem[{Puget {et~al.}(1996)Puget, Abergel, Bernard, Boulanger, Burton,
  Desert, \& Hartmann}]{Puget:1996fx}
Puget, J.~L., Abergel, A., Bernard, J.~P., {et~al.}, {Tentative detection of a
  cosmic far - infrared background with COBE}. 1996, Astron. Astrophys., 308,
  L5

\bibitem[{Reed {et~al.}(2007)Reed, Bower, Frenk, Jenkins, \&
  Theuns}]{Reed:2006rw}
Reed, D., Bower, R., Frenk, C., Jenkins, A., \& Theuns, T., {The halo mass
  function from the dark ages through the present day}. 2007, Mon. Not. Roy.
  Astron. Soc., 374, 2, \eprint{astro-ph/0607150}

\bibitem[{Saunders {et~al.}(1990)Saunders, Rowan-Robinson, Lawrence,
  Efstathiou, Kaiser, Ellis, \& Frenk}]{Saunders:1990kb}
Saunders, W., Rowan-Robinson, M., Lawrence, A., {et~al.}, {The 60-micron and
  far-infrared luminosity functions of IRAS galaxies}. 1990, Mon. Not. Roy.
  Astron. Soc., 242, 318

\bibitem[{Scoccimarro {et~al.}(2001)Scoccimarro, Sheth, Hui, \&
  Jain}]{Scoccimarro:2000gm}
Scoccimarro, R., Sheth, R.~K., Hui, L., \& Jain, B., {How many galaxies fit in
  a halo? Constraints on galaxy formation efficiency from spatial clustering}.
  2001, Astrophys. J., 546, 20, \eprint{astro-ph/0006319}

\bibitem[{Seljak(2000)}]{Seljak:2000gq}
Seljak, U., {Analytic model for galaxy and dark matter clustering}. 2000, Mon.
  Not. Roy. Astron. Soc., 318, 203, \eprint{astro-ph/0001493}

\bibitem[{Sheth \& Tormen(1999)}]{Sheth:1999mn}
Sheth, R.~K. \& Tormen, G., {Large scale bias and the peak background split}.
  1999, Mon. Not. Roy. Astron. Soc., 308, 119, \eprint{astro-ph/9901122}

\bibitem[{Smail {et~al.}(1997)Smail, Ivison, \& Blain}]{Smail:1997wn}
Smail, I., Ivison, R.~J., \& Blain, A.~W., {A Deep submillimeter survey of
  lensing clusters: A New window on galaxy formation and evolution}. 1997,
  Astrophys. J. Lett., 490, L5, \eprint{astro-ph/9708135}

\bibitem[{Tinker {et~al.}(2008)Tinker, Kravtsov, Klypin, Abazajian, Warren,
  Yepes, Gottlober, \& Holz}]{Tinker:2008ff}
Tinker, J.~L., Kravtsov, A.~V., Klypin, A., {et~al.}, {Toward a halo mass
  function for precision cosmology: The Limits of universality}. 2008,
  Astrophys. J., 688, 709, \eprint{0803.2706}

\bibitem[{Tinker {et~al.}(2010{\natexlab{a}})Tinker, Robertson, Kravtsov,
  Klypin, Warren, Yepes, \& Gottlober}]{Tinker:2010my}
Tinker, J.~L., Robertson, B.~E., Kravtsov, A.~V., {et~al.}, {The Large Scale
  Bias of Dark Matter Halos: Numerical Calibration and Model Tests}.
  2010{\natexlab{a}}, Astrophys. J., 724, 878, \eprint{1001.3162}

\bibitem[{Tinker {et~al.}(2010{\natexlab{b}})Tinker, Wechsler, \&
  Zheng}]{Tinker:2009jp}
Tinker, J.~L., Wechsler, R.~H., \& Zheng, Z., {Interpreting the Clustering of
  Distant Red Galaxies}. 2010{\natexlab{b}}, Astrophys. J., 709, 67,
  \eprint{0902.1748}

\bibitem[{Tinker \& Wetzel(2010)}]{Tinker:2009mx}
Tinker, J.~L. \& Wetzel, A.~R., {What Does Clustering Tell Us About the Buildup
  of the Red Sequence?} 2010, Astrophys. J., 719, 88, \eprint{0909.1325}

\bibitem[{Valtchanov(2017)}]{Valtchanov:2017}
Valtchanov, I.~E. 2017, {SPIRE Handbook v3.1, Herschel Explanatory Supplement
  vol. IV}, hERSCHEL-HSC-DOC-0798

\bibitem[{Viero {et~al.}(2013)}]{Viero:2012uq}
Viero, M.~P. {et~al.}, {HerMES: Cosmic Infrared Background Anisotropies and the
  Clustering of Dusty Star-Forming Galaxies}. 2013, Astrophys. J., 772, 77,
  \eprint{1208.5049}

\bibitem[{Viero {et~al.}(2019)}]{Viero:2018mdn}
Viero, M.~P. {et~al.}, {Measurements of the Cross Spectra of the Cosmic
  Infrared and Microwave Backgrounds from 95 to 1200 GHz}. 2019, Astrophys. J.,
  881, 96, \eprint{1810.10643}

\bibitem[{Wang {et~al.}(2011)Wang, Navarro, Frenk, White, Springel, Jenkins,
  Helmi, Ludlow, \& Vogelsberger}]{Wang:2010ik}
Wang, J., Navarro, J.~F., Frenk, C.~S., {et~al.}, {Assembly History and
  Structure of Galactic Cold Dark Matter Halos}. 2011, Mon. Not. Roy. Astron.
  Soc., 413, 1373, \eprint{1008.5114}

\bibitem[{Warren {et~al.}(2006)Warren, Abazajian, Holz, \&
  Teodoro}]{Warren:2005ey}
Warren, M.~S., Abazajian, K., Holz, D.~E., \& Teodoro, L., {Precision
  determination of the mass function of dark matter halos}. 2006, Astrophys.
  J., 646, 881, \eprint{astro-ph/0506395}

\bibitem[{Watson {et~al.}(2013)Watson, Iliev, D'Aloisio, Knebe, Shapiro, \&
  Yepes}]{Watson:2012mt}
Watson, W.~A., Iliev, I.~T., D'Aloisio, A., {et~al.}, {The halo mass function
  through the cosmic ages}. 2013, Mon. Not. Roy. Astron. Soc., 433, 1230,
  \eprint{1212.0095}

\bibitem[{Xia {et~al.}(2012)Xia, Negrello, Lapi, De~Zotti, Danese, \&
  Viel}]{Xia:2011dt}
Xia, J.-Q., Negrello, M., Lapi, A., {et~al.}, {Clustering of sub-millimeter
  galaxies in a self-regulated baryon collapse model}. 2012, Mon. Not. Roy.
  Astron. Soc., 422, 1324, \eprint{1111.4212}

\bibitem[{Zehavi {et~al.}(2005)}]{SDSS:2004oes}
Zehavi, I. {et~al.}, {The Luminosity and color dependence of the galaxy
  correlation function}. 2005, Astrophys. J., 630, 1, \eprint{astro-ph/0408569}

\end{thebibliography}

\end{document}